\documentclass{pasj00}

\begin{document}
\SetRunningHead{Y. Ikejiri et al.}{Photopolarimetric Monitoring of Blazars}
\Received{2011/01/19}
\Accepted{2011/03/07}
\Published{2011/06/25}

\title{Photopolarimetric Monitoring of Blazars \\
in the Optical and Near-Infrared Bands with the Kanata Telescope.I. \\
Correlations between Flux, Color, and Polarization}

\author{
Yuki \textsc{Ikejiri}\altaffilmark{1},
Makoto \textsc{Uemura}\altaffilmark{2},
Mahito \textsc{Sasada}\altaffilmark{1},
Ryosuke \textsc{Ito}\altaffilmark{1},
Masayuki \textsc{Yamanaka}\altaffilmark{1},
Kiyoshi \textsc{Sakimoto}\altaffilmark{1},
Akira \textsc{Arai}\altaffilmark{3},
Yasushi \textsc{Fukazawa}\altaffilmark{1},
Takashi \textsc{Ohsugi}\altaffilmark{2},
Koji S. \textsc{Kawabata}\altaffilmark{2},
Michitoshi \textsc{Yoshida}\altaffilmark{2},
Shuji \textsc{Sato}\altaffilmark{4},\\ and
Masaru \textsc{Kino}\altaffilmark{4}}

\altaffiltext{1}{Department of Physical Science, Hiroshima University,
Kagamiyama 1-3-1, \\Higashi-Hiroshima 739-8526}
\altaffiltext{2}{Hiroshima Astrophysical Science Center, Hiroshima
  University, Kagamiyama 1-3-1, \\Higashi-Hiroshima 739-8526}
\email{uemuram@hiroshima-u.ac.jp}
\altaffiltext{3}{Faculty of Science, Kyoto Sangyo University, Motoyama, Kamigamo, Kita-Ku, Kyoto-City 603-8555}
\altaffiltext{4}{Department of Physics, Nagoya University, Furo-cho,
Chikusa-ku, Nagoya 464-8602}

%


\maketitle

\begin{abstract}
We report on the correlation between the flux, color and polarization
variations on time scales of days--months in blazars, and discuss
their universal aspects. We performed monitoring of 42 blazars in the
optical and near-infrared bands from 2008 to 2010 using TRISPEC
attached to the ``Kanata'' 1.5-m telescope. We found that 28 blazars
exhibited ``bluer-when-brighter'' trends in their whole or a part of
time-series data sets. This corresponds to 88$\%$ of objects that were
observed for $>10$ days. Thus, our observation unambiguously confirmed
that the ``bluer-when-brighter'' trend is common in the emission from
blazar jets. This trend was apparently generated by a variation 
component with a constant and relatively blue color and an underlying
red component.  Prominent short-term flares on time scales of
days--weeks tended to exhibit a spectral hysteresis; their rising
phases were bluer than their decay phases around the flare maxima. In
contrast to the strong flux--color correlation, the correlation of the
flux and polarization degree was relatively weak; only 10 objects
showed significant positive correlations. Rotations of polarization
were detected only in three objects: PKS~1510$-$089, 3C~454.3, and
PKS~1749$+$096, and possibly in S5~0716+714. We also investigated the
dependence of the degree of variability on the luminosity and the
synchrotron peak frequency, $\nu_{\rm peak}$. As a result, we found
that lower luminosity and higher $\nu_{\rm peak}$ objects had smaller
variations in their amplitudes both in the flux, color, and polarization
degree. Our observation suggests the presence of several distinct
emitting sources, which have different variation time-scales, colors,
and polarizations. We propose that the energy injection by, for
example, internal shocks in relativistic shells is a major factor for 
blazar variations on time scales of both days and months. 
\end{abstract}

\section{Introduction}

Blazars are a subgroup of active galactic nuclei (AGN). They have
relativistic jets that are believed to be directed along the line of
sight (e.g.~\cite{bla78blazar}). The emission from blazars can be
detected in a very wide range of wavelengths from radio to TeV
$\gamma$-ray regions, with strong variability on various time-scales
(e.g.~\cite{huf92variation}). Their spectral energy distribution (SED)
is characterized by non-thermal continuum spectra that consist of low
and high-energy components. The low-energy component is believed to be
synchrotron radiation from relativistic electrons in the jet. The
emission of this component is highly polarized (\cite{Angel1980};
\cite{Mead1990}). The origin of the high-energy component is not fully
understood.  The most plausible scenario is that it is emission
via inverse Compton scattering of synchrotron emission and/or
external photons (e.g.~\cite{der92ec}; \cite{sik94ec}).

Blazars are classified into two categories: BL~Lac objects and flat
spectrum radio quasars (FSRQs) (\cite{Angel1980}). Non-thermal
continuum emission is dominant in BL~Lac objects, while broad emission
lines from AGN characterize FSRQs. BL~Lac objects are further  
classified into three subtypes based on the peak frequency of
synchrotron emission: high-, intermediate-, and low-energy peaked BL
Lac objects (HBL, IBL, and LBL).  The peak frequency lies higher than,
around, and lower than the optical band in HBL, IBL, and LBL,
respectively. 

Multi-wavelength observations are essential to study blazars
because of their emission in a wide range of wavelengths. 
The multi-wavelength study of blazars has recently entered a new era
owing to $\gamma$-ray observations obtained with the Fermi
satellite (\cite{abd10class}; \cite{abd10agn}). The origin of the
high-energy component, for example, is expected to be revealed by the
study of temporal variations in SEDs obtained with Fermi. Among the
multi-wavelength study, optical data is important to understand the
SED variations, because the optically-thin synchrotron emission in this
band allows us to estimate basic physical parameters in the emitting
region. For follow-up ground-based observations of Fermi blazars, we
executed photopolarimetric monitoring of 42 
blazars from 2008 to 2010. We obtained simultaneous optical and
near-infrared (NIR) data using the ``Kanata'' 1.5-m telescope. Our
observation provides one of the largest data sets of blazars in terms
of variations in color and polarization on a time scale of
days--months (\cite{ike09kanata}). In this paper, we report on the
results of simple correlation studies between the flux (,or
luminosity), color, polarization degree ($PD$), and polarization angle
($PA$). We also discuss the implications for the mechanism of time
variations in blazars obtained from our observations. A detailed study
of polarization and the data, itself, will be published in a forthcoming
paper. 

The behavior of colors gives us a clue to understand the mechanism
of time variations in blazars. \citet{Carini1992} report a possible 
feature that both BL~Lac and OJ~287 became bluer when they were
brighter (also see, \cite{Clements2001}). \citet{Ghisellini1997}
observed S5~0716$+$714, and found that a bluer-when-brighter trend
was seen only in its faint state. Its bluer-when-brighter trend was
later confirmed in variations on a time scale shorter than days in its
bright state (\cite{Wu2007}; \cite{Sasada2008}). \citet{Villata2002}
and \citet{Villata2004} report that the bluer-when-brighter trend in
BL~Lac was only observed in short-term variations, and not prominent
in long-term ones. \citet{Raiteri2001} performed multi-band
photometric observations of AO~0235$+$164 for four years, and found a
bluer-when-brighter trend. As reported in those past studies, the
bluer-when-brighter trend has been a well-observed feature in blazars,
while its universality has not been established. 

Systematic observations of multiple sources have also been performed 
in order to establish a characteristic feature of color variations
in blazars. \citet{Ghosh2000} performed observations of five
blazars. Their observation showed that 3C~66A only exhibited a
bluer-when-brighter trend, while the other objects showed no
significant correlation of the flux and color. \citet{Gu2006} 
investigated variations in color of eight blazars. The
bluer-when-brighter trend was confirmed in five blazars. Among the
other three objects, 3C~345 showed no correlation of the flux and
color, and 3C~454.3 and PKS~0420$-$01 showed a reddening trend when
they brightened. This ``redder-when-brighter'' trend in 3C~454.3 was 
confirmed in later observations (\cite{Raiteri2008},
\cite{Villata2006}). Thus, it is currently unclear whether the
bluer-when-brighter trend is universal in blazars.

As well as the color variation, the polarimetric variation in the
optical band has been extensively investigated in blazars.  The
temporal variation in polarization has, in general, been considered to 
be erratic in blazars (e.g.~\cite{Moore1982}). On the other hand, a
few cases have been reported in which the increase in $PD$ was
associated with flares of the total flux. For example, $PD$ of Mrk~421
increased to $\sim 14$~\% associated with its outburst in 1997
(\cite{Tosti1998}). A significant correlation of the flux and $PD$ was
also detected in AO~0235$+$164 in 2006 (\cite{Hagen-Thorn2008}). Most
recently, several rotation episodes of $PA$ have been found during 
flares (\cite{mar08bllac}; \cite{mar10pks1510}; \cite{fer103c279};
\cite{jor103c454}). In general, the number of polarimetric
observations has been much smaller than that of color observations. 
High-frequency, long-term polarimetric observations are required in
order to find universal aspects that are possibly present. 

The aim of this paper is to investigate universal aspects in blazar
variability in the optical--NIR bands. We search for them with a
simple correlation study between the flux, color, and polarization. In
section~2, we describe the observation method and reduction
processes. In section~3, first, we shortly introduce the basic
properties of our sample (subsection~3.1). Then, we report on the
results of our correlation studies of the flux and color
(subsection~3.2), and the flux and polarization (subsection~3.3). We
also search for rotation events of polarization in subsection~3.3. 
In subsection~3.4, we report on the dependence of the degree of
variability on different luminosities and synchrotron peak
frequencies. In section~4, we discuss the implication to the mechanism
of blazar variability obtained from our observation. Finally, we
summarize our findings in section~5.

\section{Sample and Observation}
\footnotesize
\begin{longtable}{p{65pt}rp{6pt}p{25pt}p{85pt}p{20pt}p{20pt}p{20pt}p{8pt}p{20pt}}
 \caption{Observation log and list of comparison
  stars.}\label{tab:comp_list} 
   \hline
   Object & Observation period & N & $T_{\rm exp}$ & 
   \multicolumn{6}{c}{Comparison star} \\
   & & & & Coordinate & $V$ & $J$ & $K_s$ & Ref. & $\sigma_V$ \\
   (1) & (2) & (3) & (4) & (5) & (6) & (7) & (8) & (9) & (10)\\

   \hline
   \endhead
   \hline
  \endfoot

 \multicolumn{10}{l}{(1)Object name. (2)Period of our monitoring.
   (3)Number of observations.}\\
 \multicolumn{10}{l}{(4)Exposure times for each $V$, $J$, and
   $K_s$-band image. (5)Coordinate of comparison
   stars.}\\
 \multicolumn{10}{l}{(6),(7),(8)$V$, $J$, and $Ks$ band magnitudes of comparison
   stars.}\\
 \multicolumn{10}{l}{(9)References for the magnitudes of comparison
   stars.}\\
 \multicolumn{10}{l}{(10) Standard deviations of the $V$-mag of the comparison
   stars during our monitoring period.  The magnitudes were measured}\\
 \multicolumn{10}{l}{with neighbor check stars in the same field. Note
 that the check stars are fainter than the blazars in some cases,}\\ 
 \multicolumn{10}{l}{and hence, $\sigma_V$ do not indicate the
   systematic error of the photometry of the blazars.  It just}\\
 \multicolumn{10}{l}{indicates the constancy level of the comparison stars.}\\ 
 \multicolumn{9}{l}{[1]\citet{Skiff2007}; [2]\citet{Villata1998};
 [3]\citet{Gonzalez2001}; [4]\citet{Adelman2008}; }\\
 \multicolumn{9}{l}{[5]\citet{Doroshenko2005}; [6]\citet{Hog2000};
 [7]\citet{McGimsey1977}.}\\
 \multicolumn{9}{l}{ $^*$We measured the $V$-magnitude of our
   comparison star using the stars listed in the reference.}\\
 \multicolumn{9}{l}{$\dag$The NIR magnitudes were referred from the
   same reference for the $V$-band magnitude.}\\
 \multicolumn{9}{l}{No symbol: The $V$-magnitude of the comparison star
  was quoted from the reference, and the NIR}\\
 \multicolumn{9}{l}{magnitudes were from the 2MASS catalog.}\\
 \endlastfoot
 QSO J0324$+$3410& 08 Nov.20 --- 09 Oct.14 &  2 & 93,11,1 &
 03 24 33.78 $+$34 10 53.7 & 13.322 & 11.232 & 10.589& [6]$^*$ & 0.010\\
 4C 14.23    & 09 Oct.15 --- 09 Dec.07 & 16 & 280,6,0 &
 07 25 20.89 $+$14 25 04.0 & 14.805 & 13.485 & 13.192& [4] & 0.014\\
 PKS 1222$+$216 & 09 Apr.22 --- 09 Apr.22 &  1 & 153,20,0 &
 12 24 41.03 $+$21 21 26.5 & 15.660 & 14.024 & 13.500 & [1] & 0.007\\
 3EG J1236$+$0457& 09 Jan.02 --- 09 Jan.08 &  5 & 123,10,1 &
 12 39 30.11 $+$04 39 52.6 & 14.095 & 12.942 & 12.638 &[4] & 0.026\\
 3C 279     & 08 Dec.09 --- 10 Jan.24 & 64 & 200,10,1 &
 12 56 16.90 $-$05 50 43.0 & 13.660 & 12.377 & 11.974 & [1]$^*$ & 0.006\\
 PKS 0215$+$015 & 08 Sep.09 --- 09 Aug.25 &  6 & 103,20,8 &
 02 17 49.22 $+$01 48 28.0 & 11.772 & 11.320 & 11.046 & [6] & 0.056\\
 QSO 0454$-$234 & 08 Oct.17 --- 09 Dec.16 & 53 & 153,5,1 &
 04 57 00.74 $-$23 26 05.9 & 12.184 & 10.849 & 10.364 & [6] & 0.034\\
 PKS 1510$-$089 & 09 Jan.12 --- 10 Jan.21 & 57 & 200,10,1 &
 15 12 53.19 $-$09 03 43.6 & 13.282 & 12.205 & 11.919 & [3]$\dag$ & 0.013\\
 PKS 1749$+$096 & 08 Jul.18 --- 09 Sep.10 & 78 & 200,5,1 &
 17 51 37.28 $+$09 39 07.1 & 11.950 & 10.252 & 9.740 & [1] & 0.037\\
 OJ 287     & 08 May 26 --- 10 Jan.31 & 174 & 103,15,1 &
 08 54 59.01 $+$20 02 57.1 & 13.986 & 12.811 & 12.445 & [3] & 0.026\\
 3C 273     & 08 Dec.11 --- 10 Jan.29 & 77 &  53,8,1 &
 12 29 08.34 $+$02 00 17.2 & 12.718 & 11.345 & 10.924 & [3] & 0.039\\
 AO 0235$+$164 & 08 Aug.12 --- 09 Jul.22 & 70 & 123,8,1 &
 02 38 32.31 $+$16 35 59.7 & 12.720 & 11.248 & 10.711 & [3]$\dag$ & 0.023\\
 OJ 49     & 08 Oct.31 --- 10 Jan.25 & 52 & 183,15,1 &
 08 32 00.74 $+$04 32 02.5 & 13.550 & 12.475 & 12.189 & [1] & 0.026\\
 MisV1436    & 08 Dec.17 --- 10 Jan.19 & 103 & 123,10,1 &
 01 36 42.49 $+$47 51 03.4 & 14.035 & 12.223 & 11.922 & [6]$^*$ & 0.034\\
 PKS 1502$+$106 & 08 Aug.09 --- 10 Jan.29 & 80 & 200,20,1 &
 15 04 36.51 $+$10 28 47.0 & 15.335 & 14.117 & 13.678 & [4] & 0.055\\
 3C 454.3    & 08 May 27 --- 10 Jan.28 & 262 & 123,10,1 &
 22 53 58.18 $+$16 09 06.9 & 13.587 & 11.858 & 11.241 & [3]$\dag$ & 0.037\\
 PKS 0754$+$100 & 08 Nov.05 --- 10 Jan.26 & 29 & 123,15,1 &
 07 57 16.12 $+$09 55 47.8 & 13.000 & 11.852 & 11.496 & [1]$\dag$ & 0.023\\
 BL Lac     & 08 May 26 --- 10 Jan.28 & 196 & 123,4,1 &
 22 02 45.45 $+$42 16 35.4 & 12.938 & 9.817 & 8.811 & [3]$\dag$ & 0.020\\
 QSO 0948$+$002 & 09 Mar.30 --- 09 Apr.10 &  3 & 183,20,1 &
 09 49 10.19 $+$00 21 39.5 & 14.944 & 13.500 & 13.139 & [4] & 0.029\\
 S4 0954$+$658 & 08 Dec.03 --- 09 Dec.14 &  5 & 183,12,1 &
 09 58 50.44 $+$65 32 09.1 & 14.610 & 12.927 & 12.455 & [1] & 0.045\\
 S5 1803$+$784 & 08 Jul.08 --- 09 Oct.18 & 35 & 123,15,3 &
 17 59 52.6 $+$78 28 50.9 & 13.052 & 11.761 & 11.381 & [1]$^*$ & 0.027\\
 RX J1542.8$+$612& 09 May 13 --- 10 Jan.28 & 65 & 203,5,0 &
 15 42 40.04 $+$61 30 25.1 & 14.000 & 10.354 & 9.593 & [4] & 0.016\\
 OQ 530     & 08 Jul.15 --- 08 Sep.10 &  3 & 123,15,3 &
 14 19 39.70 $+$54 21 55.0 & 15.961 & 13.873 & 13.131 & [4] & 0.016\\
 PKS 0048$-$097 & 08 Oct.02 --- 09 Sep.23 & 46 & 123,10,1 &
 00 50 47.23 $-$09 30 15.9 & 14.120 & 12.455 & 11.854 & [1] & 0.020\\
 ON 231     & 08 Dec.16 --- 10 Jan.28 & 14 & 153,15,1 &
 12 21 33.67 $+$28 13 04.0 & 12.080 & 10.921 & 10.597 & [1] & 0.068\\
 S2 0109$+$224 & 08 Jul.31 --- 09 Oct.29 & 74 & 123,10,1 &
 01 12 03.28 $+$22 43 26.7 & 12.510 & 11.245 & 10.886 & [1] & 0.031\\
 S5 0716$+$714 & 08 May 26 --- 10 Jan.31 & 242 & 83,15,1 &
 07 21 52.18 $+$71 18 16.1 & 12.475 & 11.320 & 10.980 & [3] & 0.021\\
 3EG 1052$+$571 & 08 Oct.31 --- 09 Oct.13 &  3 & 103,5,3 &
 10 58 37.99 $+$56 25 21.5 & 11.752 & 10.531 & 10.207 & [4] & 0.040\\
 3C 371     & 08 Jul.07 --- 09 Dec.07 & 101 &  43,8,1 &
 18 06 53.72 $+$69 45 37.4 & 12.588 & 12.219 & 11.856 & [7]$^*$ & 0.023\\
 3C 66A     & 08 Jul.09 --- 10 Jan.31 & 227 & 123,15,1 &
 02 22 55.12 $+$43 03 15.5 & 12.809 & 12.371 & 12.282 & [3] & 0.027\\
 PG 1553$+$113 & 08 Jul.08 --- 10 Jan.17 & 22 & 63,10,1 &
 15 55 52.28 $+$11 13 18.3 & 13.828 & 12.539 & 12.139 & [5] & 0.022\\
 ON 325     & 08 May 26 --- 09 Dec.26 & 47 & 133,20,1 &
 12 17 44.51 $+$30 09 43.6 & 14.960 & 13.674 & 13.232 & [1] & 0.058\\
 PKS 0422$+$004 & 08 Sep.04 --- 09 Nov.11 & 42 & 133,10,1 &
 04 24 42.42 $+$00 37 10.8 & 12.510 & 11.217 & 10.899 & [1] & 0.044\\
 H 1722$+$119  & 08 Jul.11 --- 09 Oct.17 & 28 & 63,10,1 &
 17 25 05.27 $+$11 52 11.0 & 13.210 & 11.308 & 10.710 & [1] & 0.024\\
 PKS 2155$-$304 & 08 Jul.09 --- 09 Dec.22 & 137 &  43,5,1 &
 21 59 02.47 $-$30 10 46.2 & 12.050 & 10.775 & 10.365 & [1] & 0.031\\
 1ES 2344$+$514 & 08 Jul.07 --- 09 Oct.11 & 17 &  93,6,5 &
 23 47 02.24 $+$51 43 17.6 & 12.610 & 11.421 & 11.117 & [1] & 0.030\\
 1ES 0806$+$524 & 08 Oct.30 --- 10 Jan.31 & 18 & 123,10,1 &
 08 09 40.65 $+$52 19 17.2 & 13.040 & 11.417 & 10.867 & [1] & 0.046\\
 Mrk 421    & 08 Jun.30 --- 09 Mar.31 & 42 & 63,15,1 &
 11 04 18.22 $+$38 16 30.9 & 15.570 & 14.453 & 14.106 & [2] & 0.073\\
 1ES 1959$+$650 & 08 Jul.07 --- 09 Nov.28 & 53 & 103,10,1 &
 20 00 26.51 $+$65 09 26.4 & 12.670 & 11.464 & 11.135 & [2] & 0.074\\
 Mrk 501    & 08 May 26 --- 10 Jan.17 & 46 &  63,6,1 &
 16 53 45.85 $+$39 44 08.8 & 12.598 & 10.935 & 10.399 & [3] & 0.018\\
 1ES 0647$+$250 & 08 Sep.09 --- 10 Jan.29 &  7 & 200,10,1 &
 06 50 40.57 $+$25 03 24.4 & 12.740 & 12.053 & 11.771 & [6]$^*$ & 0.015\\
 1ES 0323$+$022 & 08 Jul.25 --- 10 Jan.28 & 24 & 153,10,1 &
 03 26 13.42 $+$02 24 06.1 & 12.840 & 11.097 & 10.485 & [1] & 0.031\\
 \hline
\end{longtable}
\normalsize

\subsection{Sample} 

\tiny
\tabcolsep1pt
\begin{longtable}{rrp{12mm}rp{7mm}p{13mm}p{11mm}p{13mm}p{15mm}p{15mm}p{15mm}p{5mm}}
 \caption{Observational properties of blazars.}\\
  \hline
      Object & Class ($\log\;\nu_{\rm peak}$) & $z$ & 
      Mag.(ref.) & $A_V$ & $V$ & $V-J$ & PD & 
      $r_{color}$ & $r_{pol.}$ & $r_{col.-pol.}$ & N\\ 
      (1) & (2) & (3) & (4) & (5) & (6) & (7) & (8) & (9) & (10) &
  (11) & (12)\\
  \hline
  \endhead

  \hline
  \endfoot

  \hline
  \multicolumn{12}{l}{(1)~Object name. (2)~Blazar class and the peak
  frequency of synchrotron radiation.}\\
  \multicolumn{12}{l}{(3)~Redshift. Those which have been disputed or
    not been confirmed are indicated by parentheses.}\\
  \multicolumn{12}{l}{(4)~Variation
    range reported in previous studies.  No correction for the
    interstellar extinction was performed.}\\
  \multicolumn{12}{l}{(5)~$V$-band interstellar absorption referred from
    \citet{sch98dust}. (6)~The minimum and maximum values of $V$-band
    magnitude in this study.}\\
  \multicolumn{12}{l}{The correction for the interstellar
    extinction has been performed.  (7)~The minimum and maximum values of
    $V-J$.}\\
  \multicolumn{12}{l}{The correction for the interstellar reddening
    has been 
    performed. (8)~The minimum and maximum values of the polarization
    degree (PD) in percent.}\\
  \multicolumn{12}{l}{(9) Correlation coefficient between the $V$-band
    magnitude and $V-J$. (10)~Correlation coefficient between the
    $V$-band flux and PD.}\\
  \multicolumn{12}{l}{(11)~Correlation coefficient between the
    $V-J$ and PD.  The correlation coefficients with parentheses
    indicate that the correlation was not statistically
    significant.}\\
  \multicolumn{12}{l}{(12)~The number of observations. References:
    $^1$\citet{zho07j0324}; $^2$\citet{hea08cgrabs};
    $^3$\citet{bur66quasar}; $^4$\citet{hal03egret};}\\  
  \multicolumn{12}{l}{$^5$\citet{abd10class};$^6$\citet{nie06class};
    $^7$\citet{agncatalog}; $^8$\citet{nil08s50716}; $^9$\citet{bzcat};}\\
  \multicolumn{12}{l}{$^{10}$\citet{bra053c66a};
    $^{11}$\citet{fal90pg1553}; $^{12}$\citet{aha06pg1553};
    $^{13}$\citet{mic06rosat}; $^{14}$\citet{sba06z};}\\   
  \multicolumn{12}{l}{$^{15}$\citet{lqac}; $^{16}$\citet{con77fsrq};
  $^{17}$\citet{hau10pks1222}; $^{18}$\citet{san72qso};
  $^{19}$\citet{tra09j1236};}\\
  \multicolumn{12}{l}{$^{20}$\citet{web903c279};
  $^{21}$\citet{kik88pks0215}; $^{22}$\citet{mar82pks0215};
  $^{23}$\citet{imp88pol}; $^{24}$\citet{asiago};
  $^{25}$\citet{kat08pks1510};}\\
  \multicolumn{12}{l}{$^{26}$\citet{fan00corr};
  $^{27}$\citet{sil85oj287}; $^{28}$\citet{tak90oj287};
  $^{29}$\citet{sol083c273}; $^{30}$\citet{rai08ao0235}; $^{31}$\citet{tak98ao0235};
  $^{32}$\citet{zek81oj49};}\\
  \multicolumn{12}{l}{$^{33}$\citet{yos08misv1436};
  $^{34}$\citet{wil92pol}; $^{35}$\citet{mor08atel1661};
  $^{36}$\citet{vil063c454}; $^{37}$\citet{cor88blazar};
  $^{38}$\citet{tap77pks0754};}\\
  \multicolumn{12}{l}{$^{39}$\citet{xie02bllac};
  $^{40}$\citet{fan98bllac}; $^{41}$\citet{rai99s40954};
  $^{42}$\citet{nes02s51803}; $^{43}$\citet{mil78oq530};
  $^{44}$\citet{smi87longvar};}\\
  \multicolumn{12}{l}{$^{45}$\citet{pic88longvar};
  $^{46}$\citet{liu95on231}; $^{47}$\citet{qia02s50716};
  $^{48}$\citet{blo04j1052}; $^{49}$\citet{car983c371};
  $^{50}$\citet{mas96pks0422};}\\
  \multicolumn{12}{l}{$^{51}$\citet{bri90h1722};
  $^{52}$\citet{gri89h1722}; $^{53}$\citet{fan00pks2155};
  $^{54}$\citet{kap091es0806}; $^{55}$\citet{liu97mrk421};
  $^{56}$\citet{vil00bllacs};}\\
  \multicolumn{12}{l}{$^{57}$\citet{fei861es0323};
  $^{58}$\citet{zha081es0323}}\\  
  \endlastfoot

QSO~J0324$+$3410&FSRQ (--)$^1$  &0.063$^7$
  &$14.72V^{15}$--$15.72V^7$&0.65&$15.35$--$15.36$&$2.01$--$2.02$&$0.7$--$0.8$
  &--&--&--&2\\
4C~14.23        &FSRQ (--)$^2$  &1.038$^7$
  &$19.0p^{16}$--$17.48V^7$&0.27&$16.53$--$17.48$&$1.62$--$1.79$&$2.6$--$21.3$
  &--&$(+0.13^{+0.55}_{-0.64})$&--&16\\
PKS~1222$+$216  &FSRQ (--)$^3$  &0.435$^7$
  &$14.6R^{17}$--$17.50V^{18}$&0.07&$17.46$&--&$5.9$
  &--&--&--&1\\
3EG~J1236$+$0457&FSRQ (--)$^4$  &1.750$^7$
  &$15.98U^{19}$--$20.56V^7$&0.07&$16.44$--$17.59$&$2.45$--$2.61$&$2.5$--$9.7$
  &--&--&--&4\\
3C~279          &FSRQ (12.6)$^5$&0.538$^7$
  &$11.51V^{20}$--$17.75V^7$&0.09&$15.45$--$17.32$&$2.43$--$3.01$&$2.8$--$36.3$
  &$+0.70^{+0.13}_{-0.20}$&$+0.41^{+0.21}_{-0.27}$&$-0.38^{+0.31}_{-0.24}$&64\\
PKS~0215$+$015  &FSRQ (12.9)$^5$&1.715$^7$
  &$14.45V^{21}$--$>19.5V^{22}$&0.10&$15.65$--$17.08$&$1.88$--$2.01$&$15.2$--$26.6$
  &$(-0.40^{+1.14}_{-0.55})$&$(+0.91^{+0.09}_{-1.32})$&$(+0.13^{+0.84}_{-1.08})$&6\\
QSO~0454$-$234  &FSRQ (13.1)$^5$&1.003$^7$
  &$16.58V^{23}$--$19.84V^{23}$&0.14 &$15.16$--$16.89$&$1.85$--$2.22$&$2.2$--$12.3$
  &$(+0.29^{+0.28}_{-0.34})$&($-0.23^{+0.39}_{-0.32}$)&$(+0.23^{+0.36}_{-0.43})$&53\\
PKS~1510$-$089  &FSRQ (13.1)$^5$&0.360$^7$
  &$16.52V^{24}$--$16.88V^{25}$&0.31&$13.33$--$16.65$&$1.72$--$2.34$&$1.1$--$36.3$
  &$(-0.04^{+0.29}_{-0.28})$&$+0.83^{+0.08}_{-0.12}$&$(-0.24^{+0.30}_{-0.26})$&57\\
PKS~1749$+$096  & LBL (13.1)$^5$&0.320$^7$
  &$15.88V^{26}$--$17.88V^{15}$&0.55&$14.06$--$17.01$&$2.07$--$2.83$&$1.7$--$25.7$
  &$+0.70^{+0.10}_{-0.14}$&$(+0.15^{+0.24}_{-0.25})$&$(-0.17^{+0.25}_{-0.23})$&78\\
OJ~287          & LBL (13.4)$^5$&0.306$^7$
  &$12.2V^{27}$--$17.4V^{28}$&0.09&$13.86$--$15.65$&$1.87$--$2.45$&$10.4$--$37.8$
  &$+0.52^{+0.12}_{-0.14}$&$(-0.06^{+0.16}_{-0.16})$&$(-0.05^{+0.18}_{-0.18})$&174\\
3C~273          &FSRQ (13.5)$^5$&0.158$^7$
  &$12.3V^{29}$--$13.2V^{29}$&0.06&$12.51$--$12.75$&$0.96$--$1.19$&$0.1$--$1.4$
  &$+0.47^{+0.18}_{-0.23}$&$(+0.19^{+0.22}_{-0.24})$&$(+0.07^{+0.27}_{-0.28})$&77\\
AO~0235$+$164   & LBL (13.5)$^5$&0.940$^7$
  &$15.1V^{30}$--$19.8V^{31}$&0.25&$14.81$--$18.05$&$2.87$--$3.66$&$2.7$--$34.0$
  &$+0.68^{+0.12}_{-0.17}$&$+0.78^{+0.09}_{-0.14}$&$-0.55^{+0.22}_{-0.17}$&70\\
OJ~49           & LBL (13.5)$^6$&0.180$^7$
  &$14.2B^{32}$--$17.8B^{32}$&0.10&$14.73$--$16.26$&$1.99$--$2.30$&$1.9$--$21.0$
  &$+0.54^{+0.21}_{-0.30}$&$-0.66^{+0.22}_{-0.15}$&$(+0.28^{+0.30}_{-0.37})$&52\\
Mis~V1436       &FSRQ (13.6)$^5$&0.859$^7$
  &$14.1V^{33}$--$19.0V^{33}$&0.48&$15.53$--$17.81$&$2.66$--$3.10$&$0.5$--$40.1$
  &$(+0.07^{+0.23}_{-0.23})$&$+0.53^{+0.14}_{-0.17}$&$-0.29^{+0.24}_{-0.21}$&103\\
PKS~1502$+$106  &FSRQ (13.6)$^5$&1.839$^7$
  &$15.5V^{34}$--$19.5V^{35}$&0.10&$15.92$--$18.26$&$1.99$--$2.63$&$2.3$--$45.3$
  &$-0.35^{+0.25}_{-0.21}$&$+0.64^{+0.13}_{-0.19}$&$+0.30^{+0.22}_{-0.26}$&80\\
3C~454.3        &FSRQ (13.6)$^5$&0.859$^7$
  &$12.0R^{36}$--$16.9b^{37}$&0.33&$13.59$--$16.14$&$1.34$--$2.41$&$0.2$--$22.3$
  &$-0.67^{+0.08}_{-0.07}$ &$+0.50^{+0.10}_{-0.11}$&$+0.34^{+0.12}_{-0.13}$&262\\
PKS~0754$+$100  & LBL (13.6)$^6$&0.266$^7$
  &$13.8V^{38}$--$16.8V^{39}$&0.07&$15.08$--$17.16$&$1.60$--$2.40$&$3.7$--$21.8$
  &$+0.84^{+0.10}_{-0.26}$ &$(-0.23^{+0.45}_{-0.37})$&$(+0.30^{+0.42}_{-0.57})$&29\\
BL~Lac          & LBL (13.6)$^5$&0.069$^7$
  &$12.68B^{40}$--$17.99B^{40}$&1.02&$13.08$--$14.49$&$2.06$--$2.53$&$2.8$--$30.7$
  &$+0.64^{+0.08}_{-0.10}$ &$(-0.07^{+0.16}_{-0.15})$&$(-0.09^{+0.16}_{-0.15})$&196\\
QSO~J0948$+$0022  &FSRQ (13.8)$^5$&0.584$^7$
  &$18.61V^7$&0.24&$17.03$--$17.77$&$1.95$&$18.8$
  &--&--&--&3\\
S4~0954$+$65    & LBL (13.8)$^6$&0.367$^7$
  &$15.46V^{41}$--$17.21V^{41}$&0.37&$16.11$--$17.35$&$2.20$--$2.37$&$9.9$--$16.7$
  &$(-0.76^{+1.51}_{-0.24})$&$(-0.96^{+0.98}_{-0.04})$&$(-0.85^{+1.46}_{-0.15})$&5\\
S5~1803$+$784   & LBL (13.8)$^5$&0.680$^7$
  &$14.43V^{42}$--$17.65V^{42}$&0.16&$15.83$--$16.83$&$2.14$--$2.42$&$2.6$--$20.9$
  &$(+0.17^{+0.33}_{-0.37})$&$(+0.08^{+0.37}_{-0.39})$&$(-0.28^{+0.39}_{-0.32})$&35\\
RX~J1542.8$+$612& IBL (14.1)$^5$&--
  &$16.25V^7$--$16..97V^{15}$&0.05&$14.55$--$15.24$&$1.65$--$1.98$&$0.4$--$13.4$
  &$+0.36^{+0.23}_{-0.28}$ &$(-0.05^{+0.29}_{-0.29})$&$(+0.25^{+0.26}_{-0.30})$&65\\
OQ~530          & IBL (14.2)$^6$&0.152$^7$
  &$12.0B^{43}$--$16.8B^{44}$&0.04&$16.06$--$16.31$&$2.38$--$2.48$&$6.4$
  &--&--&--&3\\
PKS~0048$-$097  & IBL (14.3)$^5$&--
  &$15.30B^{45}$--$17.52B^{45}$&0.10&$14.92$--$16.17$&$1.69$--$2.01$&$0.5$--$21.5$
  &$(+0.23^{+0.29}_{-0.33})$&$(+0.12^{+0.30}_{-0.33})$&$(+0.10^{+0.33}_{-0.36})$&46\\
ON~231          & IBL (14.5)$^5$&0.102$^2$
  &$13B^{46}$--$17.5B^{46}$&0.07&$14.75$--$15.24$&$1.95$--$2.32$&$3.5$--$19.6$
  &$+0.78^{+0.16}_{-0.45}$&$-0.56^{+0.54}_{-0.28}$&$(+0.60^{+0.29}_{-0.65})$&14\\
S2~0109$+$224   & IBL (14.6)$^5$&$0.265^2$
  &$14.34B^{32}$--$17.41B^{32}$&0.12&$14.50$--$15.90$&$1.70$--$2.23$&$1.6$--$25.2$
  &$+0.33^{+0.20}_{-0.24}$&$(+0.04^{+0.25}_{-0.25})$&$(-0.15^{+0.27}_{-0.25})$&74\\
S5~0716$+$714   & IBL (14.6)$^5$&0.310$^8$
  &$12.43V^{47}$--$15.30V^{47}$&0.10&$12.60$--$14.71$&$1.65$--$2.19$&$0.5$--$25.7$
  &$+0.66^{+0.07}_{-0.09}$&$(-0.05^{+0.14}_{-0.13})$&$(-0.10^{+0.14}_{-0.14})$&242\\
3EG~1052$+$571  & IBL (14.6)$^5$&0.144$^2$
  &$14.2R^{48}$--$15.63R^{48}$&0.02&$15.52$--$15.56$&$2.04$--$2.11$&$3.0$--$3.9$
  &--&--&--&3\\
3C~371          & IBL (14.7)$^6$&0.050$^7$
  &$13.53V^{49}$--$14.96V^{49}$&0.11&$13.59$--$14.46$&$1.44$--$1.96$&$3.5$--$13.0$
  &$+0.92^{+0.03}_{-0.04}$&$(-0.18^{+0.22}_{-0.20})$&$(-0.06^{+0.21}_{-0.21})$&101\\
3C~66A          & IBL (15.1)$^5$&(0.444)$^{9,10}$
  &$13.50V^{26}$--$16.25V^{26}$&0.26&$13.45$--$14.84$&$1.46$--$1.90$&$1.0$--$24.9$
  &$+0.50^{+0.10}_{-0.11}$&$-0.18^{+0.14}_{-0.13}$&$(+0.03^{+0.15}_{-0.15})$&227\\
PG~1553$+$113   & HBL (15.4)$^5$&(0.360)$^{11,12}$
  &$14.62B^{45}$--$15.65B^{45}$&0.16&$13.73$--$14.49$&$1.31$--$1.47$&$0.5$--$6.9$
  &$(+0.04^{+0.45}_{-0.46})$&$(+0.36^{+0.33}_{-0.45})$&$(+0.27^{+0.39}_{-0.50})$&22\\
ON~325          & HBL (15.5)$^5$&0.130$^7$
  &$13.7B^{32}$--$17.46B^{26}$&0.07&$14.93$--$15.59$&$1.63$--$1.94$&$5.4$--$15.0$
  &$+0.55^{+0.20}_{-0.29}$ &$-0.40^{+0.30}_{-0.23}$&$(+0.26^{+0.31}_{-0.37})$&47\\
PKS~0422$+$004  & HBL (15.7)$^6$&(0.310)$^{13}$
  &$14.16B^{50}$--$17.45B^{32}$&0.32&$14.99$--$16.32$&$1.98$--$2.31$&$5.0$--$20.3$
  &$+0.86^{+0.07}_{-0.12}$ &$(-0.08^{+0.34}_{-0.32})$&$(+0.31^{+0.28}_{-0.35})$&42\\
H~1722$+$119    & HBL (15.8)$^6$&(0.018)$^{2,14}$
  &$15.77V^{51}$--$16.6V^{52}$&0.53&$14.78$--$15.33$&$1.37$--$1.77$&$1.9$--$10.3$
  &$(+0.15^{+0.38}_{-0.43})$&$+0.44^{+0.29}_{-0.43}$&$(-0.18^{+0.48}_{-0.41})$&28\\
PKS~2155$-$304  & HBL (16)$^5$  &0.116$^7$
  &$12.27V^{53}$--$14.13V^{53}$&0.07&$12.38$--$13.91$&$1.36$--$1.89$&$0.2$--$8.6$
  &$+0.41^{+0.14}_{-0.16}$&$+0.19^{+0.17}_{-0.18}$&$(+0.05^{+0.18}_{-0.18})$&137\\
1ES~2344$+$514  & HBL (16.4)$^6$&0.044$^7$
  &$15.2V^{39}$--$15.5V^{39}$&0.65&$15.08$--$15.25$&$2.58$--$2.84$&$0.7$--$5.4$
  &$+0.87^{+0.08}_{-0.22}$&$(+0.49^{+0.32}_{-0.54})$&$(-0.45^{+0.55}_{-0.34})$&17\\
1ES~0806$+$524  & HBL (16.6)$^6$&0.138$^7$
  &$14.72R^{54}$--$15.62R^{54}$&0.14&$15.45$--$15.91$&$1.86$--$2.09$&$1.1$--$8.2$
  &$+0.69^{+0.20}_{-0.42}$&$(+0.42^{+0.36}_{-0.56})$&$(-0.21^{+0.55}_{-0.44})$&18\\
Mrk~421         & HBL (16.6)$^5$&0.031$^7$
  &$11.6B^{55}$--$16B^{55}$&0.05&$13.23$--$13.69$&$1.59$--$1.94$&$0.1$--$4.9$
  &$+0.73^{+0.18}_{-0.43}$&$+0.48^{+0.28}_{-0.43}$&$(-0.49^{+0.58}_{-0.33})$&42\\
1ES~1959$+$650  & HBL (16.6)$^5$&0.047$^7$
  &$12.8V^{56}$--$16V^{56}$&0.53&$14.12$--$14.64$&$1.57$--$1.87$&$1.4$--$11.4$
  &$+0.80^{+0.08}_{-0.13}$&$(+0.03^{+0.28}_{-0.29})$&$(-0.01^{+0.28}_{-0.28})$&53\\
Mrk~501         & HBL (17.1)$^5$&0.033$^7$
  &$13.66B^{32}$--$14.93B^{32}$&0.06&$13.92$--$14.06$&$2.36$--$2.59$&$0.6$--$3.9$
  &$+0.68^{+0.14}_{-0.22}$&$(+0.03^{+0.34}_{-0.34})$&$(+0.08^{+0.31}_{-0.32})$&46\\
1ES~0647$+$250  & HBL (18.3)$^6$&0.203$^{9}$
  &$15.30V^7$&0.31&$15.50$--$16.08$&$1.25$--$1.54$&$1.9$--$5.7$
  &--&$(+0.59^{+0.40}_{-1.45})$&--&7\\
1ES~0323$+$022  & HBL (19.9)$^6$&0.147$^7$
  &$15.56V^{57}$--$17.28V^{58}$&0.34&$16.19$--$16.77$&$1.84$--$2.10$&$3.0$--$10.5$
  &$+0.78^{+0.13}_{-0.28}$&$+0.50^{+0.29}_{-0.48}$&$(-0.38^{+0.54}_{-0.37})$&24\\
\end{longtable}
\normalsize

We first selected our targets from the catalog, ``Extended list of 206
possible AGN/blazar targets for GLAST multi-frequency
analysis''\footnote{http://glastweb.pg.infn.it/blazar/} with an
apparent magnitude of $R\lesssim 16$. Then, we performed test
observations of those potential targets in early 2008. Objects were
included in our sample in the case that the test observation confirmed
that they were bright enough to monitor with the photopolarimetric mode 
($R\lesssim 16$). The number of selected objects from that catalog was
30. New objects were included in our sample when optical or
$\gamma$-ray flares of them were detected during our monitoring
period. The number of additional objects was 12. The total number of
our target is, hence, 42. Table~1 lists our targets. Bright and/or
highly variable sources were monitored more frequently than faint
and/or stationary ones. This is because one of our aims was follow-up
observations of $\gamma$-ray flares detected by Fermi.

The objects were classified into four subclasses, defined in
\citet{abd10class}. Based on high-quality SEDs obtained almost
simultaneously from radio to $\gamma$-rays, \citet{abd10class}
propose a new classification for blazars (FSRQs and BL~Lac objects): Low,
intermediate, and high synchrotron peaked blazars (LSP, ISP, and HSP
blazars, respectively). LSP, ISP, and HSP blazars are defined as those
having a synchrotron peak frequency, $\nu_{\rm peak}$, of $\nu_{\rm
  peak}\lesssim 10^{14}$~Hz, $10^{14} \lesssim \nu_{\rm peak}\lesssim
10^{15}$~Hz, and $\nu_{\rm peak}\gtrsim 10^{15}$~Hz, respectively. All
FSRQs are LSP blazars in their sample, except for one source
(J1012.9$+$2435; ISP), which is not included in our sample. BL~Lac
objects are distributed in all three subclasses. According to
\citet{abd10class}, the mean $\nu_{\rm peak}$ of LSP BL~Lac objects is
higher than that of FSRQs. This indicates that $\nu_{\rm peak}$ is
lower in the following order: FSRQs $<$ LSP-BL~Lac $<$ ISP-BL~Lac $<$
HSP-BL~Lac. We classified blazars into these four subclasses, namely
FSRQs, LSP BL~Lac, ISP BL~Lac, and HSP BL~Lac, and used abbreviated
forms of the latter three subclasses as LBL, IBL, and HBL.

We identified 27 of our monitoring objects in the catalog presented in
\citet{abd10class}. The classification of these 27 objects was based
on \citet{abd10class}. Among the remaining 15 objects, prominent
emission lines have been observed in QSO J0324$+$3410 (\cite{mar96ew};
\cite{zho07j0324}), PKS~1222$+$216 (\cite{bur66quasar}), 4C~14.23
(\cite{hea08cgrabs}), and 3EG~J1236$+$0457 (\cite{hal03egret}). Hence,
these four objects belong to FSRQs. Then, the classification of the
remaining 11 objects was based on \citet{nie06class}.  According to
\citet{nie06class}, BL~Lac objects are classified into three
subgroups: namely LBL, IBL, and HBL, defined by $\nu_{\rm peak}$
estimated from radio--optical (with X-rays in several cases)
SEDs. Their criteria for subgroups are, however, different from those
in \citet{abd10class}: $\nu_{\rm peak}\lesssim 10^{14.5}$~Hz for LBL,
$10^{14.5} \lesssim \nu_{\rm peak}\lesssim 10^{16.5}$~Hz for IBL, and
$\nu_{\rm peak}\gtrsim 10^{16.5}$~Hz for HBL. We classified the
remaining 11 objects by adopting the criteria defined in
\citet{abd10class} using $\nu_{\rm peak}$ estimated in
\citet{nie06class}. Table~2 gives $\nu_{\rm peak}$ and subclasses of
each object. Our sample includes 13 FSRQs, 8 LBLs, 9 IBLs, and 12
HBLs. We note that $\nu_{\rm peak}$ can significantly shift, depending
on the brightness of blazars (e.g.~\cite{cos01bls}). It is possible
that the classification of each object changes with time, even with the
same criterions. Table~2 also presents the red-shift, $z$, and the
variation range obtained from previous literature.

\subsection{Observation and data reduction}

We performed photopolarimetric observations of those blazars
simultaneously in the optical $V$-, and the NIR $J$-, and $K_{\rm
  s}$-bands from 2008. We used TRISPEC attached to the 1.5-m
``Kanata'' telescope at Higashi-Hiroshima Observatory. TRISPEC is
capable of simultaneous three-band (one optical and two NIR bands)
imaging or spectroscopy, with or without polarimetry
(\cite{Watanabe2005}). TRISPEC has a CCD and two InSb arrays. Table~1
presents the observation period, the number of observations, and the
typical exposure times of each image. A part of NIR images failed to
be obtained because of mechanical errors. The $K_{\rm s}$-band data
are, especially, limited, and have a low quality. In this paper, we
mainly use $V$- and $J$-band data for the color and correlation
analysis, except for in sub-subsection~3.2.1.

The photometry of blazars was performed in a standard procedure of CCD 
images; after making dark-subtracted and flat-fielded images, the
magnitudes were measured using the aperture photometry technique. 
The radius of the aperture, which depended on the seeing size of each
night, was 3--5~arcsec. These correspond to 3--4 pixels on the optical
CCD. Since the pixel scale of the NIR InSb arrays of TRISPEC is large
($\sim 1.66$~arcsec), the NIR images were slightly de-focused in
order to avoid undersampling. On the NIR arrays, the FWHM of an point
source was adjusted to be $\sim 3$ pixels, which correspond to the
aperture size of the NIR images. It is well
known that the photometry of a part of blazars is complicated by the
presence of a host galaxy component. Subtraction of 
contamination of the host galaxy is, in general, possible if the host
galaxy is well resolved on images (e.g.~\cite{kot98host};
\cite{nil07host}). Unfortunately, the angular resolution of our
instrument is too low to accurately estimate the host galaxy
contamination. Hence, we performed no correction for it in the present 
work. The obtained magnitude corresponds to the combined flux from
the jet, AGN, and the host galaxy. 

We calculated differential magnitudes of blazars using comparison
stars located in the same frame. We checked the constancy of the
brightness of the comparison stars using the differential photometry
between them and neighbor stars in the same field.  The standard
deviations of the $V$-band magnitudes, $\sigma_V$, are given in table~1. 
We confirmed that each comparison star exhibited no significant
variation.  The $V$-band magnitude of the
comparison stars were taken from previous literature. The references
are given in the ninth column of table~1. The symbol ``$^*$''
represents cases that our image includes no comparison stars listed in
the reference. In this case, we measured the $V$-magnitude of our
comparison star using the stars listed in the reference.  The $J$- and
$K_{\rm s}$-band magnitudes were referred from the 2MASS catalog
(\cite{Skrutskie2006}), except for the cases with the symbol
``$\dag$'' in table~1. The NIR magnitudes with this symbol were taken
from the same reference for the $V$-band magnitude. 

The correction for the interstellar extinction, $A_V$, and the color
excess were performed according to \citet{sch98dust}. The magnitude
and color presented in this paper are corrected values. The correction
was not done for the historical variation range given in table~2.

A set of polarization parameters was calculated from four consecutive
images, which were obtained with half-wave-plate angles of $0^\circ$,
$22.5^\circ$, $45.0^\circ$, and $67.5^\circ$. We took 12 sets of
images for each object on one night, from which three sets of
polarimetric data were obtained. We confirmed that the instrumental
polarization was smaller than 0.1~\% in the $V$-band using
observations of unpolarized standard stars. Hence, we applied no
correction for it. The zero point of the polarization angle was
corrected as the standard system (measured from north to east).

Since we present no discussion about intra-day variability in this
paper, we only use nightly-averaged photometric and polarimetric
data. Observations were sometimes carried out under bad sky
conditions. Some data obtained under such conditions have very large
errors.  They could just disrupt systematic trends, which may exist in
blazar variability. In this paper, we only use photometric data with
an error of less than 0.1~mag, color indices with an error less than 0.1,
and $PD$ with an error less than 5\%. In Appendix, we show the temporal
variation in the flux, color, $PD$, $PA$, and the color-magnitude and
magnitude-$PD$ diagrams, and the Stokes $QU$ plane for all
objects. Those data will be published in electric form in a
forth-coming paper about a detailed analysis of polarization
(Paper~II). 

\section{Results}

\subsection{Basic Properties of the Sample}

\subsubsection{Relationship between the red-shift, luminosity, and
 synchrotron peak frequency, with comments on the blazar sequence}

In this subsection, we report on basic properties of our sample.  We
investigated the relationships between the red-shift, $z$, the optical 
luminosity, $\nu L_\nu$, and the synchrotron peak frequency,
$\nu_{\rm peak}$, of our sample. We calculated the minimum, average,
and maximum $\nu L_\nu$ during our observation period for our sample
whose $z$ is known.  The observed minimum and maximum values of the
$V$-band magnitudes and $z$ are presented in table~2. No estimation of 
$\nu L_\nu$ was given in the case that $z$ has not been well
determined. Such $z$ are indicated in parentheses in table~2. 
We assumed a flat universe with $H_0=71\;{\rm km}\,{\rm s}^{-1}\,{\rm
  Mpc}^{-1}$ and $\Omega_m = 0.27$ for calculating $\nu L_\nu$. 

\begin{figure}
 \begin{center}
  \FigureFile(85mm,85mm){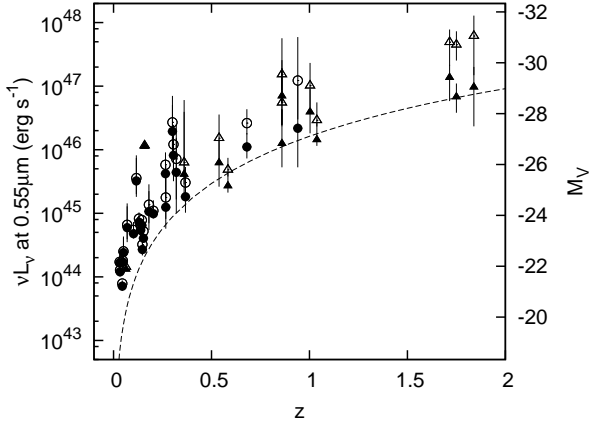}
 \end{center}
 \caption{$z$ distribution of the $V$-band luminosity ($\nu
   L_\nu$) (and absolute magnitude, $M_V$) of our sample. The circles
   and triangles represent the average $\nu L_\nu$ during our
   observation period for BL~Lac objects and FSRQs, respectively.
   The filled and open circles represent the observed and 
   $K$-corrected luminosities, respectively. The vertical bars
   associated with each point show variation ranges defined by the
   observed minimum and maximum values of the magnitude during our
   observation period. The dashed line indicates an apparent magnitude
   of $V=17.0$}\label{fig:flx2z}
\end{figure}

Figure~\ref{fig:flx2z} shows the $z$ distribution of our sample. Our
sample is distributed in $z=0.02$--$1.75$. The $\nu L_\nu$ represented
by the open circles were calculated from the observed magnitudes with
a $K$-correction. For the $K$-correction, we assumed SEDs in two power-law
forms between the $V$---$J$-band and $J$---$K_s$-band regions. The
flux densities at $0.55$, $1.22$, and $2.16\;{\rm \mu m}$ 
were obtained from our simultaneous $V$-, $J$-, and $K_s$-band
observations, respectively. Then, the rest-frame flux densities at
$0.55\;{\rm \mu m}$ were estimated with those three flux densities and
the two-power-law SEDs. The dashed line in figure~\ref{fig:flx2z}
represents an apparent magnitude of $V=17.0$. This curve indicates
that our sample had a limit magnitude of $V\sim 17$.

\begin{figure} 
 \begin{center}
  \FigureFile(85mm,85mm){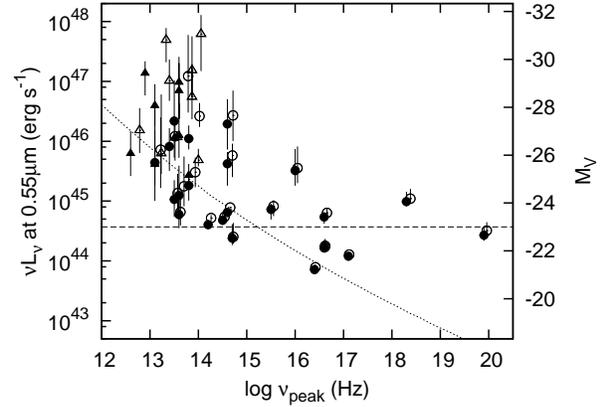}
 \end{center}
 \caption{$\nu_{\rm peak}$ distribution of the $V$-band luminosity
   ($\nu L_\nu$) (and absolute magnitude, $M_V$) of our sample. The
   circles and triangles represent BL~Lac objects and FSRQs,
   respectively. The filled and open circles represent the observed
   and the $K$-corrected luminosities, respectively. The
   redshift-correction was also performed for $\nu_{\rm peak}$ of the
   open circles. The vertical bars associated with each point show the
   minimum and maximum values of the magnitude during our observation
   period. The dashed line represents the luminosity of a typical host
   galaxy (\cite{urr00host}). The dotted line represents an apparent
   magnitude of $V=17$ obtained from the $z$--$\nu_{\rm peak}$
   relation in figure~\ref{fig:z2nu}.}\label{fig:flx2nu} 
\end{figure}

Figure~\ref{fig:flx2nu} shows the $\nu_{\rm peak}$ distribution of the 
luminosity. In addition to the $K$-correction, a red-shift correction
was performed to $\nu_{\rm peak}$ of the open circles by multiplying
$\nu_{\rm peak}$ by $(1+z)$. An anti-correlation is apparent between
$\nu L_\nu$ and $\nu_{\rm peak}$. The anti-correlation is more
emphasized between the corrected $\nu L_\nu$ and $\nu_{\rm peak}$, as
indicated by the open circles.

The anti-correlation between $\nu L_\nu$ and $\nu_{\rm peak}$ is well
known as the ``blazar sequence'', which was first proposed by
\citet{Fossati1998}. According to \citet{Ghisellini1998}, the blazar
sequence can be explained by the idea that a synchrotron cooling works 
more severely for low-energy electrons in more powerful blazars. As a
result, it leads to smaller $\nu_{\rm peak}$ in them. However, it has
been suspected that the blazar sequence scenario involves serious
problems about the selection effects (\cite{cos01bls}; \cite{pad03bls};
\cite{cac04bls}; \cite{ant05bls}; \cite{nie06class};
\cite{pad07bls}). When the blazar sequence is evaluated, a problem is
the high variability not only in $\nu L_\nu$, but also $\nu_{\rm
  peak}$ (e.g.~\cite{cos01bls}). Simultaneous measurements of $\nu
L_\nu$ and $\nu_{\rm peak}$ are, hence, required. The $\nu_{\rm peak}$ of
our sample are mostly quoted from \citet{abd10agn}, who analyzed
multi-wavelength data taken in the almost same observation periods as
those of our monitoring. Hence, our data can provide one of the most
reliable tests for the blazar sequence scenario in terms of
simultaneous measurements of $\nu L_\nu$ and $\nu_{\rm peak}$. 

As shown in figure~\ref{fig:flx2nu}, the result is apparently
compatible with the blazar sequence scenario originally proposed by
\citet{Fossati1998}. The blazars with a high $\nu_{\rm peak}$ of
$>10^{16}\,{\rm Hz}$ had $M_V\sim 6$~mag fainter than those
with a low $\nu_{\rm peak}$ of $<10^{14}\,{\rm Hz}$. It is evident that 
our observation did not cover the faintest states in several blazars,
as can be seen in table~2. The observed faintest states in a part of
LBLs and FSRQs are, in particular, $\sim 3$~mag brighter than the
previously recorded faintest states. However, the anticorrelation
between $\nu L_\nu$ and $\nu_{\rm peak}$ still appears even if $\nu
L_\nu$ are estimated from the faintest states ever recorded. This 
indicates that the flux variability actually has no major affect on
the blazar sequence scenario.

\begin{figure}
 \begin{center}
  \FigureFile(85mm,85mm){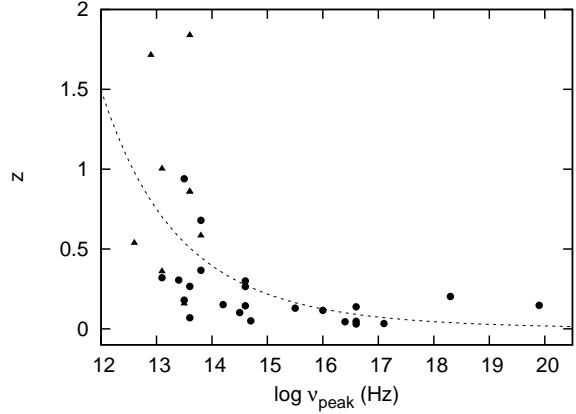}
 \end{center}
 \caption{$\nu_{\rm peak}$ distribution of the red-shift, $z$ of
   our sample. The circles and triangles represent BL~Lac objects and
   FSRQs, respectively. The dashed line represents the best-fitted
   power-law model of $z$ against $\nu_{\rm peak}$.}\label{fig:z2nu}
\end{figure} 

We comment on the selection effects. Several studies have reported
that deep radio observations found sources that have low $\nu L_\nu$ and
low $\nu_{\rm peak}$ (\cite{pad03bls}; \cite{cac04bls};
\cite{ant05bls}). Similar results were also obtained in the optical
data collected by \citet{nie06class}. The presence of sources
having low $\nu L_\nu$ and low $\nu_{\rm peak}$ contradicts the blazar
sequence scenario. Figure~\ref{fig:z2nu} shows the $\nu_{\rm peak}$
distribution of $z$ of our sample. IBLs and HBLs having $\nu_{\rm
  peak}>10^{14}\,{\rm Hz}$ can be found only in a low $z$ region of
$z<0.5$. In contrast, a part of LBLs and FSRQs can be found in a
high $z$ region of $z>0.5$, as well as in a low $z$ region. The dashed
line in figure~\ref{fig:z2nu} represents the best-fitted power-law
model of $z$ against $\nu_{\rm peak}$. While the power-law form has no
physical meaning here, this empirical model well represents a trend of
$z$ versus $\nu_{\rm peak}$. In conjunction with the curve, which
indicates the apparent magnitude of $V=17$ in figure~\ref{fig:flx2z},
this model for $z$ provides $\nu L_\nu$ corresponding to $V=17$ in
figure~\ref{fig:flx2nu}. The dotted line in figure~\ref{fig:flx2nu}
represents it. Objects are detectable over this curve in a
magnitude-limited sample.  As suggested by this curve, the apparent
lack of low $\nu L_\nu$ and low $\nu_{\rm peak}$ sources could be due
to a selection effect present in magnitude-limited samples. 

Second, the lack of high $\nu L_\nu$ and high $\nu_{\rm peak}$ blazars
is also suspected to be due to a selection effect. \citet{gio05bls}
propose that such blazars, if exist, would have featureless optical
spectra, which make it difficult to estimate $z$ and luminosity (also
see, \cite{gio02bls}). \citet{urr00host} have reported that host 
galaxies of BL~Lac objects have an absolute magnitude of $M_R=-23.7\pm
0.6$. Assuming $V-R=0.7$ for elliptical galaxies, we show the $V$-band
absolute magnitude of the typical host galaxy in
figure~\ref{fig:flx2nu} by the dashed line. It has been reported that
the luminosity of the host galaxy is not severely dependent on $z$ in
a region of $z<0.5$ (\cite{sca00host}; \cite{urr00host}). Our sample
includes only a few IBLs and HBLs having luminosities one order higher
than that of the typical host galaxy. A significant part of BL~Lac
objects has such high luminosities in $\nu_{\rm peak}\lesssim
10^{14}\;{\rm Hz}$. They are, however, LBLs in $z>0.5$, where the
absolute magnitudes of the host galaxy are poorly known. Thus, our
observation is compatible with the scenario proposed by
\citet{gio05bls}. 

\subsubsection{Relationship between the color and synchrotron peak
 frequency} 

\begin{figure} 
 \begin{center}
  \FigureFile(85mm,85mm){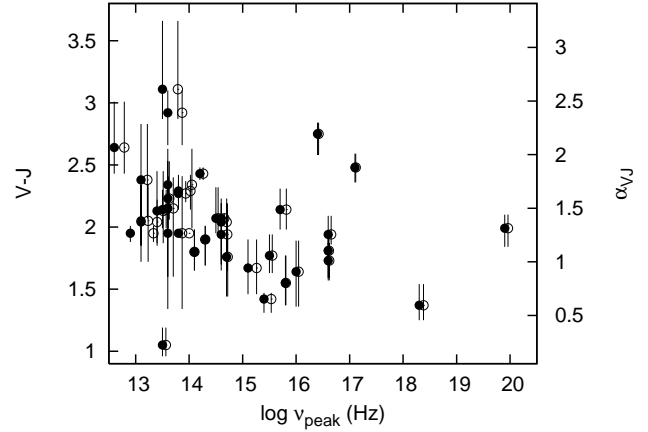}
 \end{center}
 \caption{$\nu_{\rm peak}$ distribution of the $V-J$ color index
   (, and the spectral index, $\alpha_{VJ}$) of our sample. The filled
   and open circles represent the observed and the red-shift corrected
   data, respectively. The vertical bars associated with each point
   show the minimum and maximum values of $V-J$ during our observation
   period.}\label{fig:col2nu}
\end{figure} 

Figure~\ref{fig:col2nu} shows the $\nu_{\rm peak}$ distribution of
$V-J$. The spectral index between the
$V$- and $J$-band region, $\alpha_{VJ}$, defined with 
$f_\nu\propto \nu^{-\alpha_{VJ}}$, is also shown in the right
side of the figure. We calculated $\alpha_{VJ}$ from the flux density
at 0.55 and $1.22\;{\rm \mu m}$ estimated from the $V$- and $J$-band
observations, assuming a power-law form of spectra. The figure
indicates an anti-correlation between $V-J$ and $\nu_{\rm peak}$ in a
region of $\nu_{\rm peak}\lesssim 10^{16}\;{\rm Hz}$. This
anti-correlation can be naturally expected if the optical--NIR
emission is dominated by synchrotron emission from jets. The
overall trend of $V-J$ (, or $\alpha_{VJ}$) is qualitatively analogous
to that of the spectral index between the radio and optical regions
(\cite{Fossati1998}; \cite{nie06class}).

In the lowest $\nu_{\rm peak}$ regime, objects are distributed around
$\alpha_{VJ}=1.5$. This is consistent with the model of synchrotron
self-Compton emission in blazar jets; the spectral index is predicted
to be $1.5$ in a high energy region where Compton cooling is efficient
(\cite{chi02ssc}). The optical band definitely corresponds to this
high energy region in low $\nu_{\rm peak}$ objects. On the other
hand, most objects can reach $\alpha_{VJ}>1.5$, as can be seen from
the figure. This might suggest that the Compton cooling actually
worked more efficiently than the model predicted. Alternatively, it
may suggest that the number of high-energy electrons is much smaller
than that expected from the standard power-law form of the energy
distribution of electrons. 

The anti-correlation of $V-J$ and $\nu_{\rm peak}$ appears to be weak
in a high $\nu_{\rm peak}$ region of $\nu_{\rm peak}\gtrsim
10^{15}\;{\rm Hz}$. In this $\nu_{\rm peak}$ region, the spectral
index is distributed around $\sim 1.0$. There is no object in our
sample in which fully self-absorbed synchrotron emission ($\alpha
\lesssim -1$) is dominant in the optical--NIR band.  These colors of
high $\nu_{\rm peak}$ sources imply that blazars have nearly flat SEDs
($\alpha\sim 1$) in a region even 4--5 orders of magnitude below
$\nu_{\rm peak}$. Alternatively, the contamination of the red host
galaxy may be significant, particularly in low-luminosity HBLs. In
fact, we found that two low luminosity HBLs, 1ES~2344$+$514 ($\nu_{\rm
  peak}=10^{16.4}\;{\rm Hz}$) and Mrk~501 ($\nu_{\rm
  peak}=10^{17.1}\;{\rm Hz}$), had quite red colors of $V-J \sim 2.8$
and $\sim 2.5$, respectively, which suggest a large contribution of
their host galaxies in the observed fluxes. In addition, it has been
reported that the UV--IR emission of Mrk~501 can be reproduced with a
strong contribution of an elliptical galaxy (\cite{kat01mrk501}). 

Before closing this subsection, we mention the anomalies in a low
$\nu_{\rm peak}$ region in figure~\ref{fig:col2nu}. First, the reddest
source is AO~0235$+$164 ($V-J=3.1$). This object is suggested to be
affected by absorption by foreground galaxies
(e.g.~\cite{yan89ao0235}). The quite red color of AO~0235$+$164 is,
hence, due to color excess arising from the foreground
galaxies. According to \citet{rai05ao0235}, the color excess, $E(V-J)$,
arising from them is estimated to be $0.673$. The
mean corrected $V-J$ is, then, calculated as 2.44, which is rather
normal for FSRQs and LBLs having $\nu_{\rm peak}\sim 10^{13-14}\;{\rm
  Hz}$. The second reddest source is MisV1436 ($V-J=2.9$), whose
violent variability has been first reported in 2008 by the MISAO
project. \footnote{$\langle$http://www.aerith.net/misao/variable/MisV1436.html$\rangle$}
No detailed study has been performed for this object. Its atypically
red color suggests that this blazar is a noteworthy object. Third, the
bluest source is 3C~273 ($V-J=1.1$), one of the most famous QSOs. A large
contamination of the big blue bump originated in its AGN component has
been confirmed in optical spectroscopy and polarimetric observations
(\cite{imp893c273}; \cite{smi933c273}).  The blue color of 3C~273 is
due to the strong contribution of the AGN component.

\subsection{Light Curve and Color Variation}
\subsubsection{Time lag between variations in the optical and NIR
 bands} 

Multi-wavelength studies of blazar variability have shown that
variations in different wave-bands occasionally correlate with 
significant time-lags (e.g.~\cite{ulr97review}). It has been reported
that the variation in low-energy synchrotron photons tends to lag
behind that in high-energy ones. In HBLs, for example, variations in
soft X-rays lag those of hard X-rays by hours (\cite{tak96mrk421};
\cite{kat00pks2155}) and variations in UV photons lag those of X-rays
or $\gamma$-rays by 1--2~d (\cite{buc96mrk421}; \cite{urr97pks2155}). 
A time lag of low-energy photons indicates a spectral hysteresis in
flares; for example, a spectrum becomes hardest before the flare
maximum (\cite{kat00pks2155}). Such spectral hystereses have also been
reported in optical flares (\cite{cip03gc0109};
\cite{wu07s50716}). The time lag and spectral hysteresis are expected
in the case that synchrotron cooling governs temporal variations in a
flare region. High-energy electrons diminished more quickly than 
low-energy ones by cooling (\cite{tak96mrk421}; \cite{Kirk1998}).

We searched for universal time lags between the optical and NIR
variations for each object. We calculated the correlation function
between the $V$-, $J$- and $K_{\rm s}$-band light curves. We used the
discrete correlation function ($DCF$) method \citep{Edelson1988}. The
$DCF$ method was proposed to estimate a correlation function from
unevenly sampled data. For two data streams, $a_{i}$ and $b_{i}$, the
$DCF$ method first calculates a set of unbinned discrete correlations, 
\begin{eqnarray}
UDCF_{ij} = \frac{(a_{i} - \overline{a})(b_{j} - \overline{b})}
 {\sqrt{(\sigma_{a}^{2} - e_{a}^{2})(\sigma_{b}^{2} - e_{b}^{2})}},
\end{eqnarray}
where $\overline{a}$ and $\overline{b}$ are the averages of the data
values, $\sigma_{a}$ and $\sigma_{b}$ are their standard deviations,
and $e_{a}$ and $e_{b}$ are the observation error of each data. $DCF$
for each time lag, $\tau$, is defined as an average of $UDCF$ having the
same $\tau$: Averaging over the M pairs for which $\tau - \Delta \tau
/2 \le \Delta \tau_{ij} < \tau + \Delta \tau /2$,
\begin{eqnarray}
DCF(\tau) = \frac{1}{M}\sum UDCF_{ij}.
\end{eqnarray}
The standard error for each bin is:
\begin{eqnarray}
\sigma(\tau) = \frac{1}{M-1} \{\sum [UDCF_{ij}-DCF(\tau)]^2 \}^{1/2}.
\end{eqnarray}

\begin{figure} 
 \begin{center}
  \FigureFile(85mm,85mm){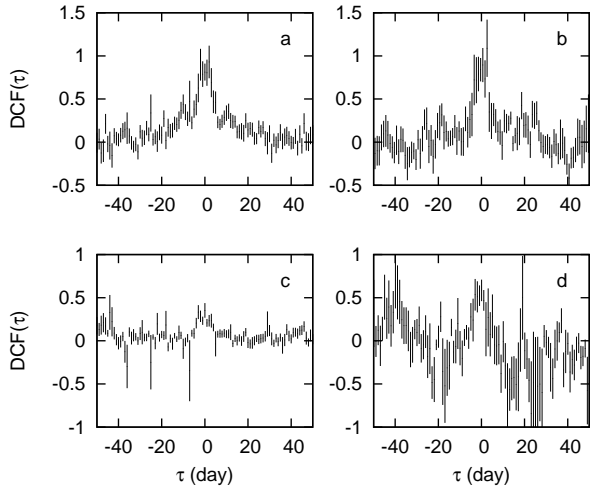}
 \end{center}
 \caption{$DCF$ between the light curves in the $V$- and $J$-bands
 (panel~a), the $V$- and $K_{\rm s}$-bands (panel b), the $V$-band
   light curve and the $V-J$ color (panel c), and the $V$-band light
   curve and the $V-K_{\rm s}$ color (panel d) calculated from the data
   of BL~Lac.}\label{fig:BLLac_lag}  
\end{figure} 

The upper panels in figure~\ref{fig:BLLac_lag} show the $DCF$ between
the light curves in the $V$- and $J$-bands (panel~a) and the $V$- and 
 $K_{\rm s}$-bands (panel b) calculated from the data of BL~Lac. 
We also calculated the $DCF$ between the $V$-band light curve and the
$V-J$ or $V-K_{\rm s}$ color, as shown in the lower panels in
figure~\ref{fig:BLLac_lag}. We found no significant time-lag in these 
results for BL~Lac. As well as BL~Lac, we calculated and checked $DCF$
for all blazars we observed.  We confirmed that no object
exhibited a universal significant time-lag between variations in the
different bands and the colors. 

\begin{figure} 
 \begin{center}
  \FigureFile(85mm,85mm){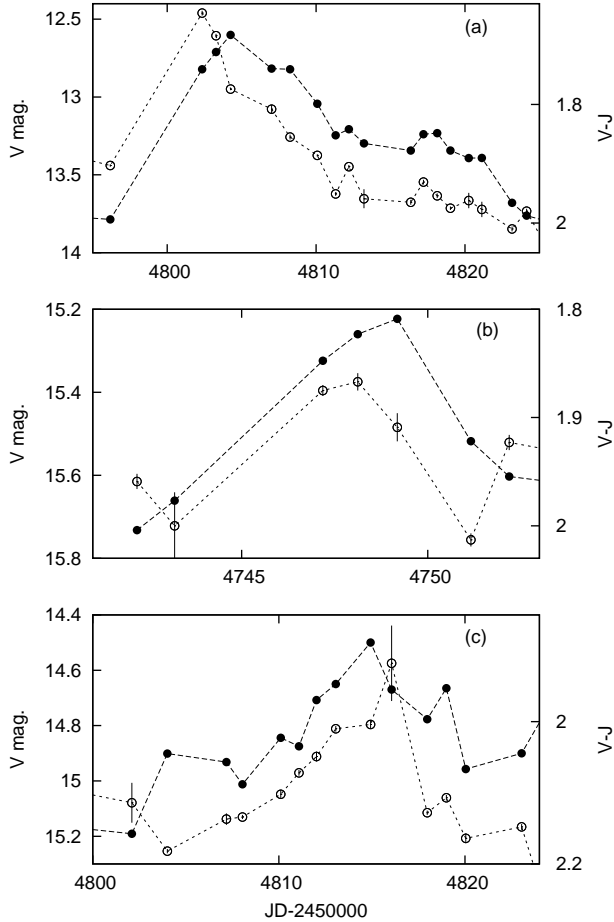}
 \end{center}
 \caption{Examples of flares showing time lag between
 different wave-bands. Panel~a, b, and c: The $V$-band light curves
 (filled circles) and $V-J$ color (open circles) during flares in
 S5~0716$+$714, PKS~0048$-$097, and S2~0109$+$224,
 respectively.}\label{fig:vvjdcf}
\end{figure} 

Then, we searched for time lags in the optical and NIR light
curves for each prominent flare, while no clear lag was found between
$V$- and $J$-band light curves of flares in all objects. 
On the other hand, we found several cases that the bluest phase in
$V-J$ preceded the flare maximum. Panels~(a) and (b) of
figure~\ref{fig:vvjdcf} present examples
for S5~0716$+$714 and PKS~0048$-$097, respectively. The lags were 2
and 1~d for S5~0716$+$714 and PKS~0048$-$097, respectively. We
systematically searched for time lags between the flux and 
color during such flares. It is expected that a time lag possibly
present in small flares would readily disappear by the composition of
another small flares. Hence, prominent flares should be selected. We
selected such flares with the following criteria: (i) A flare maximum
was defined as the brightest point in a period of 15~d before and after
the maximum date.  (ii) A peak-to-peak amplitude of the flare must be
larger than $V=1.0$ within 30~d. (iii) A flare maximum must be recorded
within 5~d after and before the neighboring observations. The selection
with these criteria yielded 17 flares. Among those 17 events, the
peak of the flux coincided with the bluest time in nine
events. Seven events showed the precedence of the bluest phase against
the flare maximum. One event in S2~0109$+$224 possibly showed a lag of the
bluest time behind the flare maximum, which is shown in 
panel~(d) of figure~\ref{fig:vvjdcf}. All lags were 1--2~d. These
results suggest that the bluest phase of flares tended to precede the
flare maximum by $\lesssim 1$~d. 

The spectral hysteresis associated with flares is again discussed in
sub-subsection~3.2.3. In the next subsection, we investigate the
correlation between the flux and color, neglecting time lags between
them. 

\subsubsection{Correlation between the $V$-band light curve and the
 $V-J$ color index}

\begin{figure*} 
 \begin{center}
  \FigureFile(170mm,170mm){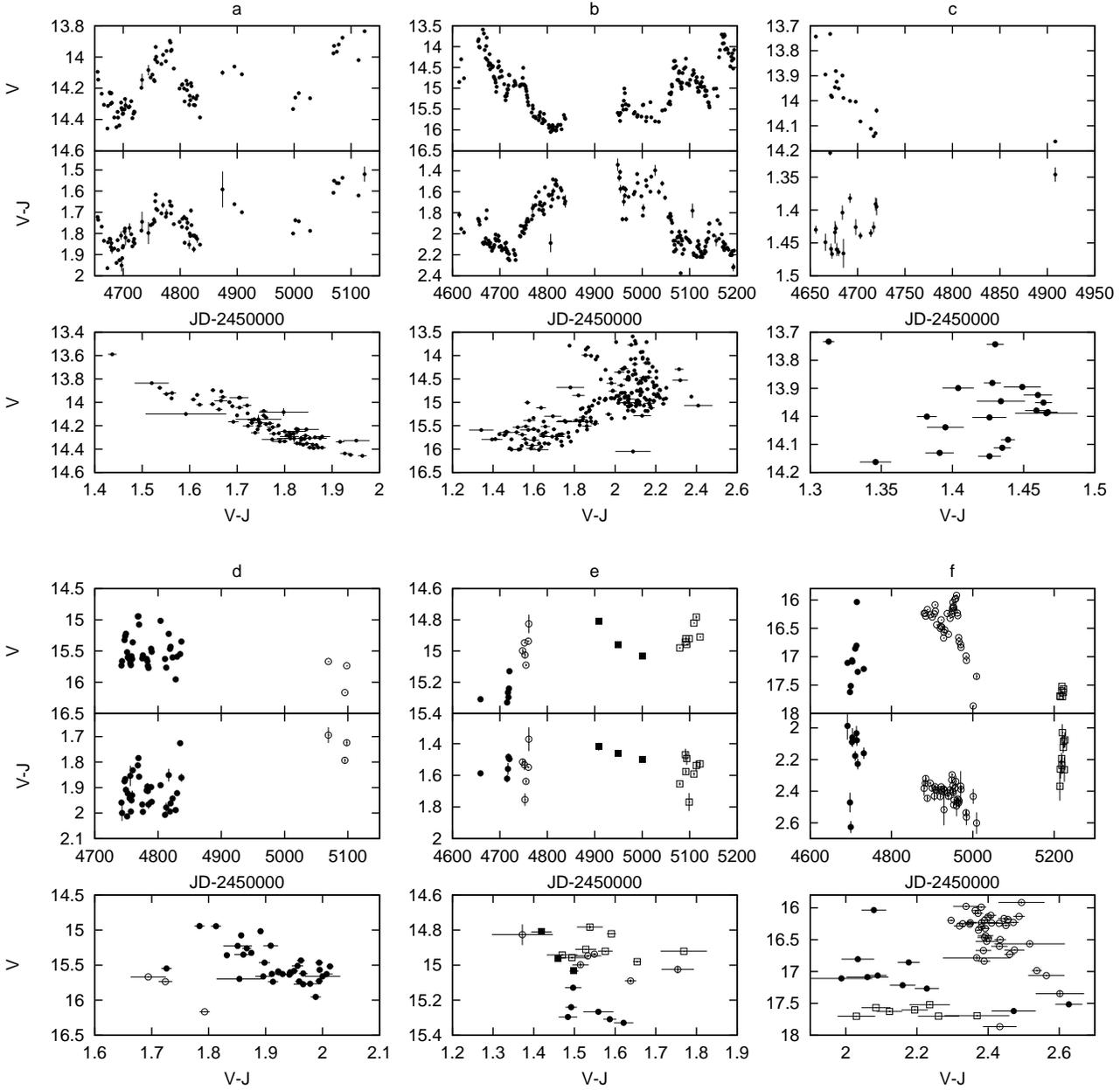}
 \end{center}
 \caption{Examples of the $V$-band light curve (top), $V-J$ color
 variation (middle), and their color--magnitude diagram 
 (bottom) of blazars. Panel a: 3C~371, b: 3C~454.3, c: PG~1553$+$113,
 d: PKS~0048$-$097, e: H~1722$+$119, and f: PKS~1502$+$106. In the
 lower three panels, different symbols represent observations taken in 
 different periods of time.}\label{fig:exam} 
\end{figure*}

We present examples of the light curves, color variations, and 
color-magnitude diagrams in figure~\ref{fig:exam}. Panel~(a) shows
those of 3C~371. The color-magnitude diagram clearly indicates that
the object became bluer when it was brighter. In this paper, we
conclude that an object had a bluer-when-brighter trend in the case
that it showed a significant positive correlation between the $V$-band
magnitude and the $V-J$ color index. We used the Pearson
product-moment correlation coefficient. The significance of the
correlation was tested using the Student's $t$-test. The correlation
coefficient and its 95~\% confidence interval of 3C~371, for example,
were calculated as $0.92^{+0.03}_{-0.04}$. Thus, we conclude that
3C~371 had a bluer-when-brighter trend. We calculated the correlation
coefficients using the whole data sets of 42 blazars, as well as
3C~371. The results are listed as $r_{\rm color}$ in table~2. We found
that 23 blazars had significant bluer-when-brighter trends. The
correlation coefficients that are not statistically significant are
shown with parentheses in table~2.

Two blazars, 3C~454.3 and PKS~1502$+$106, had significant 
negative correlations between the light curve and color, in other
words, a redder-when-brighter trend ($r_{\rm
  color}=-0.67^{+0.08}_{-0.07}$ and $-0.35^{+0.25}_{-0.21}$,
respectively, as shown in table~2). Panel~(b) of figure~\ref{fig:exam}
shows the light curve and color-magnitude 
diagram of 3C~454.3. As reported in \citet{Sasada2010}, this object
exhibited a redder-when-brighter trend when it was fainter than $V\sim
15$, while a bluer-when-brighter trend was observed when it was
brighter. This feature can be confirmed in figure~\ref{fig:exam}. A
redder-when-brighter trend was detected in the whole data because the
object had almost stayed in the faint state during our observation
period. As well as 3C~454.3, PKS~1510$-$089 and PG~1553$+$113 also
exhibited redder-when-brighter trends in their faint states (see,
figures~21 and 28). No significant correlation of the flux and color
was detected in those two objects when their whole data were used. 
Bluer-when-brighter trends were significantly detected in 3C~454.3 and
PKS~1510$-$089 in their bright states ($V<15.0$ for 3C~454.3 and
$<15.5$ for PKS~1510$-$089).

It is widely accepted that the redder-when-brighter trend in the faint
state is attributed to a strong contribution of thermal emission from
an accretion disk (also see, \cite{Smith1986}; \cite{Hagen-Thorn1994},
\cite{Pian1999}). The contribution of the thermal disk emission would
definitely be small when the flux from the jet increases. Hence, the
bluer-when-brighter trends in the bright states of 3C~454.3 and
PKS~1510$-$089 probably have the same origin as those in ordinary
objects.  Including 3C~454.3 and PKS~1510$-$089, the number of objects
that showed a bluer-when-brighter trend is 25.

PG~1553$+$113 exhibited a significant redder-when-brighter trend in its
faint state ($V>13.8$), as well as 3C~454.3 and PKS~1510$-$089. No
bluer-when-brighter trend was significantly detected in its bright
state. Panel~(c) of figure~\ref{fig:exam} shows the color-magnitude
diagram of this object. The redder-when-brighter trend in the faint
state, however, possibly weakens above $V=13.8$. This weakening of the
redder-when-brighter trend may suggest a transition of a
bluer-when-brighter trend in a bright state of PG~1553$+$113.

There are seven blazars that showed no significant correlation
between the $V$-band magnitude and the $V-J$ color, in addition to
PKS~1510$-$089 and PG~1553$+$113. The numbers of observations are
quite small for two objects: PKS~0215$+$015 and S4~0954$+$65. Only 6
and 5 observations were made, respectively.  They are too small to
conclude anything about the correlation. There are five blazars that
was observed for $>10$~d and showed no significant correlation of the
light curve and color: QSO~0454$-$234, MisV1436, S5~1803$+$784,
PKS~0048$-$097, and H~1722$+$119.

Even in these five objects, bluer-when-brighter trends were
occasionally seen in a short period of time. Panel~(d) of
figure~\ref{fig:exam} shows the variation in the magnitude and the
color for PKS~0048$-$097. Our observation of PKS~0048$-$097 covered
two seasons in 2008 (JD~2454741--2454851) and 2009
(JD~2455068--2455097). A significant bluer-when-brighter trend was
detected in the 2008 observation shown in the filled circles in
panel~(d). A possible bluer-when-brighter trend can be seen also in
the 2009 observation shown in the open circles, while the number of
observations is too small to conclude it. It is important to note
that, in the color-magnitude diagram, the 2009 data apparently shows a
different bluer-when-brighter slope from the 2008 data. This is the
reason why no significant correlation was detected when we performed a
correlation analysis using all data. 

A similar behavior can also be seen in H~1722$+$119 [panel e of figure 
\ref{fig:exam}]. We divided our observation period into four periods
of time (JD~2454659--2454720, JD~2454747--2454760,
JD~2454908--2455001, and JD~2455079--2455121), and show them by
different symbols in panel~(e). Bluer-when-brighter trends are
apparently seen in the first (filled circles), second (open circles),
and third (filled squares) epochs, while the number of observations is 
small.  A sign of bluer-when-brighter trends can also be seen in
QSO~0454$-$234, MisV~1436, and S5~1803$+$784 in their bright states
(see figure~22, 25, and 27, respectively).  QSO~0454$-$234 exhibited a
significant bluer-when-brighter trend in its bright state ($V<16.0$),
while the trend is not statistically significant in MisV~1436 and
S5~1803$+$784. 

We found that PKS~1502$+$106 also exhibited a flux--color behavior
similar to those of PKS~0048$-$097 and H~1722$+$119, although a
redder-when-brighter trend was detected in this object in its whole
data from 2008 to 2010. Panel~(f) of figure~\ref{fig:exam}
shows the variations in the magnitude and color of PKS~1502$+$106.
We calculated correlation coefficients for each year, and found
significant bluer-when-brighter trends in both the 2008 and 2009 data
sets. In 2010, this object remained faint ($V > 17.2$) and showed no
significant correlation between the light curve and color.

The sample reduces to 32 blazars once we exclude 10 objects, in which
the number of $V-J$ observations is $<10$. In these 10 objects,
observations are too few to conclude whether there is a significant
correlation between the light curve and color. Among those 32 blazars,
in summary, 23 blazars showed significant bluer-when-brighter trends
in their total data. Five blazars showed temporary significant
bluer-when-brighter trends in short periods of time (QSO~0454$-$234,
PKS~1510$-$089, PKS~1502$+$106, 3C~454.3, and PKS~0048$-$097). Then,
we conclude that a total of 28 blazars, corresponding to 88~\% of the
well-observed sample, had the bluer-when-brighter trend. Four blazars 
(PG 1553$+$113, H~1722$+$119, S5~1803$+$784, and MisV~1436) possibly
showed a sign of bluer-when-brighter trends, although the trends were
not statistically significant in our data. 

\subsubsection{$J$- and $V$-band Flux Diagram}

\begin{figure} 
 \begin{center}
  \FigureFile(85mm,85mm){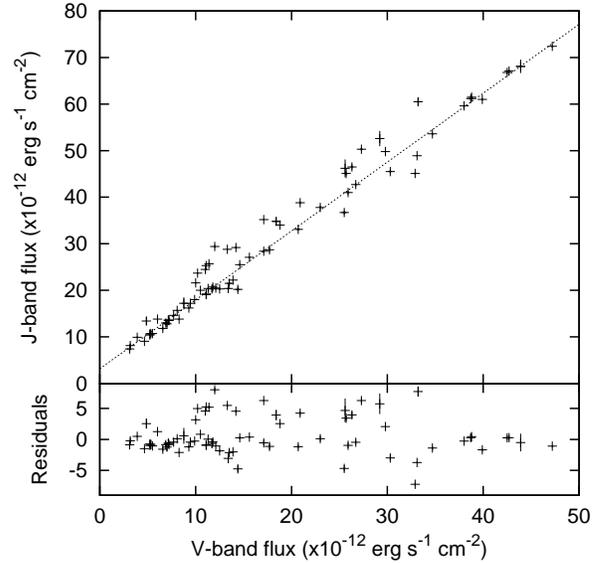}
 \end{center}
 \caption{Upper panel: $J$- and $V$-band flux variations observed
   in PKS~1749$+$096.  The lower panel: Residuals of the $J$-band flux
   from the best-fitted linear model.}\label{fig:pks1749ff} 
\end{figure} 

\begin{figure} 
 \begin{center}
  \FigureFile(85mm,85mm){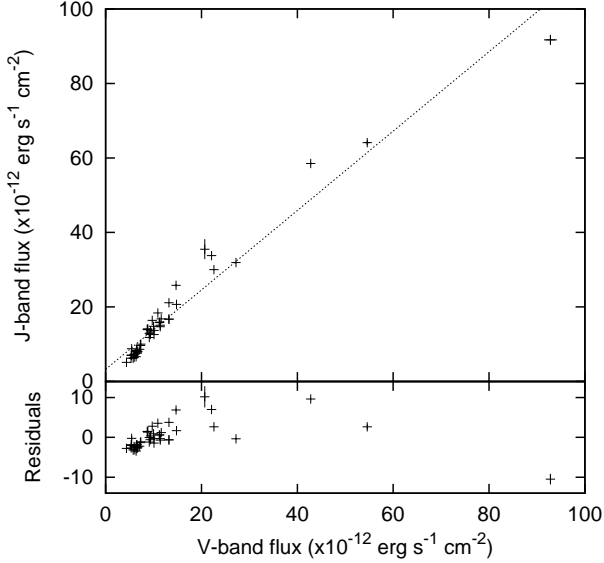}
 \end{center}
 \caption{As for figure~\ref{fig:pks1749ff}, but for
   PKS~1510$-$089.}\label{fig:pks1510ff}
\end{figure}

\begin{figure} 
 \begin{center}
  \FigureFile(85mm,85mm){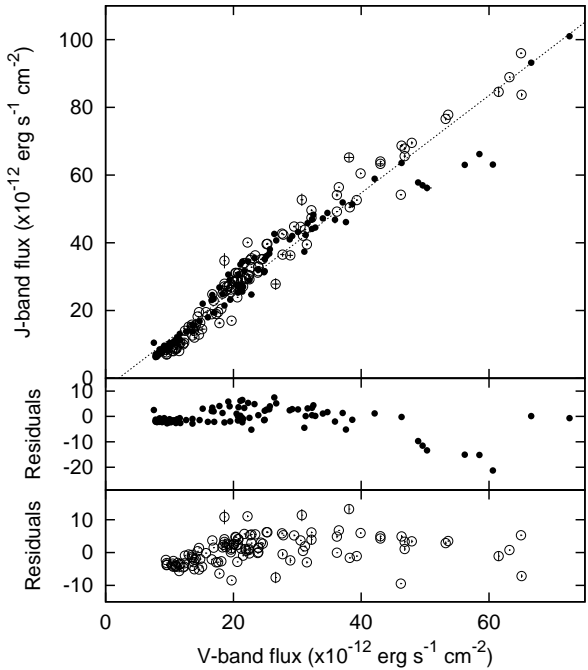}
 \end{center}
 \caption{Top panel: As for figure~\ref{fig:pks1749ff}, but
   for 3C~454.3.  The filled and open circles represent the
   observations in 2008 and 2009, respectively. The middle and bottom
   panels: Residuals of the 2008 and 2009 data,
   respectively.}\label{fig:3c454ff}
\end{figure}

Most blazars exhibited the bluer-when-brighter trend, as mentioned
in the last subsection.  Two scenarios can be considered to generate the
bluer-when-brighter trend. The first scenario is that the variation
component, itself, had a bluer-when-brighter trend.  The second scenario
is that the color of the variation component was constant with time,
while an underlying component was redder than the flare component. We
can evaluate those two scenarios using the $F_J$--$F_V$ diagrams,
where $F_J$ and $F_V$ denote the $J$- and $V$-band fluxes,
respectively. Examples of the $F_J$--$F_V$ diagrams are presented in
figures~\ref{fig:pks1749ff}, \ref{fig:pks1510ff}, and
\ref{fig:3c454ff}. In these figures, achromatic variations are 
described in a form of $F_J=cF_V$, where $c$ is a constant. 

$F_J$ can be described with a linear function of $F_V$ in most 
objects. A typical example is PKS~1749$+$096, as shown in
figure~\ref{fig:pks1749ff}. The linear relationship holds over one 
order of magnitude of the flux. We confirmed that linear regression
models for $F_J$ and $F_V$ yielded significant positive intersections
of $F_J$ at $F_V=0$ in all bluer-when-brighter objects. These results
support that there was an underlying component which was redder than
the variation components and the color of the variation components was
roughly constant.

We note that there are two possible exceptions of our 28
bluer-when-brighter blazars in terms of the $F_J$--$F_V$ relation: 
PKS~1510$-$089 and 3C~454.3.  Their $F_J$--$F_V$ diagrams are shown in
figures~\ref{fig:pks1510ff} and \ref{fig:3c454ff}, respectively. In
both objects, we can see a sign of systematic deviations from their
best-fitted linear models; the slope is steep in a low flux regime,
but shallow in a high flux one. The 2008 and 2009 observations of
3C~454.3 are represented by the filled and open circles,
respectively. PKS~1510$-$089 and 3C~454.3 had a common feature that 
they showed redder-when-brighter trends when they were faint, as
mentioned in sub-subsection~3.2.2. The low-flux regimes in the
$F_J$--$F_V$ diagram correspond to the redder-when-brighter phase.

\begin{figure*} 
 \begin{center}
  \FigureFile(170mm,170mm){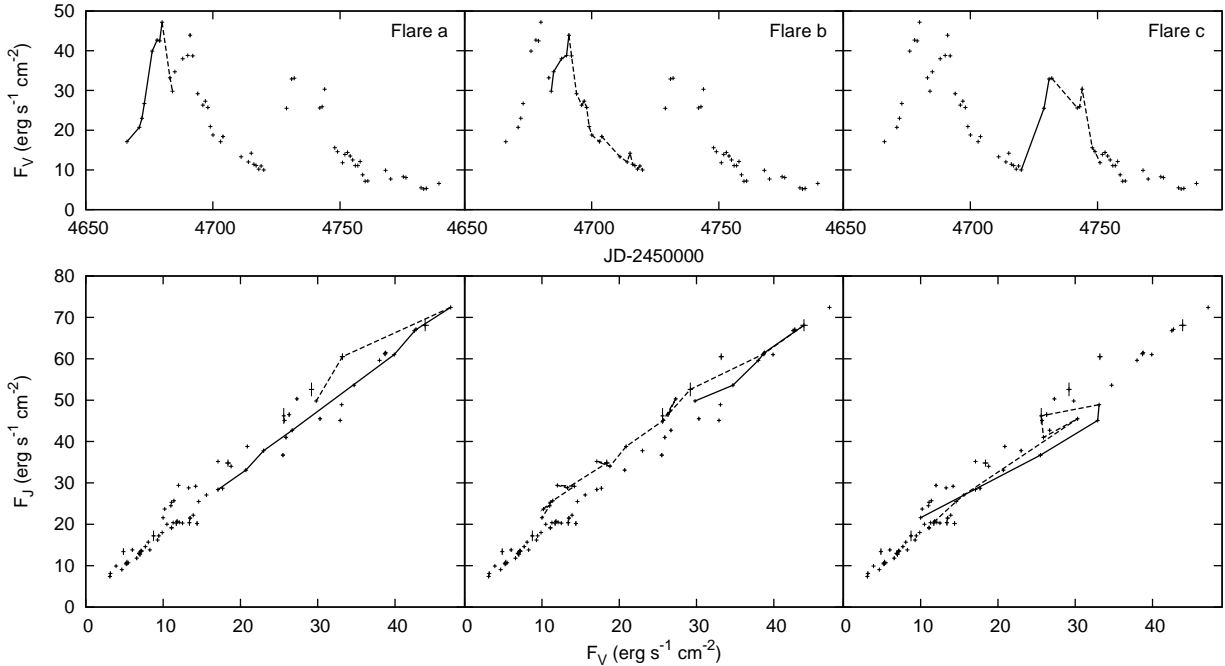}
 \end{center}
 \caption{$V$-band light curves (upper panels) and the
   $F_J$--$F_V$ diagrams (lower panels) of PKS~1749$+$096. The
  rise and decay phases of three flares (flare~a, b, and c) are
  represented by the solid and dashed lines in each panel,
  respectively.}\label{fig:pks1749}
\end{figure*} 

Next we focus on the behavior of short-term flares on time scales
of days--weeks in the $F_J$--$F_V$ diagram. In sub-subsection~3.2.1,
we describe our search for spectral hysteresis associated with flares through
time-lags between different band fluxes, or between the flux and the color.
This approach could be effective if the observed colors directly represent
the color of the variation components. However, as mentioned above, the
observed color is probably a composition of an underlying red
component and variable ones. It could be better to search for a sign
of hysteresis using the $F_J$--$F_V$ diagram.

Figure~\ref{fig:pks1749} shows the $F_J$--$F_V$ diagram (lower
panels) and the $V$-band light curves (upper panels) for three
short-term flares (flare~a, b, and c) observed in
PKS~1749$+$096. Flares~a, b, and c are shown in left, middle, and right
panels, respectively. In flare~a, the rising
phase appears to draw a different path from the decay phase in the
$F_J$--$F_V$ diagram. As a result, we can see hysteresis in this
diagram. The object appears to draw a ``loop'' in the $F_J$--$F_V$
diagram, moving in the anti-clockwise direction in the loop during the
flare. This behavior means that the rising phase of the flare was
bluer than the decay phase around the flare maximum. We can also see
similar hystereses in flares~b and c, while the hysteresis in
flare~c may be just an erroneous one because a detailed behavior was
overlooked around the peak of flare~c.

We searched for such hystereses for all short-term flares that we
observed. It is expected that a hysteresis of a small flare in the
$F_J$--$F_V$ diagram is disturbed by the superposition of other
small flares. Hence, we again used the 17 prominent flare events that 
are selected in sub-subsection~3.2.1. Then, we calculated the slopes in
the $F_J$--$F_V$ diagram for rise and decay phases of those flares.
The slopes were defined as: 
\begin{eqnarray}
(\Delta F_J/\Delta F_V)_{\rm rise} &=& (F_{J,p}-F_{J,p-1})/(F_{V,p}-F_{V,p-1})\\
(\Delta F_J/\Delta F_V)_{\rm decay} &=& (F_{J,p}-F_{J,p+1})/(F_{V,p}-F_{V,p+1}),
\end{eqnarray}
where $F_{J(V),p}$ and $F_{J(V),p\pm1}$ denote the $J$-($V$-)band
fluxes at the flare maximum and its neighboring observations. The case
of $(\Delta F_J/\Delta F_V)_{\rm rise}=(\Delta F_J/\Delta F_V)_{\rm
  decay}$ means that there is no hysteresis. The case of $(\Delta
F_J/\Delta F_V)_{\rm rise}>(\Delta F_J/\Delta F_V)_{\rm decay}$
indicates that a rising phase is bluer than a decay phase, as can be seen in 
figure~\ref{fig:pks1749}. The case of $(\Delta F_J/\Delta F_V)_{\rm
  rise}<(\Delta F_J/\Delta F_V)_{\rm decay}$ indicates that a rising
phase is redder than a decay phase.

\begin{figure} 
 \begin{center}
  \FigureFile(85mm,85mm){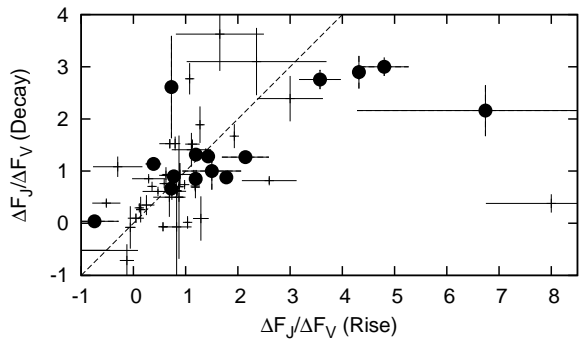}
 \end{center}
 \caption{Slope in the $F_J$--$F_V$ diagram of the rise and
  decay phases around the maxima of short-term flares. The filled
  circles and crosses represent the flares which were selected with
  peak-to-peak amplitudes of $>1.0$~mag and $>0.1$~mag,
  respectively. The dashed line represents $(\Delta F_J/\Delta
  F_V)_{\rm rise}=(\Delta F_J/\Delta F_V)_{\rm decay}$. }\label{fig:flares}
\end{figure} 

Figure~\ref{fig:flares} shows $\Delta F_J/\Delta F_V$ of the rise
and decay phases of the 17 flares, as indicated by the filled
circles. There are 11 flares showing $(\Delta F_J/\Delta F_V)_{\rm
 rise}>(\Delta F_J/\Delta F_V)_{\rm decay}$, five flares showing
$(\Delta F_J/\Delta F_V)_{\rm rise}<(\Delta F_J/\Delta F_V)_{\rm
 decay}$, and one flare showing no significant hysteresis. This result
suggests that the spectral hysteresis of $(\Delta F_J/\Delta F_V)_{\rm
  rise}>(\Delta F_J/\Delta F_V)_{\rm decay}$ was
more preferable for the prominent short-term flares. The crosses
in the figure represent $\Delta F_J/\Delta F_V$ for small flares
selected with a peak-to-peak amplitude of 0.1--1.0. The fraction of
these small flares showing $(\Delta F_J/\Delta F_V)_{\rm rise}>(\Delta
F_J/\Delta F_V)_{\rm decay}$ was comparable to that showing $(\Delta
F_J/\Delta F_V)_{\rm rise}<(\Delta F_J/\Delta F_V)_{\rm decay}$. This
is probably due to the superposition of multiple flares that disturb
the hysteresis pattern of each flare. 

The spectral hysteresis was detected in the sample of prominent
flares whose peaks were recorded within 5 d after and before
the neighboring observations (see sub-subsection~3.2.1). Together with
the time-lag analysis described in subsection~3.2.1, the time-scale of the
hysteresis events that we detected is, hence, less than a few
days. This is comparable with or possibly shorter-than-average
sampling rate of our monitoring. It is possible that short-term
hysteresis events could be overlooked in the flares of $(\Delta F_J/\Delta
F_V)_{\rm rise}\leq (\Delta F_J/\Delta F_V)_{\rm decay}$.

\subsection{Light Curve and Polarization}
\subsubsection{Time lag between variations in the light curve and the
 polarization degree} 

\begin{figure} 
 \begin{center}
  \FigureFile(85mm,85mm){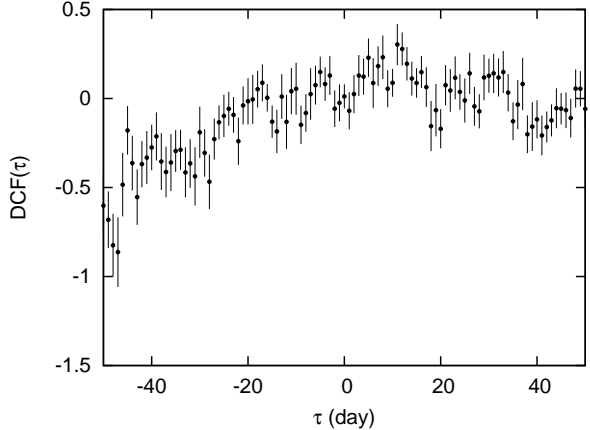}
 \end{center}
 \caption{$DCF$ calculated with the $V$-band light curve and the
 polarization degree of BL~Lac.}\label{fig:BLLac_pol} 
\end{figure}

We searched for time lags between the light curve and the polarization
degree ($PD$) using the $DCF$ method, as described in sub-subsection~3.2.1. 
Figure~\ref{fig:BLLac_pol} presents an example of $DCF$, which was
calculated with all data of the $V$-band flux and $PD$ of
BL~Lac. As shown in this figure, we found no significant time-lag
between them. We calculated $DCF$ for all blazars that we observed, and
confirmed that no objects exhibited a significant time-lag between the
light curve and $PD$. As mentioned in the next subsection, the
correlation, itself, was weak between the flux and $PD$.

\begin{figure} 
 \begin{center}
  \FigureFile(85mm,85mm){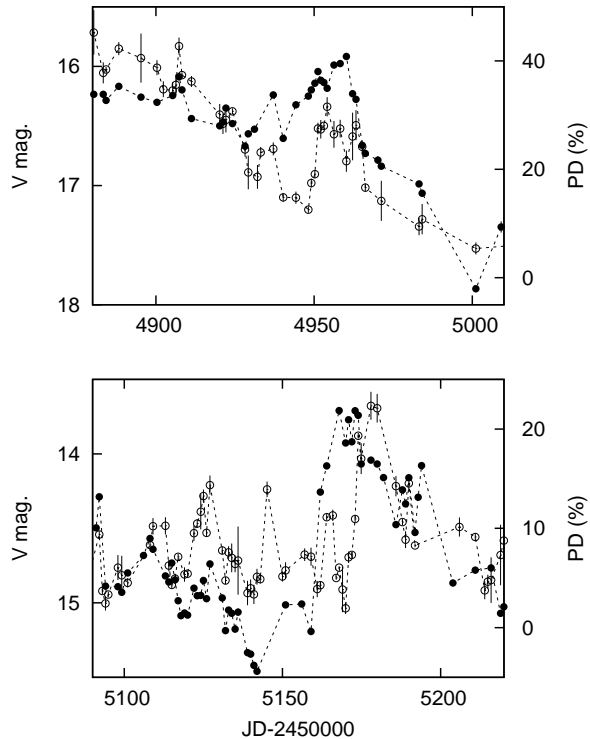}
 \end{center}
 \caption{$V$-band light curves and $PD$ variations of flares in
   PKS~1502$+$106 (upper panel) and 3C~454.3 (lower panel). The filled
   and open circles represent the light curve and $PD$,
   respectively.}\label{fig:pdlag} 
\end{figure}

We selected 17 prominent flare events in sub-subsection~3.2.1. The
peak timings of their light curves were coincident with those of $PD$, or
there was no correlation between them at all, except for the flares
observed in PKS~1510$-$089, AO~0235$+$164, and
PKS~1502$+$106. \citet{sas10ao0235} report that the increases in $PD$
were associated with prominent flares in PKS~1510$-$089 and
AO~0235$+$164 using the same data as that in the 
present paper. The $PD$ peaks preceded the flux peaks by 1~d in
PKS~1510$-$089 and 4~d in AO~0235$+$164. On the other hand,
figure~\ref{fig:pdlag} shows the cases that the $PD$ maximum lagged the
flux one. The upper panel shows the light curve and $PD$ in
PKS~1502$+$106. A large flare having double peaks can be seen in
JD~2454940--2454970. The double-peaked feature was also seen in the
variation in $PD$, while both peaks of $PD$ lagged behind those of the
light curve by 3~d. Similar events were also found in 3C~454.3. A
large flare in $\sim$JD~2455170 was followed by a subsequent increase
in $PD$, as shown in the lower panel of figure~\ref{fig:pdlag}. The lag
was about 5~d. All of those time-lags were not universal in each
object; there were flares showing no or shorter time-lags in all objects.

Thus, there was no universal time-lag between the flux and $PD$.  As
well as the flux--color correlation, the flux--$PD$ correlation is also
discussed without respect to time lags in the next subsection.

 \subsubsection{Correlation between the $V$-band light curve and the
 polarization degree} 

\begin{figure*} 
 \begin{center}
  \FigureFile(170mm,170mm){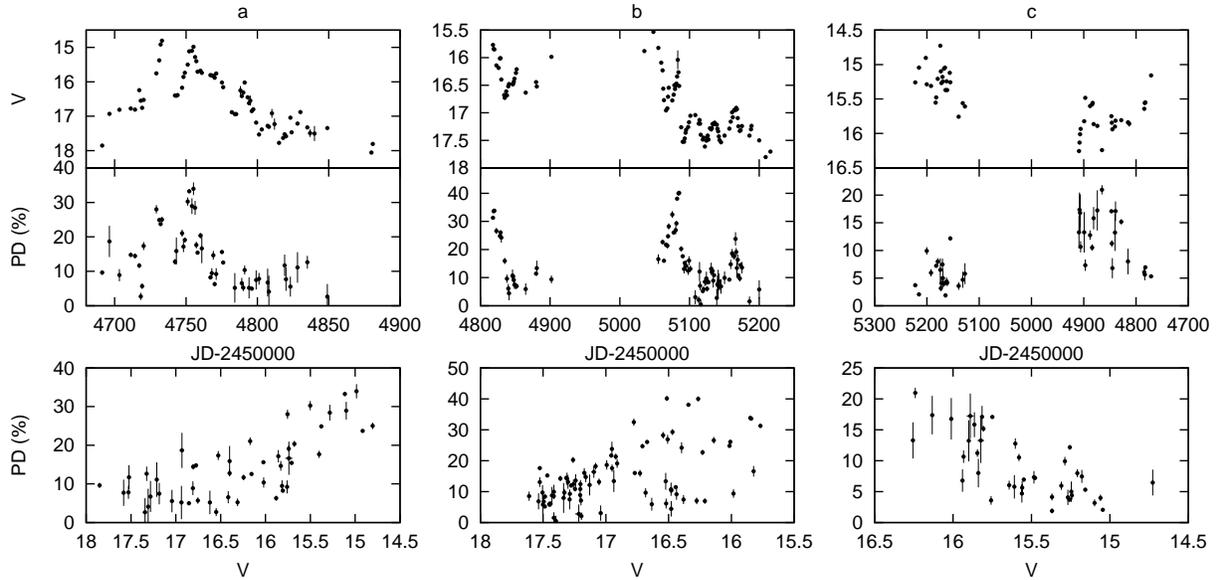}
 \end{center}
 \caption{Examples of the time variation in the $V$-band magnitude
   (top) and polarization degree (middle), and the flux--polarization
   degree diagram (bottom). Panel a: AO~0235$+$164, b: MisV1436, and
   c: OJ~49.}\label{fig:polari}
\end{figure*}

We report on the correlation of the flux and $PD$ in this subsection. 
We present examples of the light curves and variations in $PD$ in
figure~\ref{fig:polari}. Panel~(a) of figure~\ref{fig:polari} shows
those of AO~0235$+$164. $PD$ of this object became higher when it was
brighter. As described in subsection~3.2.2, we calculated the Pearson
correlation coefficients with the $V$-band flux and $PD$, and tested
them using the Student's $t$-test. The correlation coefficients are
presented as $r_{\rm pol.}$ in table~2. We identified that 10 blazars
exhibited significant positive correlations between the $V$-band flux
and $PD$. 

The flux--$PD$ correlation was not simple in several objects, even for 
$r_{\rm pol.}>0$. For example, panel~(b) of figure~\ref{fig:polari}
shows the $V$-band light curve and $PD$ of MisV~1436. This is one of
objects that showed a significant positive correlation between the
flux and $PD$. However, $PD$ in JD~2454817--2454901 were much lower than
those in JD~2455034--2455085, although the $V$-band magnitudes in
these periods were comparable. Similar behavior was also observed in
PKS~1749$+$096. Significant positive correlations between the flux and
$PD$ were detected when we analyzed the flares between
JD~2454720--2454760 and JD~2454894--2455084 in PKS~1749$+$096
(figure~23). However, no significant correlation
was detected when we used all data. This is due to the presence
of flares without an increase in $PD$. Hence, the correlation was not
universal in those objects.

We identified that four blazars (OJ~49, ON~231, 3C~66A, and ON~325) showed 
a significant negative correlation of the flux and $PD$, namely $r_{\rm
 pol.}<0$. Panel~(c) of figure~\ref{fig:polari} shows the light
curve and $PD$ of OJ~49. $PD$ were lower when the object was brighter. In
addition to those four blazars, OJ~287 and BL~Lac also exhibited
flares in which negative correlations between the flux and $PD$ were
observed: A flare of OJ~287 in JD~2454612--2454983 and a flare of
BL~Lac in JD~2454612--2454864 (figures~23 and 26). 

The sample reduces to 33 blazars once we exclude the objects in which
available polarization data were less than 10. The numbers of 
objects with significant positive and negative correlation
correspond to 30~\% and 12~\% of the sample, respectively. 

In addition, we calculated the correlation coefficients between the
$V-J$ color and $PD$.  The results are shown in table~2 as $r_{\rm
  col.-pol.}$. A positive (or negative) $r_{\rm col.-pol.}$ means that
an object becomes redder (or bluer) when $PD$ increases (or
decreases). The numbers of objects with significant positive and
negative correlation, and non-significant correlation are, 2, 3, and
29, respectively. The correlation between $V-J$ and $PD$ was quite weak.

\subsubsection{Rotation of polarization angle}

Possible rotations of polarization angle ($PA$) have been reported and
discussed both in optical and radio observations. However, a part 
of them was so sparsely sampled that it can be explained by a result
of a random walk in the Stokes $QU$ plane (e.g.~\cite{all81rotation};
\cite{jon85random}; \cite{kik88random}). Hence, polarimetric
observations with high observation density are required to establish real
rotation events of polarization. Recently, smooth rotations of optical
polarization have been reported in several blazars based on long-term
polarimetric observations (\cite{mar08bllac}; \cite{Sasada2010};
\cite{mar10pks1510}; \cite{fer103c279}; \cite{jor103c454}). Our
polarimetric observation allowed us to perform a systematic search for
such rotations.  

A search for polarization rotation events is, however, not
straightforward. Two issues should be taken into account; one is an
ambiguity of $180^\circ$ in $PA$, and the other is the presence of
multiple polarization components. First, we resolved the $180^\circ$
ambiguity by assuming that objects favor gradual changes in 
$PA$ rather than rapid changes. In other words, a difference in $PA$
between neighboring two observations should be smaller than
$90^\circ$. We added $-180^\circ$ (or $+180^\circ$) to the observed $PA$
in the case that $PA_n > PA_{n-1} +90^\circ$ ( or ${\rm
  PA}_n < PA_{n-1} -90^\circ$), where $PA_i$ is the $i$-th 
observation. The corrected $PA$ by this method is shown in the
left-bottom panel in the figures in Appendix.

Second, there are blazars in which a stationary or long-term variation
component of polarization has been proposed (e.g.~\cite{Moore1982};
\cite{vil10oj287}). Polarization rotations in such objects should be
discussed for each component, i.e., long- or short-term ones. In this
paper, we define the average $QU$ as a stationary polarization
component. The averages of $Q$ and $U$ were calculated iteratively; we
first calculated an average of $Q$ (or $U$) using all data, and
after discarding outliers ($>3\sigma$) from the sample, re-calculated
it. The iteration continued until no outliers remained. 

Even after those two-types of corrections were performed, we could not
find any clear universal correlations between the light curve and any
sets of corrected $PA$. However, we found several noteworthy
objects or flares in which possible rotations of polarization were
seen. We describe them below. 

\begin{figure} 
 \begin{center}
  \FigureFile(85mm,85mm){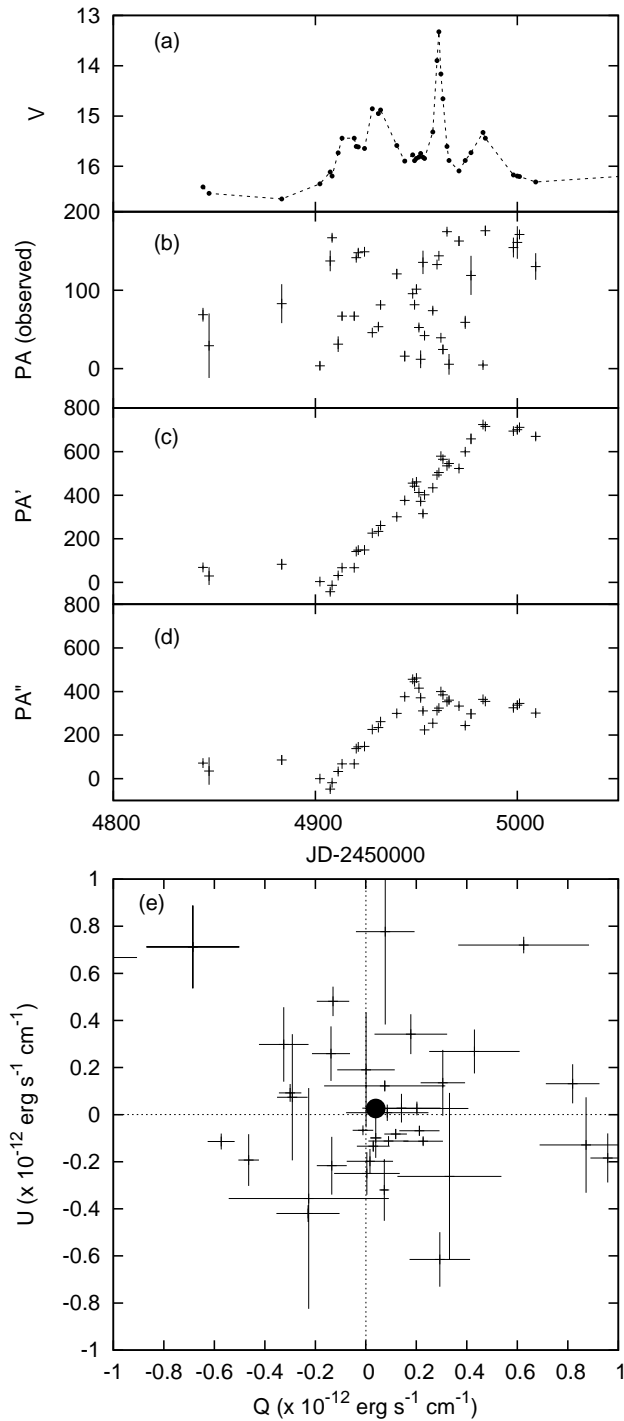}
 \end{center}
 \caption{Variation in $PA$ of PKS~1510$-$089 observed in
   2009. Panels~(a) the $V$-band light curve, (b) observed $PA$, (c)
   corrected $PA$ for the $180^\circ$ ambiguity, (d) $PA$ measured from
   the average $(Q,U)$, and (e) the $QU$ plane around the origin of
   $QU$.  In panel~(e), the average $(Q,U)$ is indicated by the filled
   circle.}\label{fig:PKS1510_pa} 
\end{figure} 

Figure~\ref{fig:PKS1510_pa} shows the temporal variation in the
observed and corrected $PA$ and the $QU$ distribution of
PKS~1510$-$089. The observed $PA$, shown in panel~(b), apparently
varies randomly. However, as shown in panel~(c), we can see an
increasing trend in the $180^\circ$-ambiguity corrected $PA$ during
an active phase of the object. This rotation event has also been
reported in \citet{mar10pks1510}, and confirmed in
\citet{sas10ao0235}. The rotation event continued for $\sim 90\;{\rm
  d}$ and its amplitude in $PA$ was over $700^\circ$. The long duration
and large amplitude of the rotation event are definitely difficult to
be explained by the result of a random walk in the $QU$ plane
(e.g.~\cite{Moore1982}).

Similar rotation events having a long duration and a large amplitude
were seen only in two objects: 3C~454.3 (figure~25) and PKS~1749$+$096
(figure~23).  All of those rotation events
have a common feature, that they were associated with active phases or
flares. This implies that the rotation events were related to the
variation in the magnetic field direction in the emitting region of
the flares. Another possible rotation event can be seen in
S5~0716$+$714 (figure~29). The change in $PA$ of this object was,
however, not as smooth as observed in the above three objects, but rather
had a step-like profile.  It continued for a very long time of $\sim
400\;{\rm d}$.

It should be noted that the $180^\circ$-ambiguity correction is
sensitive to the sampling density of observations.  Furthermore, the
results should be considered carefully if there are multiple
polarization components. The behavior of $PA$ in PKS~1510$-$089
apparently changes if the origin of $(Q,U)$ is shifted to their
observed averages, as shown in panel~(d) of 
figure~\ref{fig:PKS1510_pa}. The increasing trend in $PA$ is terminated
earlier than that shown in panel~(c). The average $(Q,U)$ was
$(3.9\times 10^{-14},2.6\times 10^{-14})$ (in
erg~s$^{-1}$~cm$^{-1}$~Hz$^{-1}$), indicated by the filled circle in
the $QU$ plane (panel~e). This average $(Q,U)$ corresponds to a $PD$ of
$\sim 0.6$\% at $V=16.0$. This result demonstrates that the 
$180^\circ$-ambiguity correction could be affected by another
polarization component, even if it has a quite small contribution to
the total flux. The polarization rotation in PKS~1510$-$089 probably
started at the onset of the active phase. It was then interrupted by
a sharp spike-like flare in $\sim$JD~2454960. After this spike, it is
unclear whether the rotation continued or not. 

\begin{figure} 
 \begin{center}
  \FigureFile(85mm,85mm){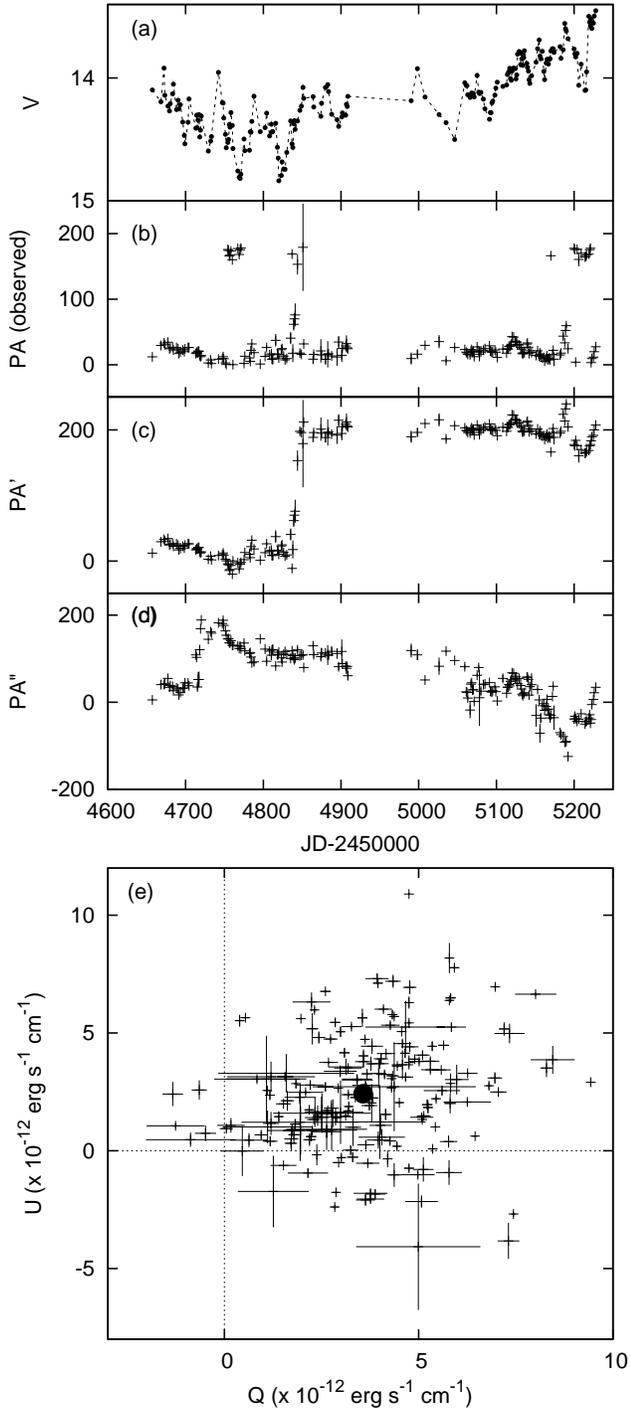}
 \end{center}
 \caption{Variation in $PA$ of 3C~66A. Panels~(a) the
  $V$-band light curve, (b) observed $PA$, (c) corrected $PA$ for the 
  $180^\circ$ ambiguity, (d) $PA$ measured from the average
  $(Q,U)$, and (e) the $QU$ plane in which all observations are
   included. In panel~(e), the average $(Q,U)$ is indicated by the
   filled circle.}\label{fig:3C66A_pa}
\end{figure} 

Another noteworthy case is 3C~66A. Figure~\ref{fig:3C66A_pa} is the
same as figure~\ref{fig:PKS1510_pa}, but for 3C~66A. Our 
$180^\circ$-ambiguity correction yields a short-term rotation
event between JD~2454840--2454850, as can be seen in panel~(c).
However, this feature cannot be seen in panel~(d), in which $PA$ are
measured from the observed average of $(Q,U)$. As shown in the $QU$
plane in panel (e), the observed $QU$ are distributed not around the
origin of $(Q,U)$, but around the average point indicated by the
filled circle. Hence, this observation suggests that the observed
polarization in 3C~66A can be interpreted as a composition of two
polarization components, namely, long- and short-term variation
components. If this is the case, the apparent $PA$ variation in
panel~(c) means no real variation in the magnetic field direction in
the emitting area. A long-term component should be subtracted from
observed polarization when the rotation of polarization is discussed
for objects like 3C~66A.

Similar cases to 3C~66A can be found in RX~J1542.8$+$612 (figure~27)
and QSO~0454$-$234 (figure~24); short-term rotation events appeared in
their corrected $PA$.  Furthermore, their average $(Q,U)$ were
significantly deviated from their origin of $(Q,U)$. 
\citet{mar08bllac} have reported the detection of a polarization rotation
in BL~Lac. The rotation event was well sampled, continuing for about
one week and having an amplitude in $PA$ of $240^\circ$. This object
has, however, been proposed to have two polarization
components. It is known that the observed average of $(Q,U)$ is
significantly deviate from the origin of $(Q,U)$
(\cite{Hagen-Thorn2008}). Our observation confirmed it, as can be seen
in figure~26. If short-term variation components are superimposed on
the stationary or long-term polarization component, the model for the
rotation event reported in \citet{mar08bllac} should be
reconsidered.

\begin{figure} 
 \begin{center}
  \FigureFile(85mm,85mm){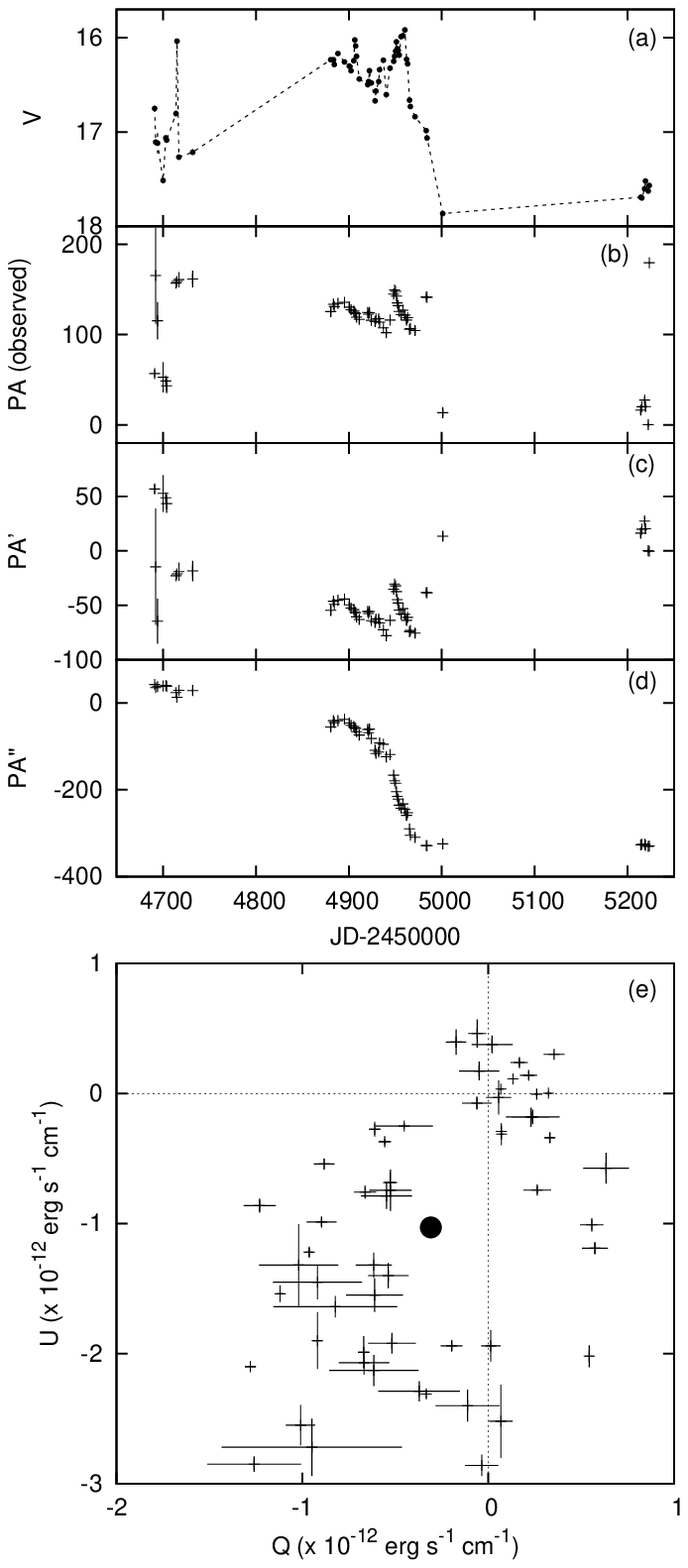}
 \end{center}
 \caption{As for figure~\ref{fig:3C66A_pa}, but for
  PKS~1502$+$106.}\label{fig:PKS1502_pa}
\end{figure} 

Finally, we show the case for PKS~1502$+$106 in
figure~\ref{fig:PKS1502_pa}. A rotation of polarization is not seen in
the observed and $180^\circ$-ambiguity corrected $PA$ [panels (b) and (c)],
but can be seen in the origin-shifted $PA$ [panel (d)]. The validity of this
correction is, however, brought into question. The observed $(Q,U)$ is
expected to be distributed centered around the stationary component in
the two-component picture. However, there is only few points at the
vicinity of the average $(Q,U)$, as shown in panel~(e). This result is 
inconsistent with the picture of the two-component model. This example
tells us that the shift of the origin of $(Q,U)$ could generate false
variations in $PA$ in some cases.  

In summarizing this subsection, we offer several cautions to avoid
misinterpretation of observed $PA$ variations in terms of the polarization
rotations. First, it is required that a rotation episode is well
sampled when it is proposed to be a real one. A rotation episode
indicated by only a few data points, cannot be distinguished from
results of a random walk in the $QU$ plane. Second, a rotation event
would more likely to be real when it has a larger variation
amplitude in $PA$. An apparent rotation having a small amplitude of
$<180^\circ$ can be readily produced by a composition of a few flares
having different $PA$. Third, the appearance of a rotation event should
not be changed by a small shift of the origin of $(Q,U)$. As mentioned
above, a part of the rotation event in PKS~1510$-$089 is quite
sensitive to a small change of the origin of $(Q,U)$. Finally, the
temporal behavior in the $QU$ plane should be carefully checked not
only during the rotation event, but also before and after it. If the
observed average of $(Q,U)$ is significantly deviated from the origin,
a stationary or long-term variation component in polarization could be
present. Then, polarization variations should be discussed after the
long-term component is subtracted from the observation. 

\subsection{The $\nu_{\rm peak}$ and luminosity dependence of variability}

\begin{figure*} 
 \begin{center}
  \FigureFile(170mm,170mm){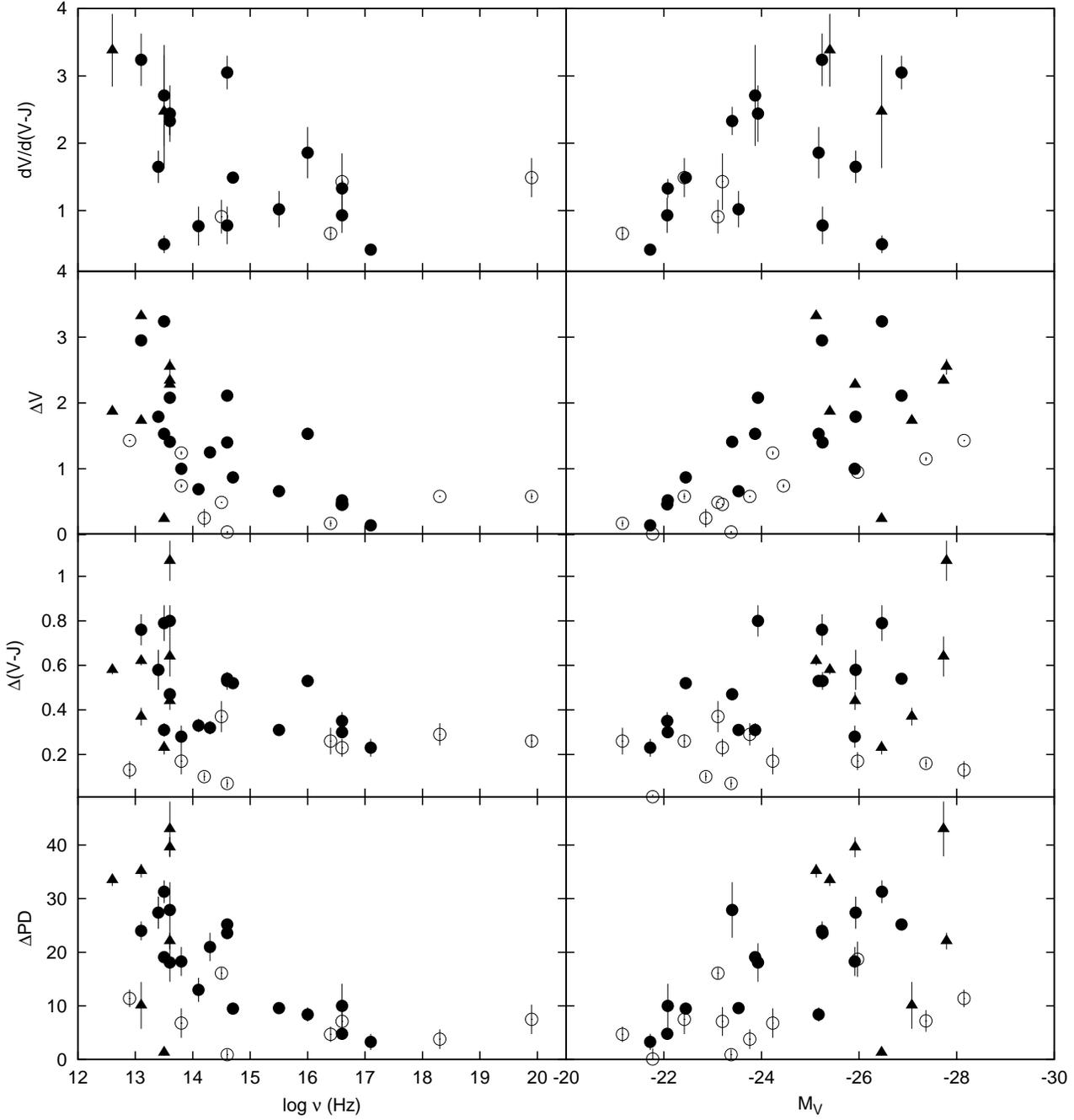}
 \end{center}
 \caption{The $\nu_{\rm peak}$ and $M_V$ distribution of the degree of
   variability. From top to
   bottom, the panels show $dV/d(V-J)$, the peak-to-peak variation
   amplitudes of the $V$-band flux ($\Delta V$), the $V-J$ color
   [$\Delta (V-J)$], and the $V$-band $PD$ ($\Delta PD$). The filled
   circles, triangles, and open circles represent BL~Lac objects,
   FSRQs, and poorly observed objects in which the number of
   observations are less than 20, respectively. }\label{fig:var2}
\end{figure*} 

In this subsection, we report on the dependence of the degree of
variability in the flux, color, and $PD$ on $\nu_{\rm
  peak}$. The $\nu_{\rm peak}$ is related to the luminosity of the
blazars in our sample 
(see subsection~3.1). We note that variation amplitudes may be
underestimated because blazars generally have a variation time-scale
longer than our observation period. To minimize this observation bias,
we calculated the slope in the color-magnitude diagram, $dV/d(V-J)$,
as well as ordinary peak-to-peak amplitudes.  We estimated the slope,
$dV/d(V-J)$, by fitting a linear function to the observed $V$ and
$V-J$. $dV/d(V-J)$ can be considered as an index of the degree of the
flux variability against the color variation. 

Figure~\ref{fig:var2} shows the dependence of the degree of variations
on $\nu_{\rm peak}$ and the average absolute magnitude, $M_V$, in the
left and right panels, respectively. From top to bottom, the panels
show $dV/d(V-J)$, the peak-to-peak variation amplitudes of the
$V$-band flux ($\Delta V$), the $V-J$ color [$\Delta (V-J)$], and the
$V$-band $PD$ ($\Delta PD$). In the case of 3C~454.3 and PKS~1510$-$089,
$dV/d(V-J)$ were calculated using the data in their bright states
($V<15$ for 3C~454.3 and $<15.5$ for PKS~1510$-$089). This is for
eliminating the redder-when-brighter stage (see, sub-subsection~3.2.2).

All panels in figure~\ref{fig:var2} suggest that the degrees of
variation were low in high $\nu_{\rm peak}$, or faint objects. 
They can be high in low $\nu_{\rm peak}$, or bright objects. The
poorly observed objects represented by the open 
circles tend to be distributed in low variability regions compared
with the others. This is naturally interpreted as underestimations of
the variation amplitudes in those objects due to the short period of
observations. Several FSRQs, particularly, 3C~273 and QSO~0454$-$234, 
exhibit a small variation amplitude, although they have a quite low
$\nu_{\rm  peak}$. This atypical feature can be understood by large
contamination of the thermal emission from accretion disks. Those
invariable component probably made the variability of the jet
component small, apparently. Thus, the variation amplitudes of poorly
observed objects (the open circles) and FSRQs (the filled triangles)
can be considered as lower limits of the amplitudes. Then, we can see
good correlations of $dV/d(V-J)$, $\Delta V$, $\Delta(V-J)$ , and
$\Delta PD$ with $\nu_{\rm peak}$ and $M_V$. Those characteristics of
the degree of variation suggest that the variability is stronger in a
low $\nu_{\rm peak}$ object, and hence in the emission from higher
energy electrons.

The fact that the variability in the flux and color is larger in 
lower $\nu_{\rm peak}$ objects has been reported in previous studies
(e.g.~\cite{ulr97review}). The present study confirmed this feature
with a large sample. In addition, the same feature can be also seen in
the variability of $PD$. This implies that the flares in blazars are
generally associated with the variation of $PD$ (see also,
\cite{sas10ao0235}). 
 
\section{Discussion}

\subsection{``Bluer-When-Brighter'', as a Universal Aspect of
 Blazars}

Our $V$- and $J$-band photometric observation showed that the
``bluer-when-brighter'' trend was detected in a high fraction of 
blazars we observed: 28/32 well-observed sources. The fraction of
bluer-when-brighter objects is quite high compared with those reported
in previous studies: \citet{Gu2006} investigated variations in $V-I$
of three FSRQs and five BL~Lac objects.  They found that three BL~Lac
objects exhibited significant bluer-when-brighter trends, while two
FSRQs exhibited redder-when-brighter trends. \citet{ran10color} 
performed an analogous analysis for six FSRQs and six BL~Lac objects.
They found that three BL~Lac objects exhibited significant
bluer-when-brighter trends, and four FSRQs exhibited 
redder-when-brighter trends. Those studies thus provide lower
fractions of bluer-when-brighter objects than our results. On the
other hand, several studies have indicated that the
bluer-when-brighter trend is common in BL~Lac objects:
\citet{dam02alpha} have investigated variations in the optical spectral
index of eight BL~Lac objects, and found that all objects showed
bluer-when-brighter trends. \citet{fio04color} report on their
photometric study of 37 blazars, and propose that bluer-when-brighter
trends are more prominent in objects showing stronger variability. In
those two studies, no evaluation of the significance of the
correlation has been given. \citet{vag03alpha} also analyzed the
spectral slope variability for eight BL~Lac objects, and reported that
all objects showed bluer-when-brighter trends. However, the sampling 
rate of their data was $\sim 25$~d, and it is too large to study
variability on a time scale of days. Thus, our present study, for the
first time, yields an unambiguous confirmation that the
bluer-when-brighter trend is a common feature in variations on time
scales of days--months in blazars.

We propose that there are two factors that can disturb the general
bluer-when-brighter trend. First, a strong contribution of a blue
thermal emission from the accretion disk leads to a
redder-when-brighter trend. A typical case is 3C~454.3, as shown in
panel~(b) of figure~\ref{fig:exam}. The observed color variation
suggests that the redder-when-brighter trend was observed only when
the objects were faint. This color behavior is consistent with the
scenario that the contribution of the thermal emission from the
accretion disk becomes strong when the synchrotron emission from the
jet weakened (\cite{Villata2006}). Even in the redder-when-brighter
objects, our observation revealed that the bluer-when-brighter trend
appeared when they were bright. 

\begin{figure} 
 \begin{center}
  \FigureFile(85mm,85mm){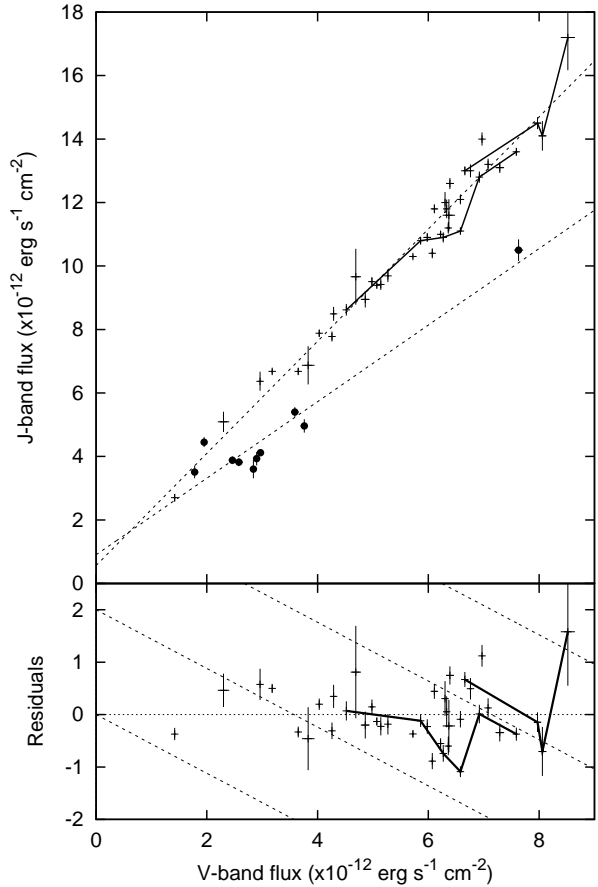}
 \end{center}
 \caption{Upper panel: $J$- and $V$-band flux observed in
  PKS~1502$+$106. The filled circles and crosses represent the
  observations in 2008 and 2009, respectively. The dashed lines
  indicate the best-fitted linear regression models for each year. The
  thick solid lines represent rising phases of large flares. 
  Lower panel: Residuals of the $J$-band flux from the best-fitted
  model for the 2009 data. The dashed lines represent the slope of the
  best-fitted model for the 2008 observations.}\label{fig:pks1502col2} 
\end{figure} 

Second, the bluer-when-brighter trend could be hidden if there are
multiple variation components having different bluer-when-brighter
sequences. We see the color behavior of long-term variation 
components, taking the case of PKS~1502$+$106 as an example. Panel~(f)
of figure~\ref{fig:exam} shows the light curve, color, and
color-magnitude diagram of PKS~1502$+$106. A bluer-when-brighter
trend was not detected in the whole data of this object, but
significantly detected in each year, as reported in sub-subsection~3.2.2. 
Figure~\ref{fig:pks1502col2}
shows the $F_J$--$F_V$ diagram of PKS~1502$+$106. The filled circles
and crosses represent data obtained in 2008 and 2009,
respectively. As can be seen in this figure, $F_J$ can be described
with a linear function of $F_V$ both in the 2008 and 2009 data. The
dashed lines in the figure represents the best-fitted linear
regression models for each year. The slope of the 2008 data in the
$F_J$--$F_V$ diagram is clearly different from that of 2009; a steep
slope in the 2009 data compared with the 2008 one. As can be seen from
figure~\ref{fig:exam}, short-term flares are apparently superimposed
on a long-term outburst component in the 2009 light curve. The
long-term component reached the maximum at the beginning of our 2009
observation, and then started a gradual decay through that
year. Hence, the general trend seen in the $F_J$--$F_V$ diagram of the
2009 data actually originates from the long-term outburst
component. In contrast, the object almost remained fainter than $V=17$
in 2008, while several short-term flares brighter than $V=17$ were
observed. There was no prominent outburst component in 2008.  
 
It is interesting to note that, in the 2009 season, the $F_J$--$F_V$
slopes of the early phase of the short-term flares were possibly
analogous to those in 2008.  The thick solid lines in  
figure~\ref{fig:pks1502col2} represent the rising phases of the
prominent short-term flares peaked at JD~2454951 and 2454960. The
lower panel shows the residuals of $F_J$ from the best-fitted model
for the 2009 data. The dashed line in the lower panel indicates the
$F_J$--$F_V$ slope of the 2008 data. The figure suggests that there
were stages of shallow $F_J$--$F_V$ slopes during the early rising
phases of the flares in 2009. After these stages, the slope appears to
become very steep near the flare peaks. This may be a sign of spectral
hysteresis, which was frequently observed in short-term flares, as
mentioned in sub-subsection~3.2.3. The early shallow $F_J$--$F_V$ slopes
look close to the slope of the short-term flare in 2008.

Those observations give us an idea that the color behavior of the
short-term flares in 2008 was the same as that in 2009.  In addition,
the color of the outburst component in 2009 was probably redder than
the short-term flares. As well as PKS~1502$+$106, PKS~0048$-$097 and
H~1722$+$119 clearly showed different color behavior in each year, as
shown in the lower panels of figure~\ref{fig:exam}.  Those phenomena
may also be due to the presence of a long-term component, whose color was
different from that of short-term flares, and moreover, variable with
time. This idea is supported by the fact that correlation coefficients
were low even in the bluer-when-brighter sources; as shown in table~2,
$r_{\rm col}$ was less than $0.70$ in more than half of objects. The low
correlation-coefficients can be interpreted by the scenario that the
observed color is a composition of short- and long-term components
that have different bluer-when-brighter sequences. The two-component
scenario has also been proposed to explain the color behavior of
BL~Lac (\cite{Villata2002}; \cite{Villata2004}).

The above two factors, namely the presence of the thermal disk
emission and multiple synchrotron variation components, can disturb
the general bluer-when-brighter trend in blazar variability. Our
observation overcomes those factors by covering a wide range of
wavelengths, namely, the simultaneous optical and NIR observations and
a high sampling rate. Those two advantages
of our observation led to the high fraction of bluer-when-brighter
objects compared with previous studies. In other words, one should
carefully discuss apparent redder-when-brighter trends or achromatic
variations observed in blazars, while taking into account the above two
factors that can disturb the universal bluer-when-brighter trend. 

It has been suspected that the bluer-when-brighter trend is common
in BL~Lac objects, but not in FSRQs because of a large contribution of 
thermal emission in FSRQs (\cite{Gu2006}; \cite{ran10color}). Our
objects that were observed more than 10~times include seven FSRQs and
seven LBLs. The correlation analysis with all data sets showed that 
significant bluer-when-brighter trends were detected in two FSRQs
and six LBLs (see table~2). This result confirms the previous
studies, which suggest that the color--flux correlation of FSRQs is
different from that of BL~Lac objects. However, the bluer-when-brighter
trend was significantly detected even in the other four FSRQs when
they were bright (3C~454.3, PKS~1510$-$089, PKS~1502$+$106, and  
QSO~0454$-$234), as mentioned in sub-subsection~3.2.2.  Hence, the jet
emission in FSRQs presumably has the same color behavior as that in
BL~Lac objects.

The relationships of $F_J$ and $F_V$ of 3C~454.3 and PKS~1510$-$089
suggest that their variation components did not have a constant
color, but a blueing trend with an increase in flux (see
figure~\ref{fig:pks1510ff} and \ref{fig:3c454ff}). Alternatively,
it can be interpreted with the scenario that two variation components
had different time-scales and colors, as proposed above. A red and
long-term outburst component would be dominant in their low flux
regime.  The redder-when-brighter trend was observed probably because
an underlying source was the blue thermal emission. The emission from
short-term flares would be dominant in the high-flux regime, resulting
in the standard bluer-when-brighter trend. This scenario is supported
by the light curve structure; short-term flares, making the objects
brightest, were superimposed on long-term outburst components (see
figures~25 and 23).

In general, variations on time-scales of months--years are expected to
have larger amplitudes than those of days--weeks in blazars
(\cite{huf92variation}; \cite{tak00mrk421}; \cite{kat01variation};
\cite{cha083c279}). Hence, the variation amplitudes reported in
subsection~3.4 are considered to be those of the long-term variation
component. In subsection~3.2.3, we reported that the color of the
variation component was constant in most objects. Since this
constant-color feature is a result from the whole data set obtained
during our observation period, it is probably a feature of the
long-term variation component.

Thus, the implications for the blazar variability obtained from our
observation are summarized below: The short-term flares on time scales
of days--weeks tended to exhibit spectral hysteresis; their 
rising phase was bluer than the decay phase around the flare maxima.
They were occasionally superimposed on the long-term variations on
time scales of months--years.  The color of the long-term component
was apparently constant. In addition to those short- and long-term
synchrotron components, a stationary component was probably
embedded. The bluer-when-brighter trend is commonly observed because
this stationary component had a redder color than the variation
components. A part of FSRQs has a blue and thermal emission as an
underlying component.

\subsection{Physical Implication from the Bluer-When-Brighter Trend}

Several models have been proposed to explain the bluer-when-brighter
trend in blazars. When we consider two distinct
synchrotron components, the bluer-when-brighter trend could be
explained if a flare component has a higher $\nu_{\rm peak}$ than an
underlying component. Even when we consider a single
synchrotron component, it could be explained by a shift of its 
$\nu_{\rm peak}$ to a higher region. In either case, high $\nu_{\rm
  peak}$ can be obtained by the injection of energy into emitting
regions, and thereby, the increase of the number of high energy
electrons. The most plausible scenario for the origin of
the energy injection is internal shocks in relativistic shells
(e.g.~\cite{spa01shock}; \cite{zha02lag}). Besides the energy
injection scenario, the bluer-when-brighter trend can be explained by
a change in the beaming factor, $\delta$, of the emitting region.
This is because it leads to not only an apparent increase in the
observed flux, $\nu f_\nu \propto \delta^4$, but also a shift of the
observed frequency, $\nu_{\rm obs} \propto \delta$
(e.g. \cite{Villata2004}; \cite{pap07doppler}; \cite{lar10bllac}). 
 
The short-term variations on a time scale of days--weeks tended to
exhibit the spectral hysteresis in the $F_J$--$F_V$ diagram, as
reported in sub-subsection~3.2.3. This
hysteresis presumably has the same nature as the hysteresis which have
been seen between the flux and color, or spectral index
(\cite{kat00pks2155}; \cite{cip03gc0109};
\cite{wu07s50716}). \citet{Kirk1998} reproduce the hysteresis between 
the flux and spectral index using their model for the acceleration of
electrons in a shock region. The fact that the short-term variations
tended to exhibit the spectral hysteresis supports the energy
injection scenario. It is difficult to explain such a hysteresis
pattern with a change in $\delta$.

The long-term variation components on a time scale of months--years
apparently had a constant color. The bluer-when-brighter trend was
observed because the underlying source was redder than the variation
component, as mentioned in sub-subsection~3.2.3 and the last
subsection. The emission of the underlying source is presumably
synchrotron emission, because the thermal emission from AGN should be
much bluer than the synchrotron emission of the long-term variation
component. The red color of the underlying source suggests that
its $\nu_{\rm peak}$ is lower than that of the long-term component.

In subsection~3.4, we showed that a higher $\nu_{\rm peak}$ objects
exhibited smaller variations all in the flux, color, and $PD$.
Based on the discussions in the last subsection, it suggests that the
variation amplitude of the long-term component was small in a region
of $\nu<\nu_{\rm peak}$ and vice versa. This is preferable for the
scenario that the long-term variation is also attributed to the energy
injection events. If changes in $\delta$ would be the major cause, it
means an eccentric situation that $\delta$ changes only occur in
components having high $\nu_{\rm  peak}$. Otherwise, large variations
would be observed in a region of $\nu<\nu_{\rm peak}$ when low
$\nu_{\rm  peak}$ components are amplified by an increase in
$\delta$. The energy injection scenario can explain the $\nu_{\rm
  peak}$ dependence of the variation amplitude without such an
eccentric situation.

\subsection{Implications from Polarization Variations}

The high fraction of bluer-when-brighter objects implies that
there is a major mechanism common to blazar variations. The most
plausible explanation is the energy injection scenario, as mentioned
in the last subsection. The internal shock is a promising candidate
for the mechanism of the energy injection to the emitting region. If
the internal shock results in an aligned magnetic field in the emitting
region, a universal aspect of polarization variations can be expected. 

In contrast to the strong correlation between the flux and color, the
correlation between the flux and $PD$ was weak; 30~\% and 12~\% of the
sample exhibited significant positive and negative correlations,
respectively. On the other hand, these fractions of objects are too
high to conclude that the polarization variations are completely
random. It is possible that there is no universal law of polarization
associated with blazar variations. If this is the case, our
observation indicates that flares could occur independent of the
structure of the magnetic field in the flaring source.

Alternatively, a universal law could be hidden in the observed
polarization, although it was present. Polarization can be described
with two parameters, for example, $PD$ and $PA$. It can also be treated like a 
two-dimension vector with Stokes $QU$ parameters. This property of
polarization is an important difference from that of color. The weak
correlation of the flux and $PD$ may be caused by the presence of two or more
polarization components. Even if a flare has specific $PD$ and $PA$, they
would disappear in the observed $PD$ and $PA$ variations because of the
superposition of other flares having different $PD$ and $PA$. Based on
this picture, we have developed a method to separate the observed
polarization into two components: A short-term variation component
having a positive correlation between the total and polarized fluxes,
and a long-term one (\cite{Uemura2010}). Applying this method to our
blazar data, we found that the behavior of polarization in OJ~287 and
S2~0109$+$224 can be explained with this two-component picture. The
two-component picture is also supported by the color behavior as
described in subsections~4.1 and 4.2. 

There are three well-observed blazars in our sample which showed 
a significant negative correlation between the flux and $PD$: OJ~49,
ON~231, and 3C~66A. In addition, temporary negative correlations were
observed in BL~Lac and OJ~287, as mentioned in subsection~3.3.2.  
It is noteworthy that the observed $QU$ of all those five objects do
not center at the origins of $QU$, but concentrate around regions
significantly apart from the origins. Such distributions are also seen
in their $(Q/I,U/I)$ plane. These distributions strongly suggest the
presence of long-term polarization components. The negative
correlation means that the flux variations were associated with the
systematic changes in polarization.  If the long-term component is
subtracted, the variations might change to those showing positive
correlation.  Hence, the negative correlation can also be explained by
the two-component scenario. 

\section{Summary}

We performed photopolarimetric monitoring of 42 blazars in the optical
 and near-infrared band using the Kanata telescope.
Our findings are summarized below:
 \begin{itemize}
  \item The bluer-when-brighter trend was observed in 28 blazars, 
   which correspond to 88~\% of our well-observed sample. Our
   observation unambiguously confirmed that it is a universal aspect
   in blazars.
  \item The redder-when-brighter trend was observed in three objects
    (3C~454.3, PKS~1510$-$089, and PG~1553$+$113) when they were
    faint. Even in those three objects, the bluer-when-brighter trends
    can be seen when they were bright.
  \item The bluer-when-brighter trend was, in general, generated by a
   variation source apparently having a constant color and an
   underlying red source. 
  \item Prominent short-term flares tended to exhibit a spectral
   hysteresis; their rising phase was bluer than their decay phase
   around the flare maxima.
  \item Significant correlations of the flux and the polarization
   degree were observed only in 10 blazars, which correspond to 30~\%
   of our well-observed sample. The correlation was weak compared
   with that between the flux and color.
  \item The rotations of polarization were observed in
   PKS~1510$-$089, 3C~454.3, and PKS~1749$+$096, and possibly in
   S5~0716+714. We demonstrated that false events of polarization
   rotations could be detected in the case that multiple polarization
   components are present. 
  \item We found that the variation amplitudes were smaller in low
    $\nu_{\rm peak}$ objects all in the flux, color, and polarization
    degree.
 \end{itemize}
Based on those results, we propose that there are several distinct
variation sources on different time-scales, colors, and
polarizations in blazars. Both short- and long-term variation
components are probably attributed to the energy injection into
emitting sources, for example, with the internal shock in relativistic
shells. 

This work was partly supported by a Grand-in-Aid from the Ministry of
Education, Culture, Sports, Science, and Technology of Japan
(22540252 and 20340044).

\appendix
\section{Temporal Variations in Flux, Color, and Polarization}

Figures~21--34 show temporal variations in flux, color, and
polarization of blazars obtained by our observation. The left
panels are temporal variations in five parameters. From top to bottom,
the panels show the $V$-band light curve, the $V-J$ color, the
polarization degree ($PD$) in percent, the polarization angle ($PA$) in
degree, and the corrected $PA$. The last one was corrected for the
$180^\circ$ ambiguity of $PA$ (for a detail, see sub-subsection~3.3.3). The 
upper and lower middle panels show the color-magnitude diagram and the
$PD$-magnitude diagram, respectively. The upper and lower right panels
show the distribution of $(Q/I,U/I)$ and $(Q,U)$, respectively.

\begin{figure*}
 \FigureFile(160mm,170mm){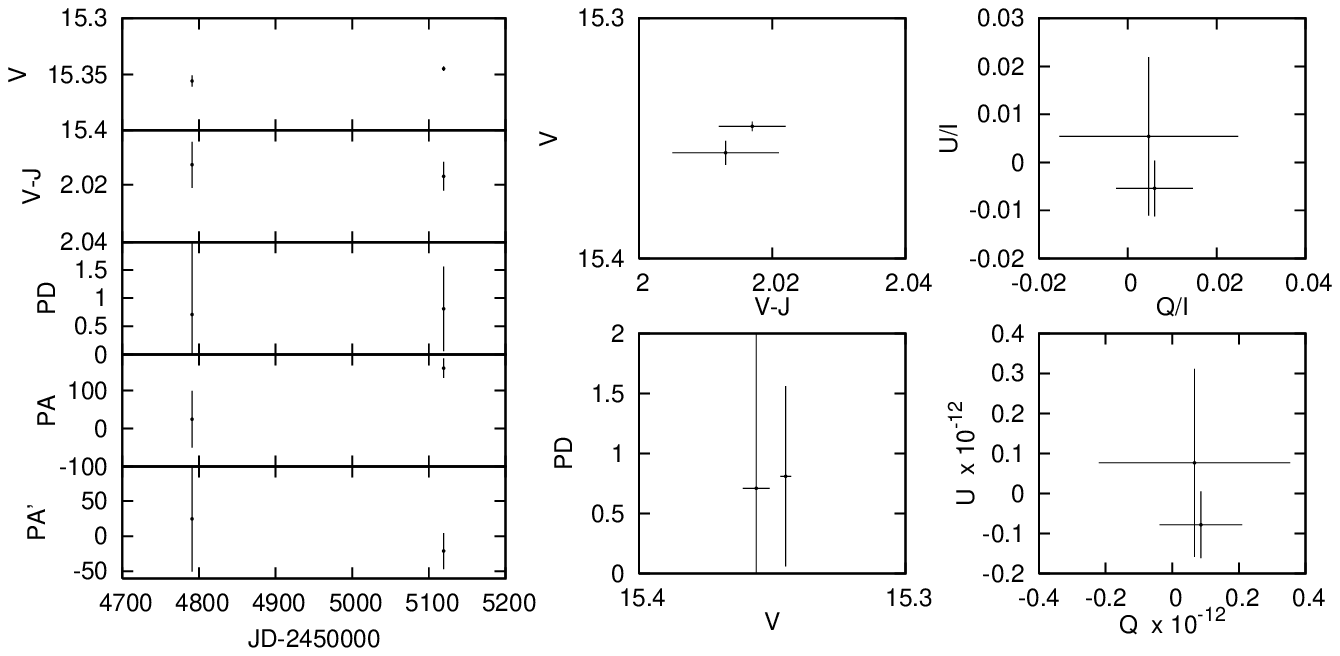}
 \FigureFile(160mm,170mm){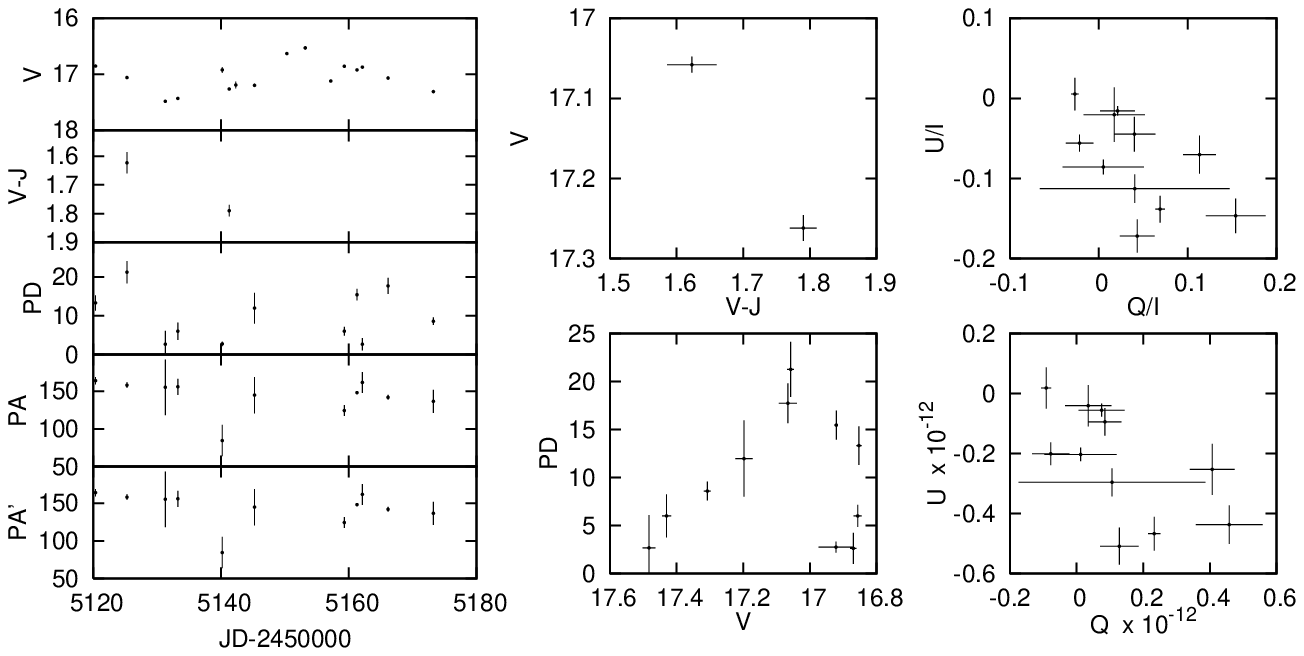}
 \FigureFile(160mm,170mm){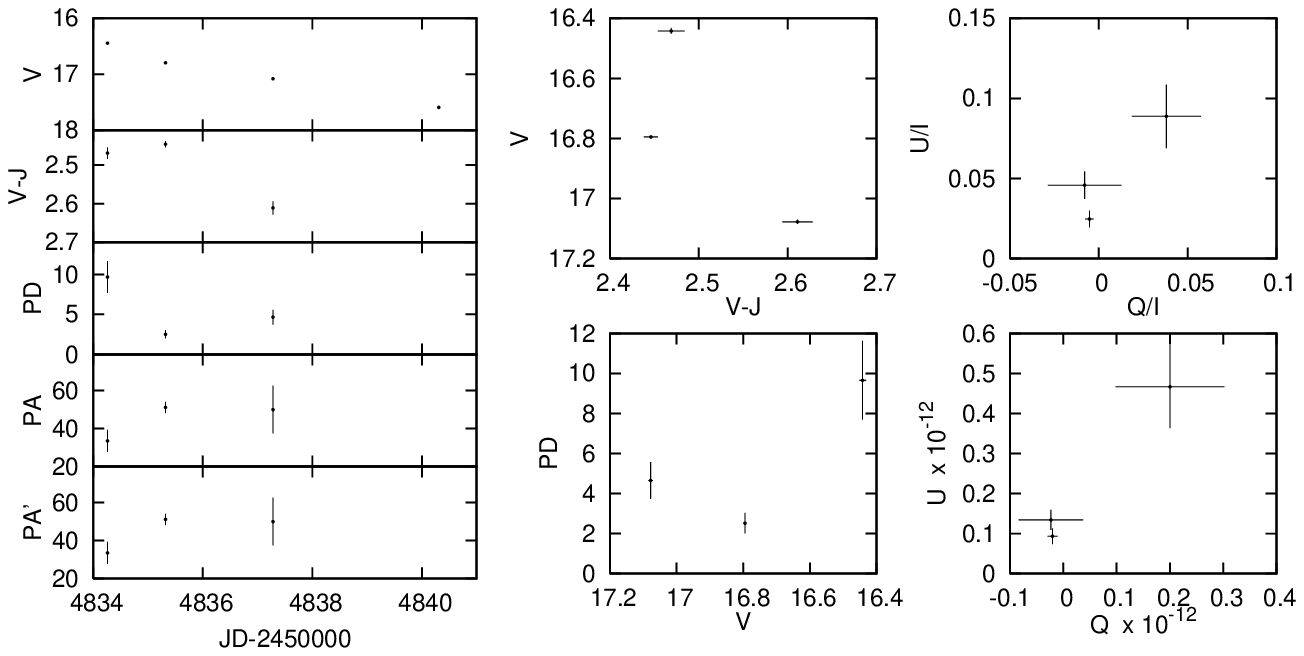}
 \caption{QSO~0324$+$341 (top), 4C~14.23 (middle), and QSO~1239$+$044
  (bottom).}
\end{figure*}
\begin{figure*}
 \FigureFile(160mm,170mm){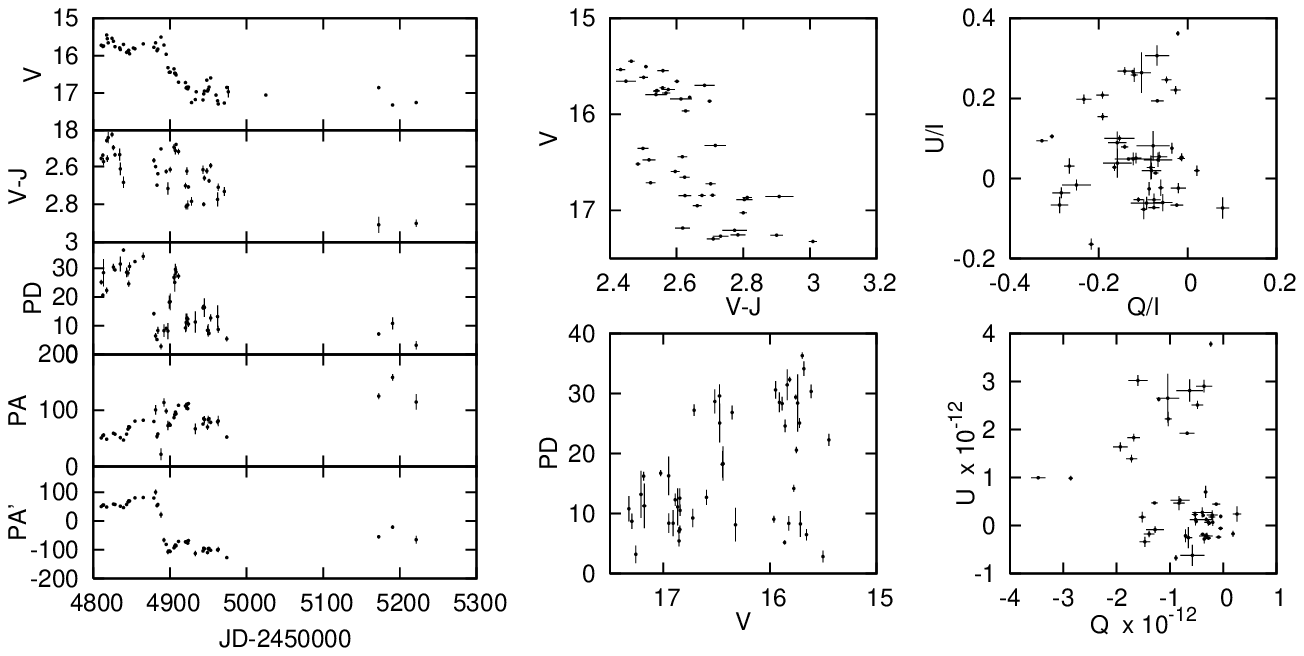}
 \FigureFile(160mm,170mm){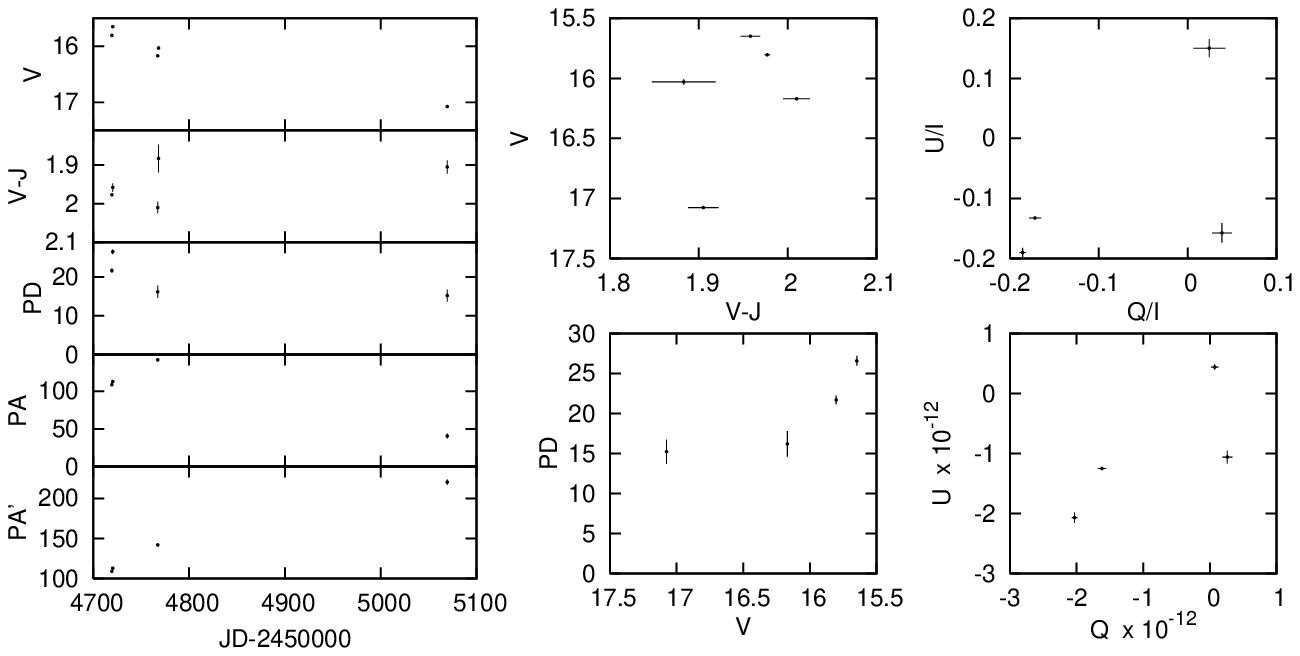}
 \FigureFile(160mm,170mm){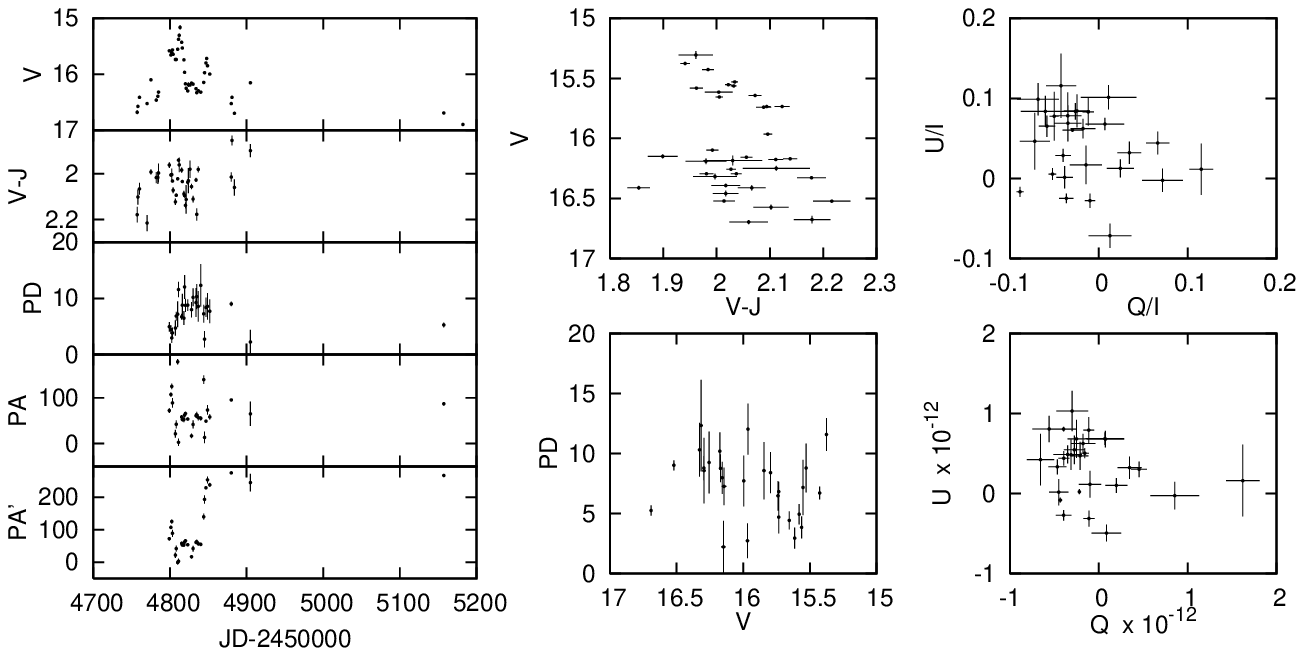}
 \caption{3C~279 (top), PKS~0215$+$015 (middle), and QSO~0454$-$234
  (bottom).}
\end{figure*}
\begin{figure*}
 \FigureFile(160mm,170mm){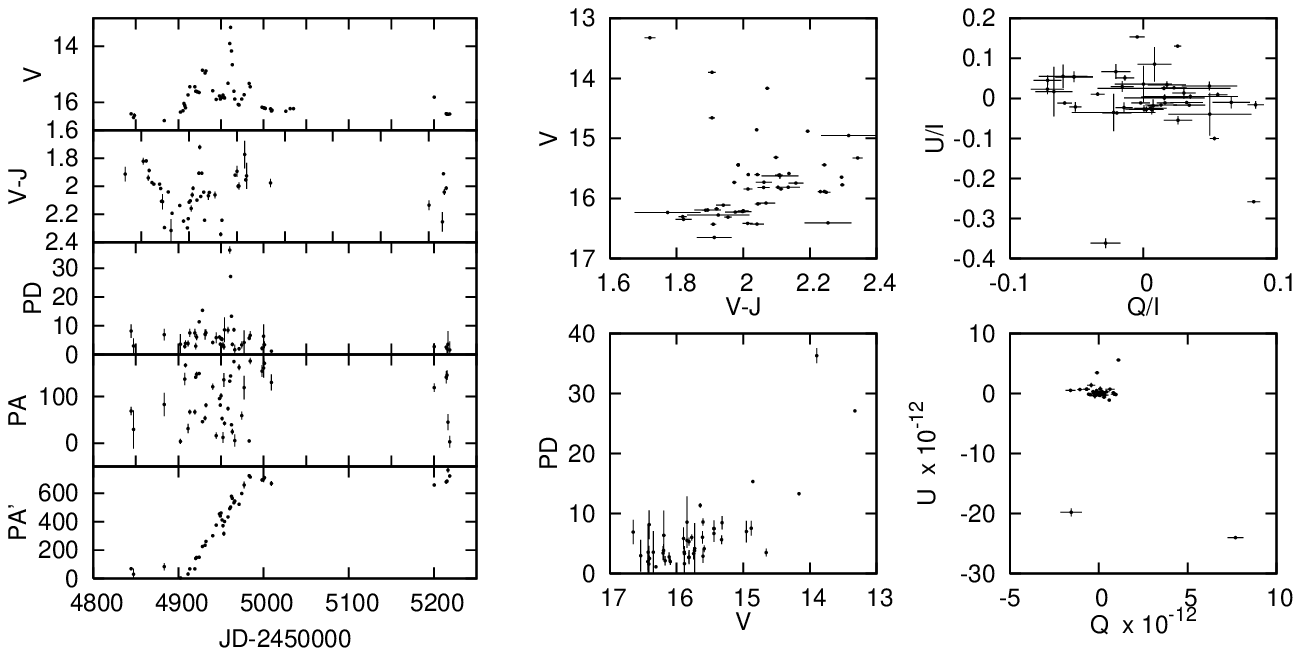}
 \FigureFile(160mm,170mm){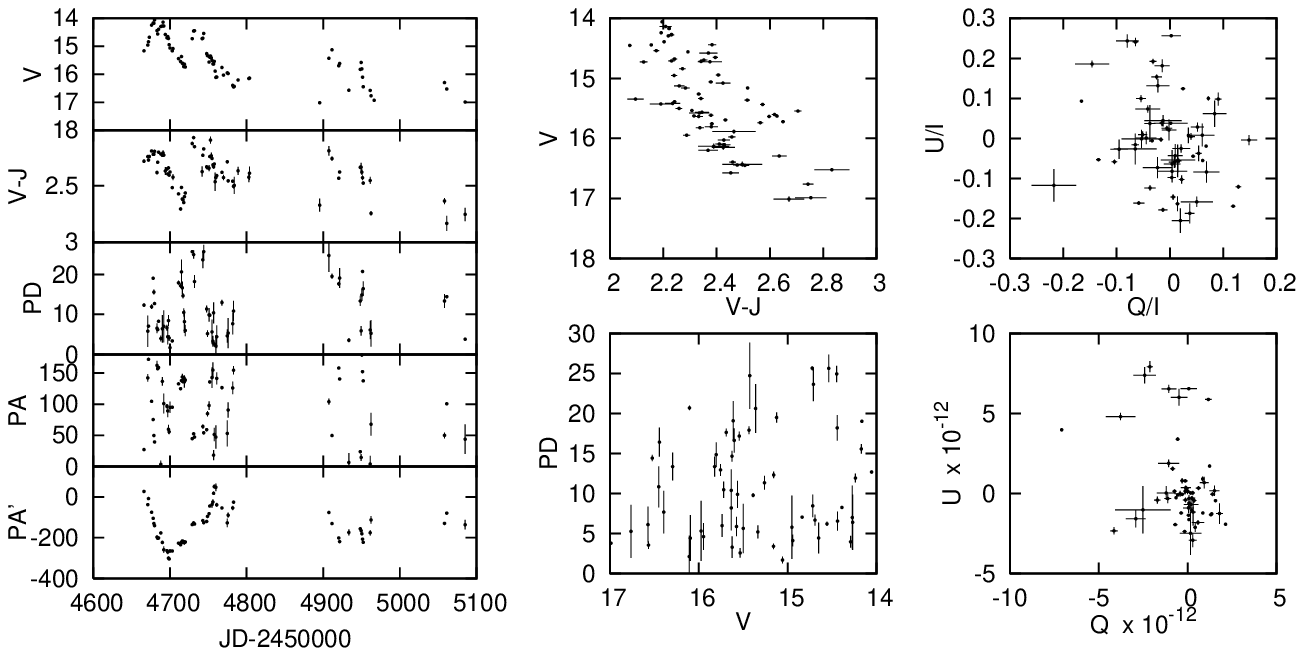}
 \FigureFile(160mm,170mm){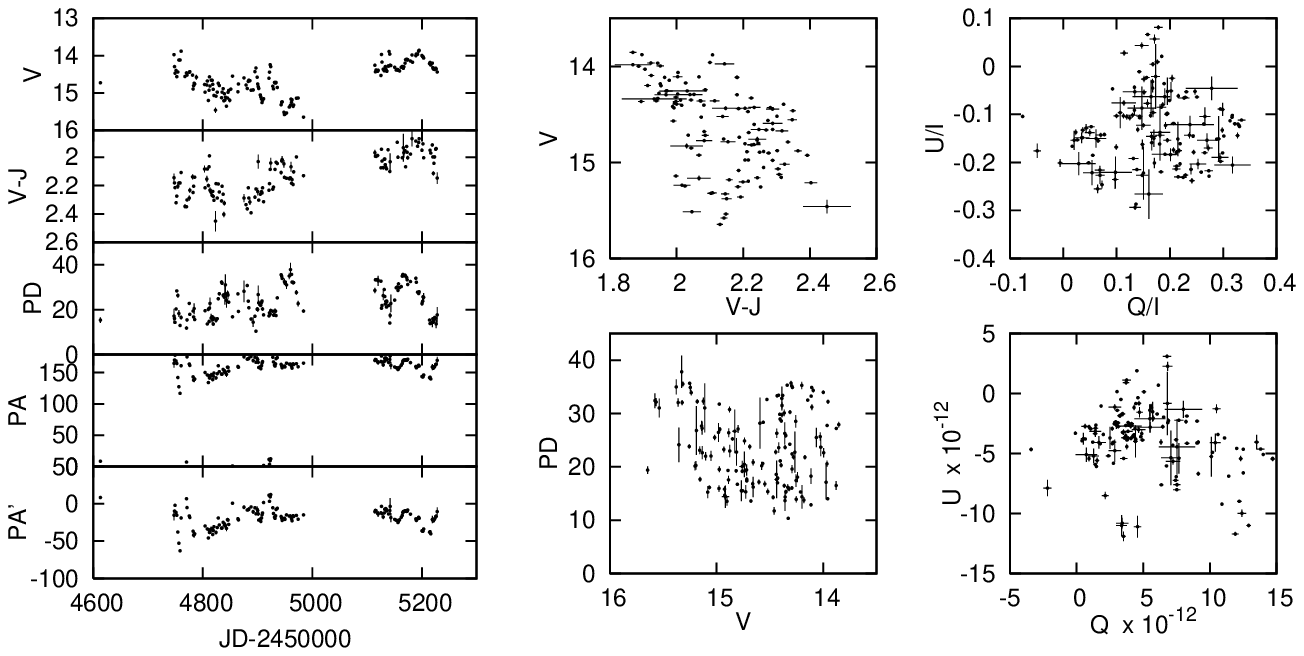}
 \caption{PKS~1510$-$089 (top), PKS~1749$+$096 (middle), and OJ~287
  (bottom).}
\end{figure*}
\begin{figure*}
 \FigureFile(160mm,170mm){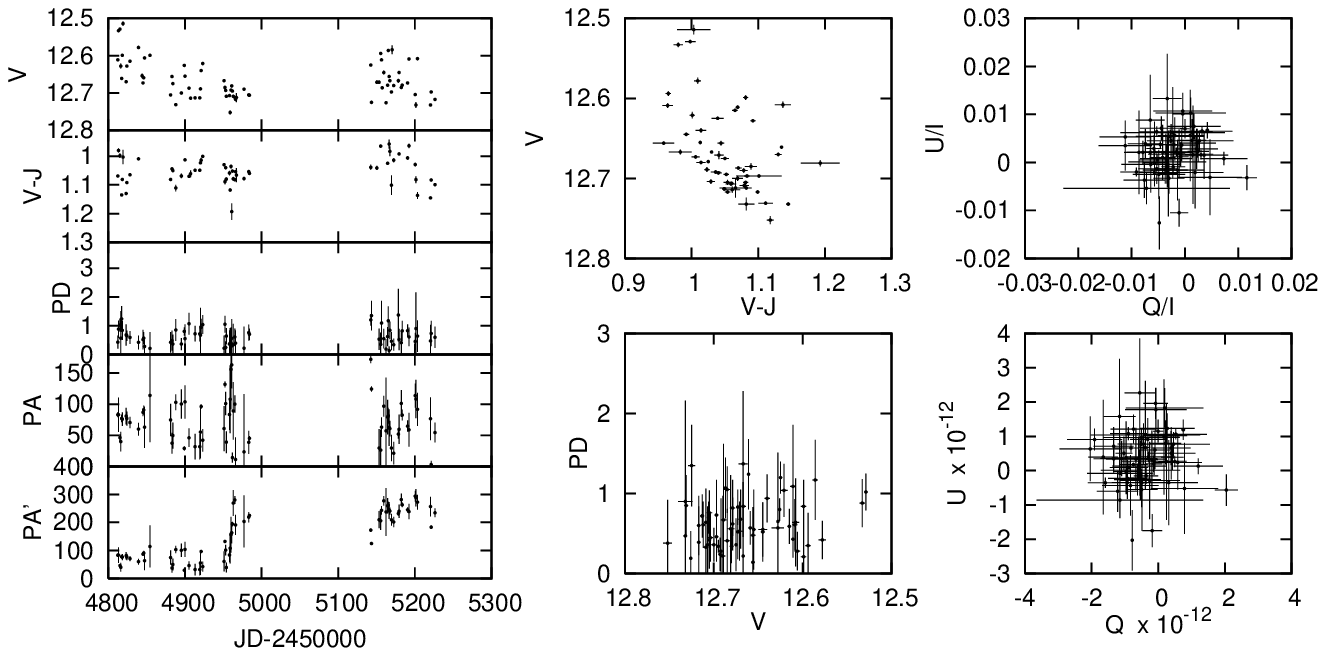}
 \FigureFile(160mm,170mm){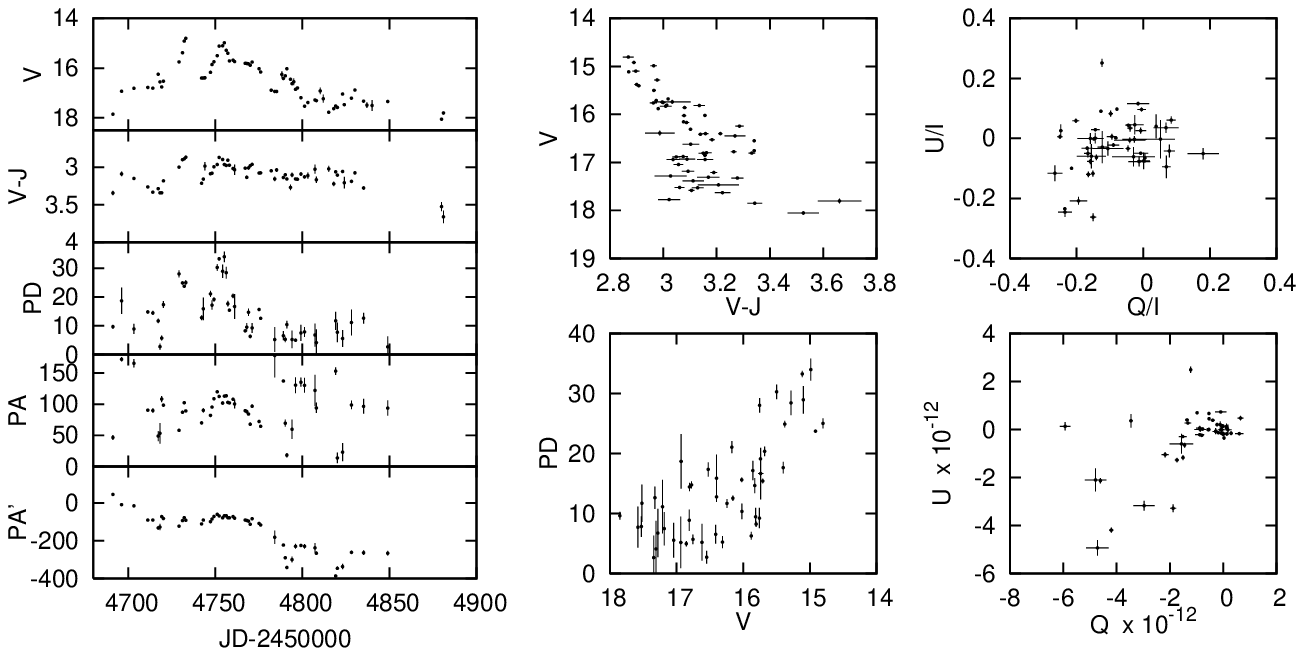}
 \FigureFile(160mm,170mm){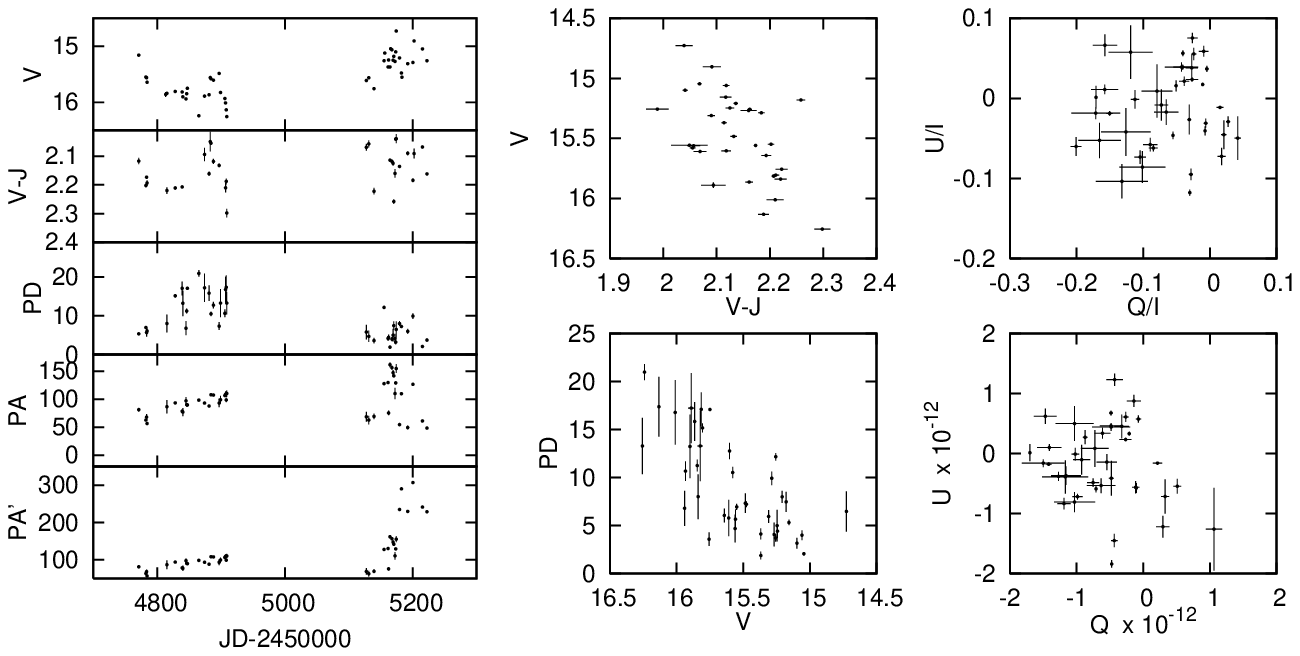}
 \caption{3C~273 (top), AO~0235$+$164 (middle), and OJ~49
  (bottom).}
\end{figure*}
\begin{figure*}
 \FigureFile(160mm,170mm){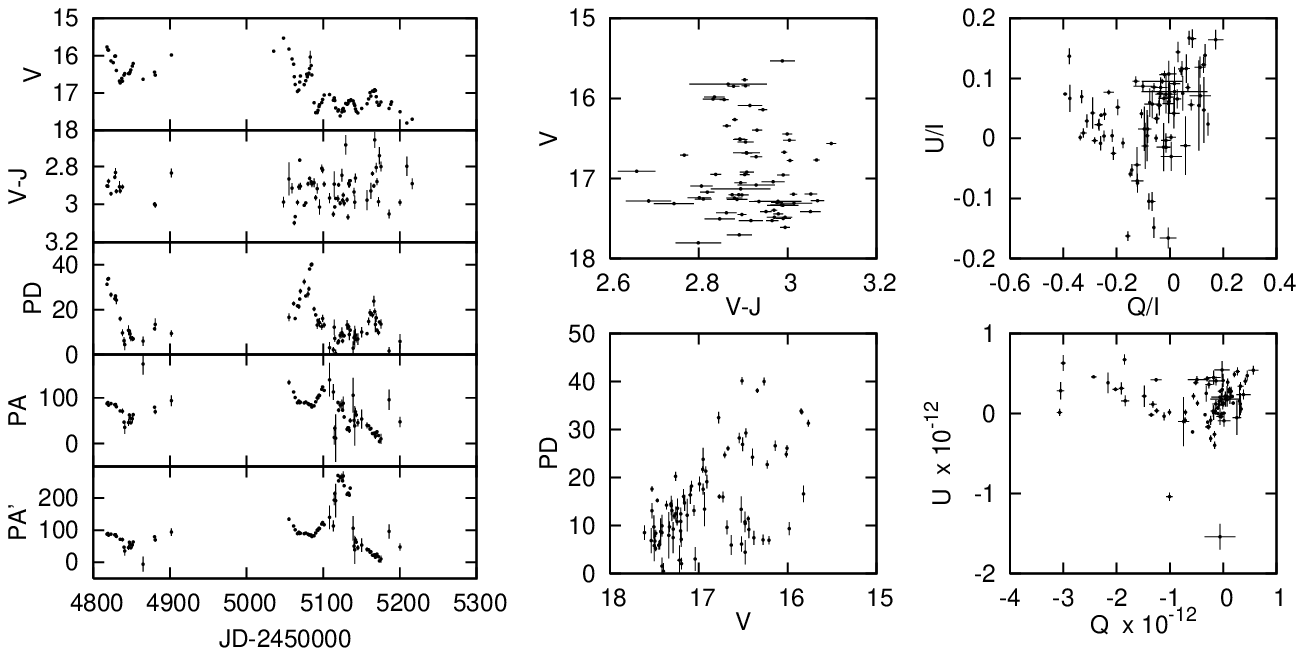}
 \FigureFile(160mm,170mm){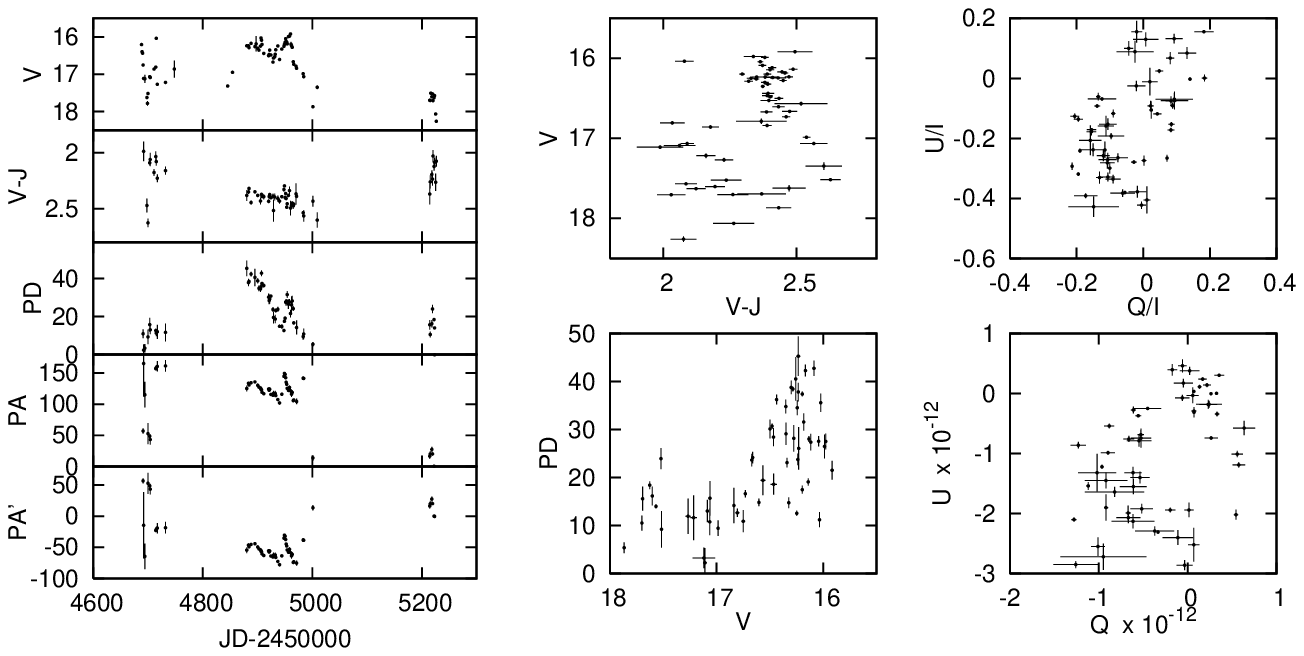}
 \FigureFile(160mm,170mm){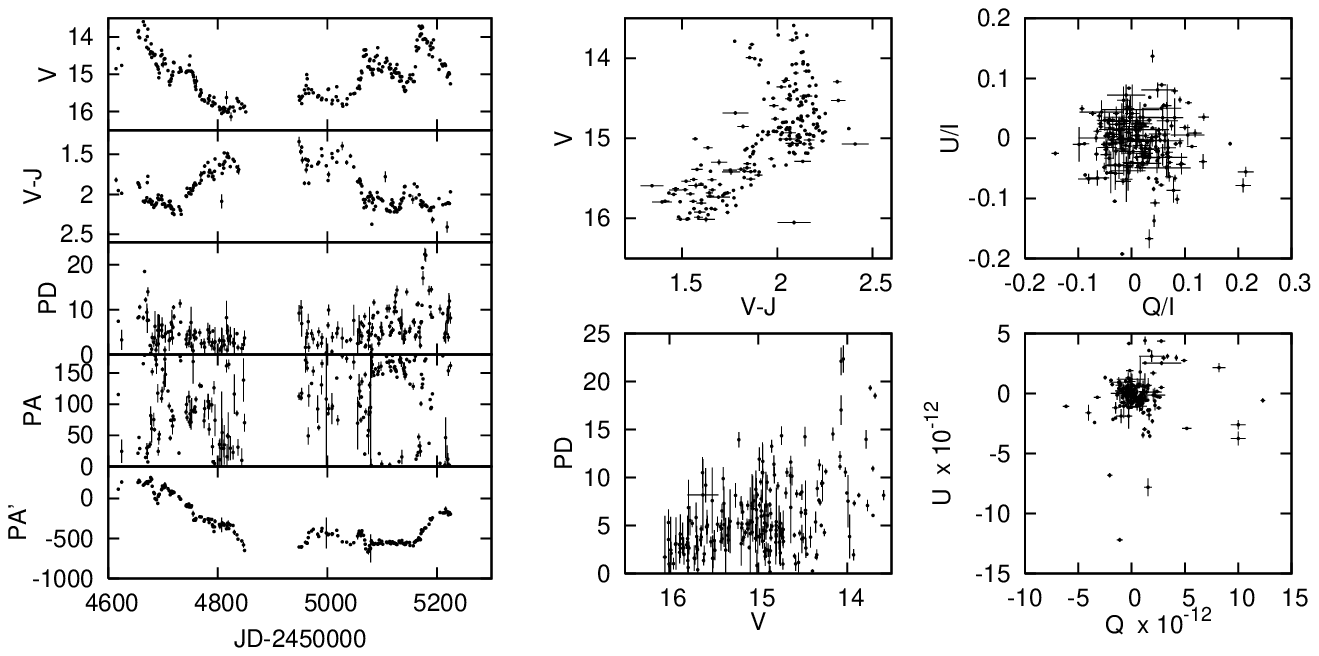}
 \caption{MisV1436 (top), PKS~1502$+$106 (middle), and 3C~454.3
  (bottom).}
\end{figure*}
\begin{figure*}
 \FigureFile(160mm,170mm){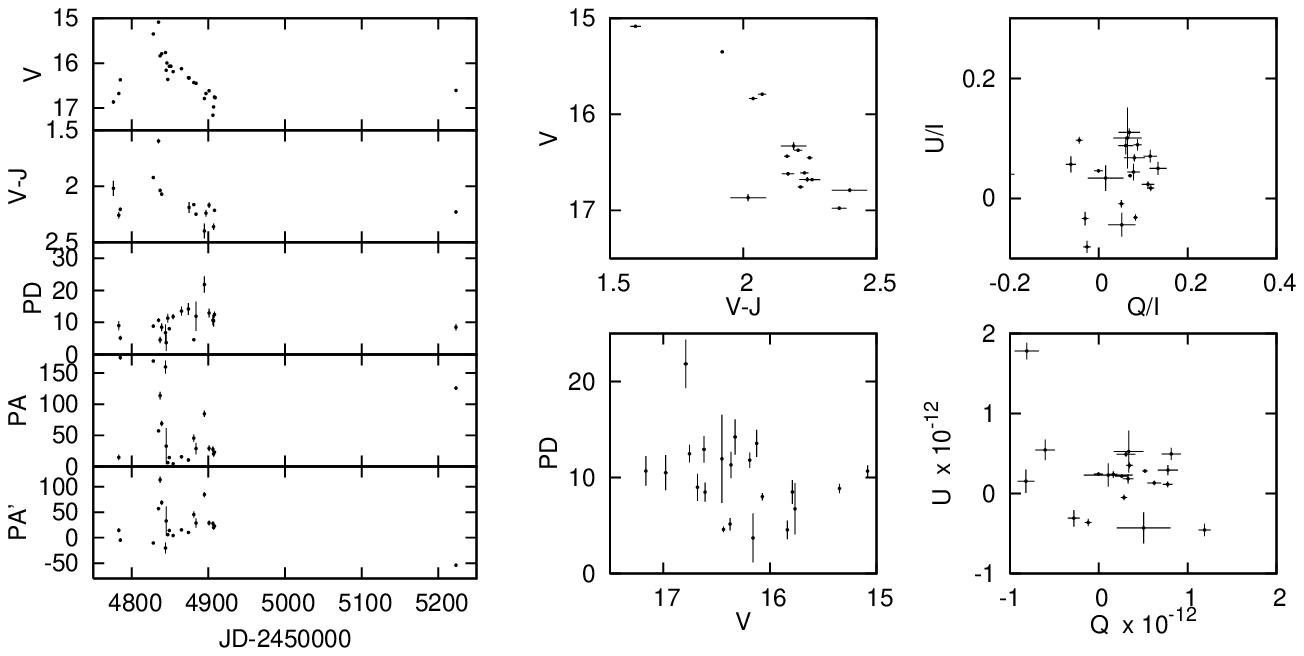}
 \FigureFile(160mm,170mm){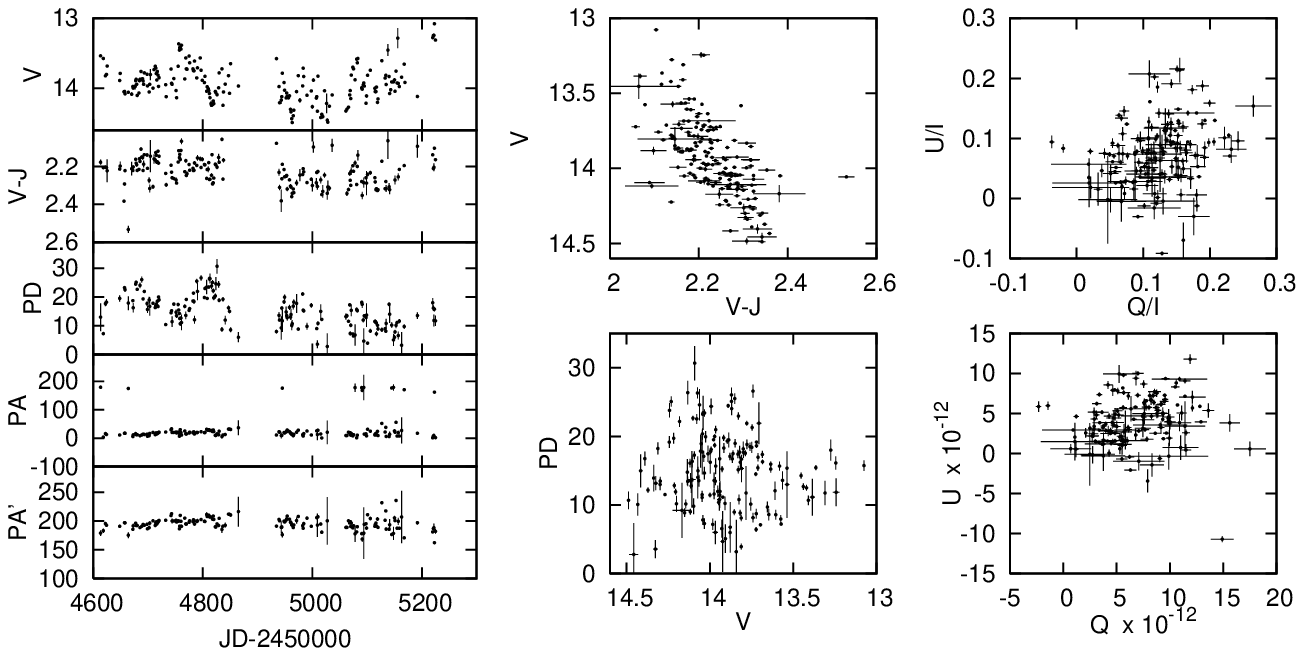}
 \FigureFile(160mm,170mm){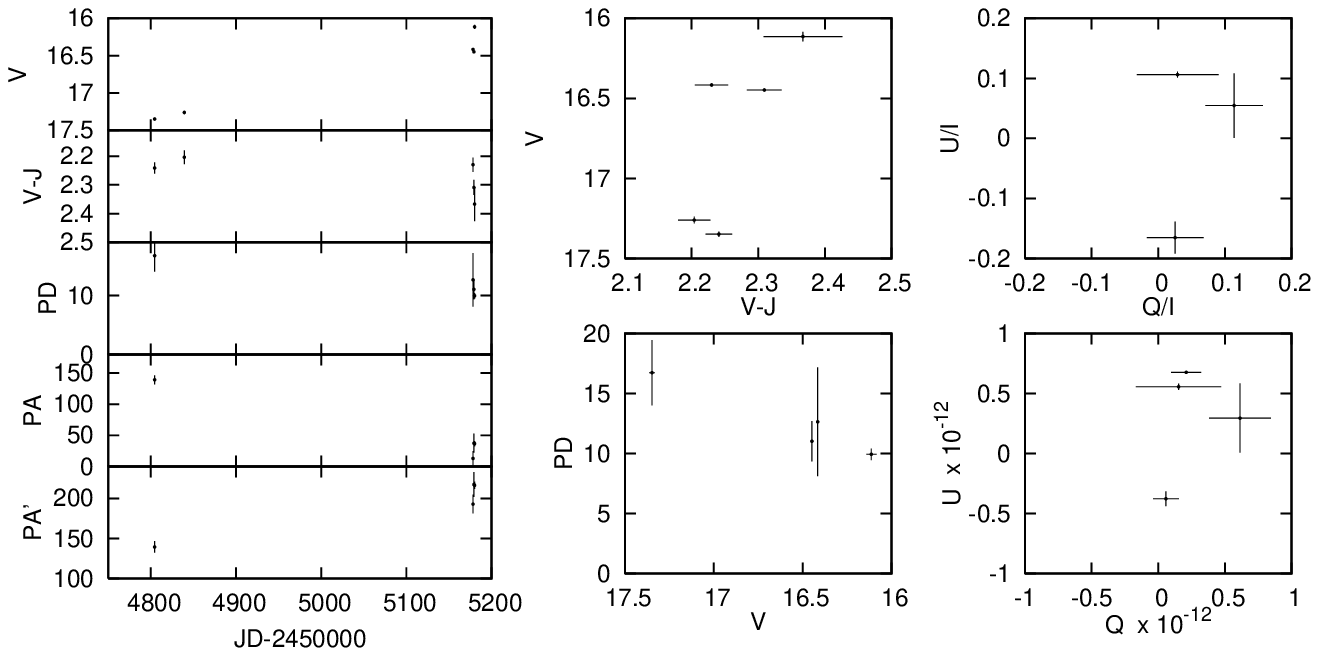}
 \caption{PKS~0754$+$100 (top), BL~Lac (middle), and S4~0954$+$65
  (bottom).}
\end{figure*}
\begin{figure*}
 \FigureFile(160mm,170mm){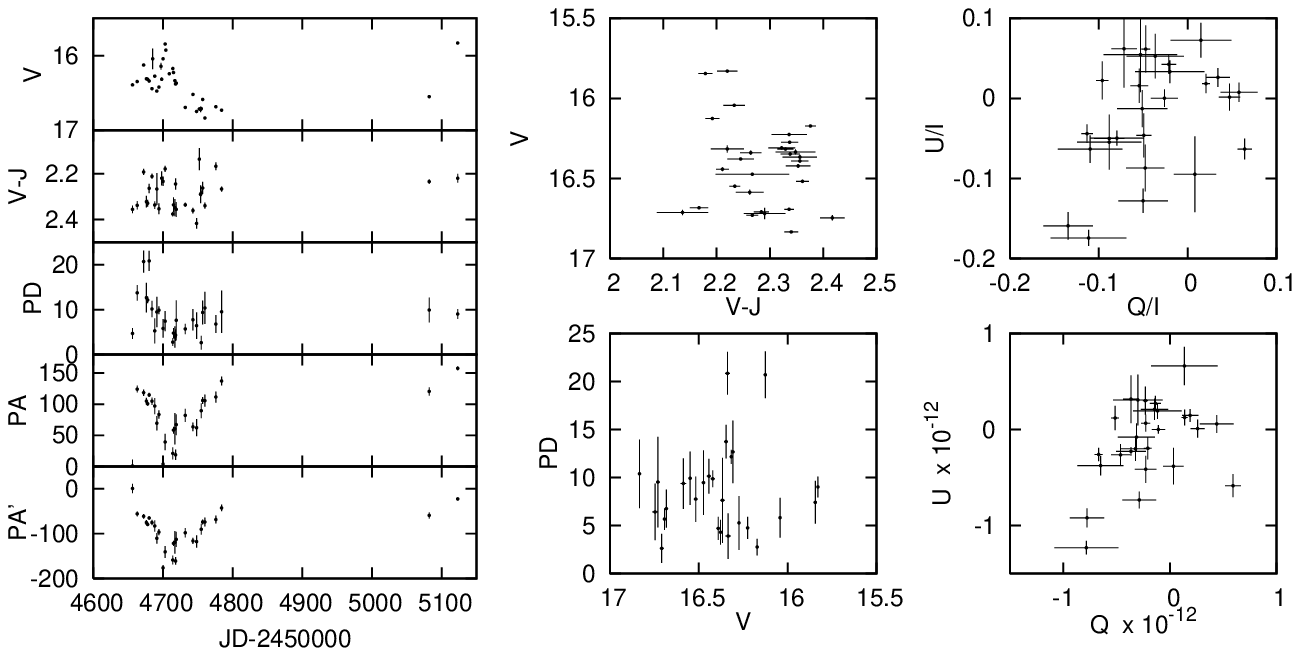}
 \FigureFile(160mm,170mm){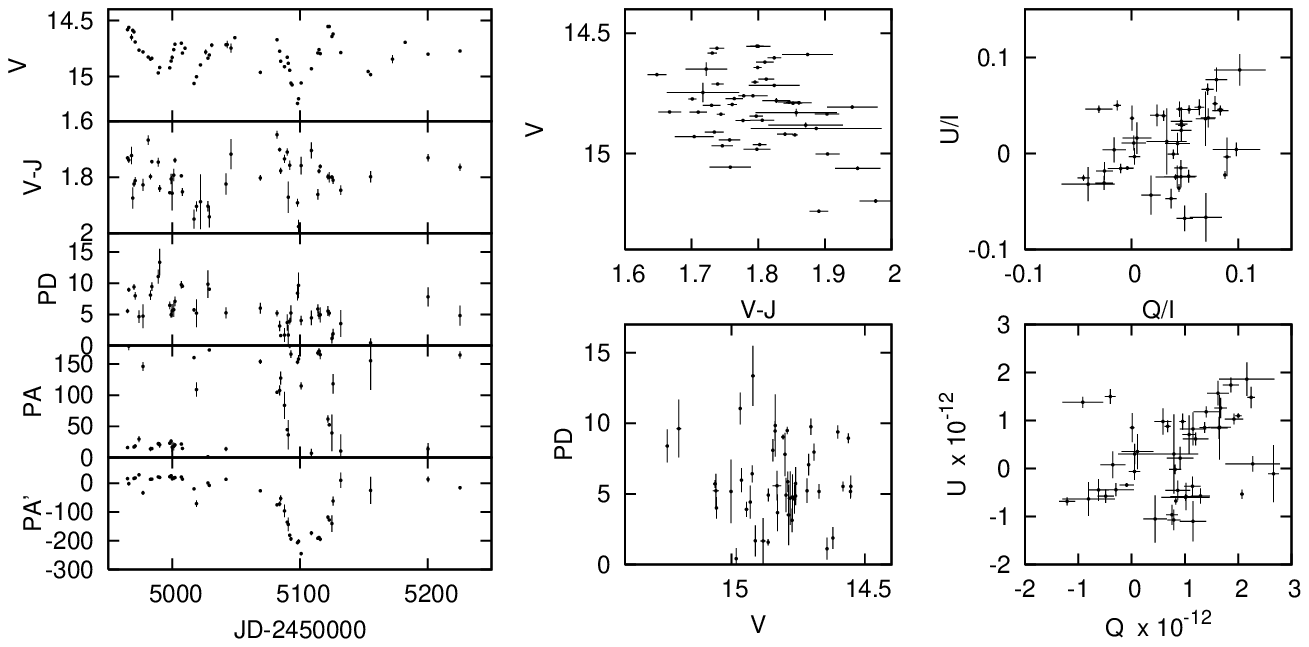}
 \FigureFile(160mm,170mm){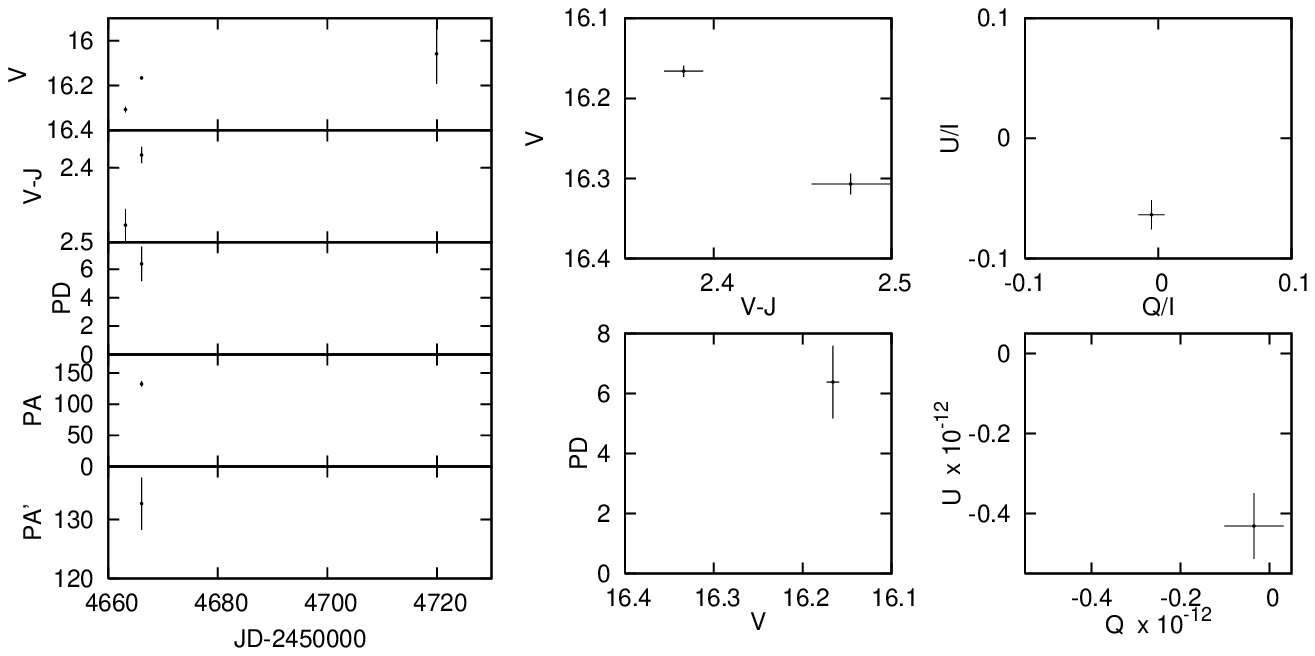}
 \caption{S5~1803$+$784 (top), RX~J1542.8$+$612 (middle), and OQ~530
  (bottom).}
\end{figure*}
\begin{figure*}
 \FigureFile(160mm,170mm){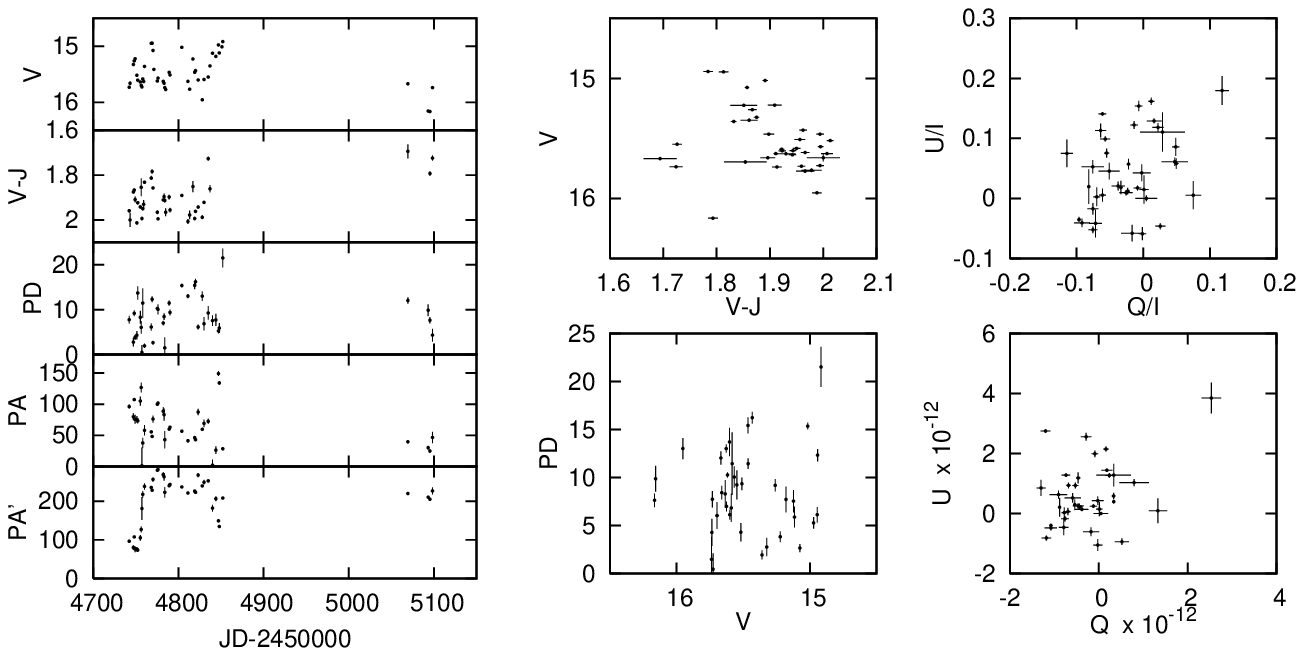}
 \FigureFile(160mm,170mm){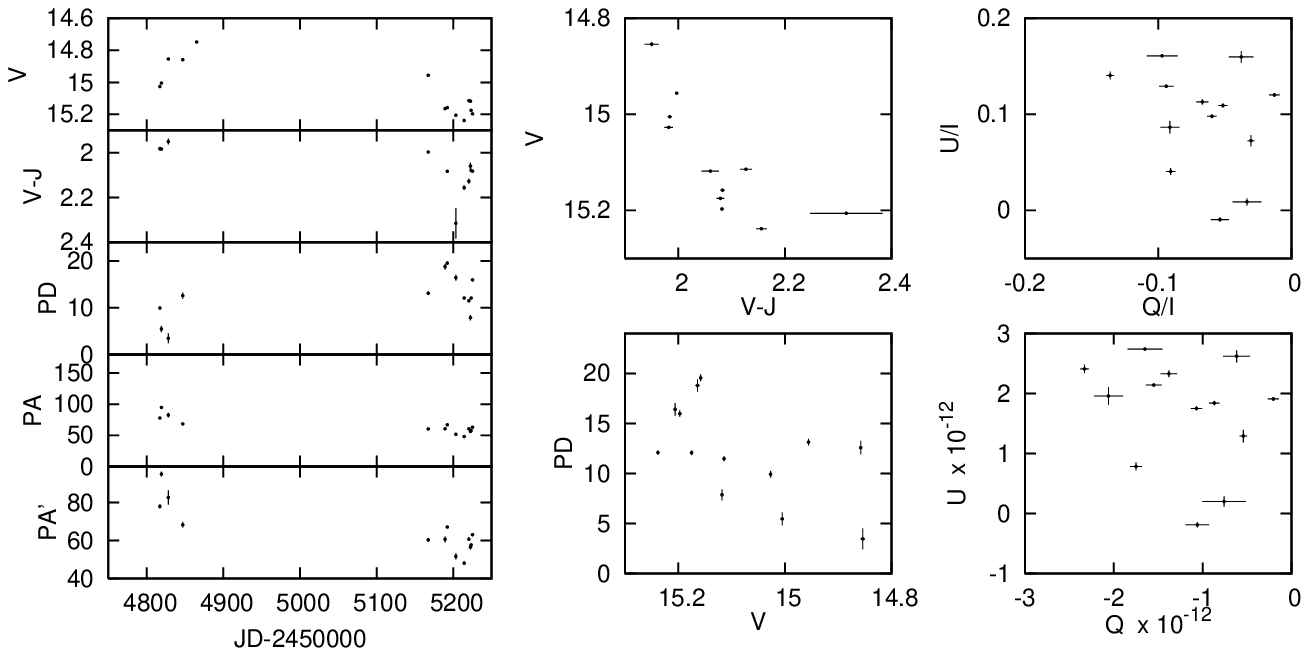}
 \FigureFile(160mm,170mm){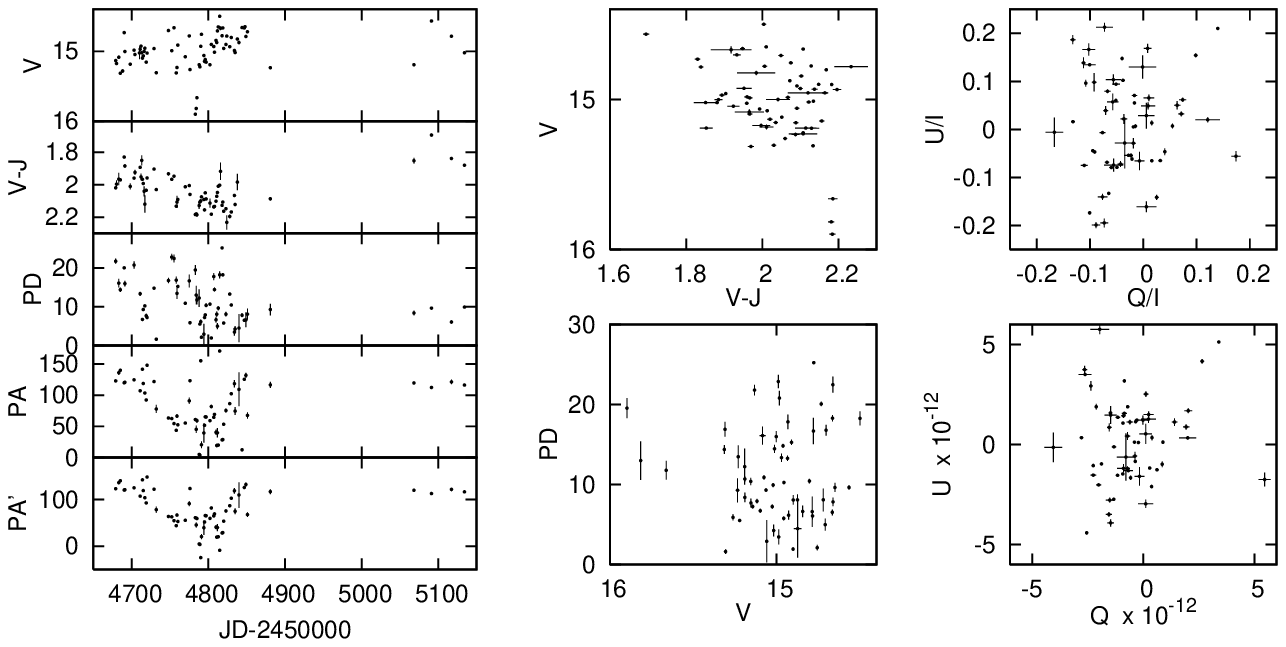}
 \caption{PKS~0048$-$097 (top), ON~231 (middle), and S2~0109$+$224
  (bottom).}
\end{figure*}
\begin{figure*}
 \FigureFile(160mm,170mm){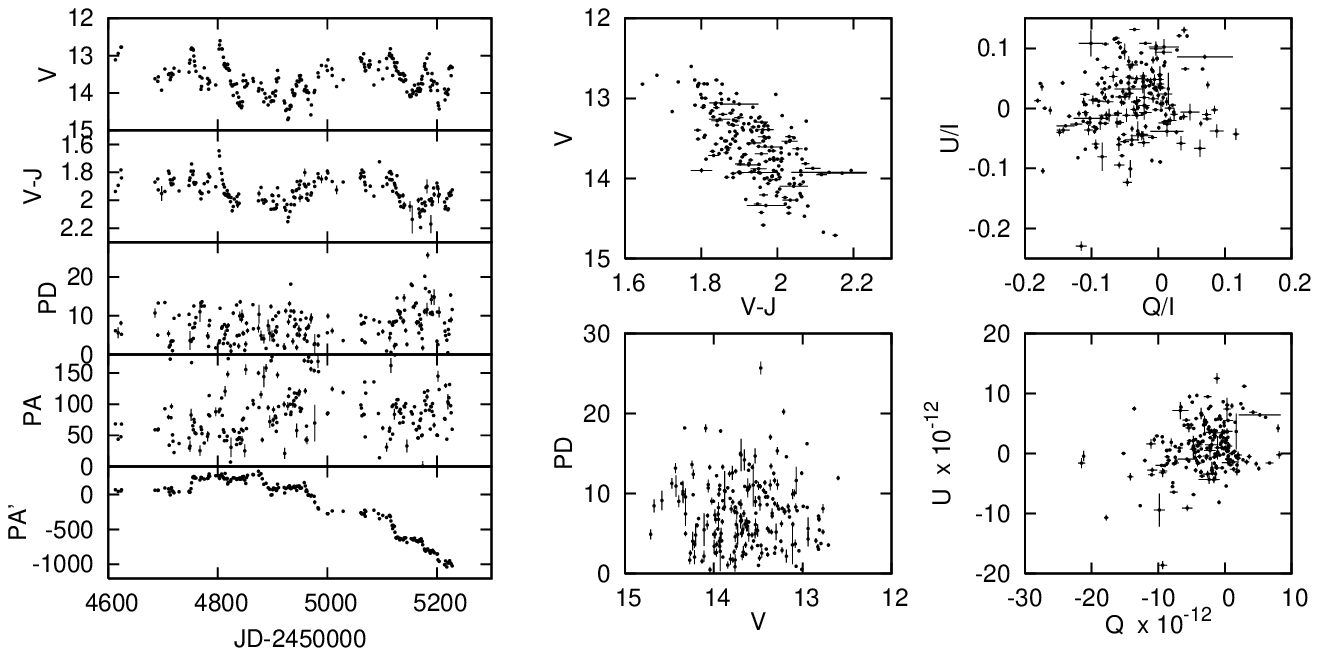}
 \FigureFile(160mm,170mm){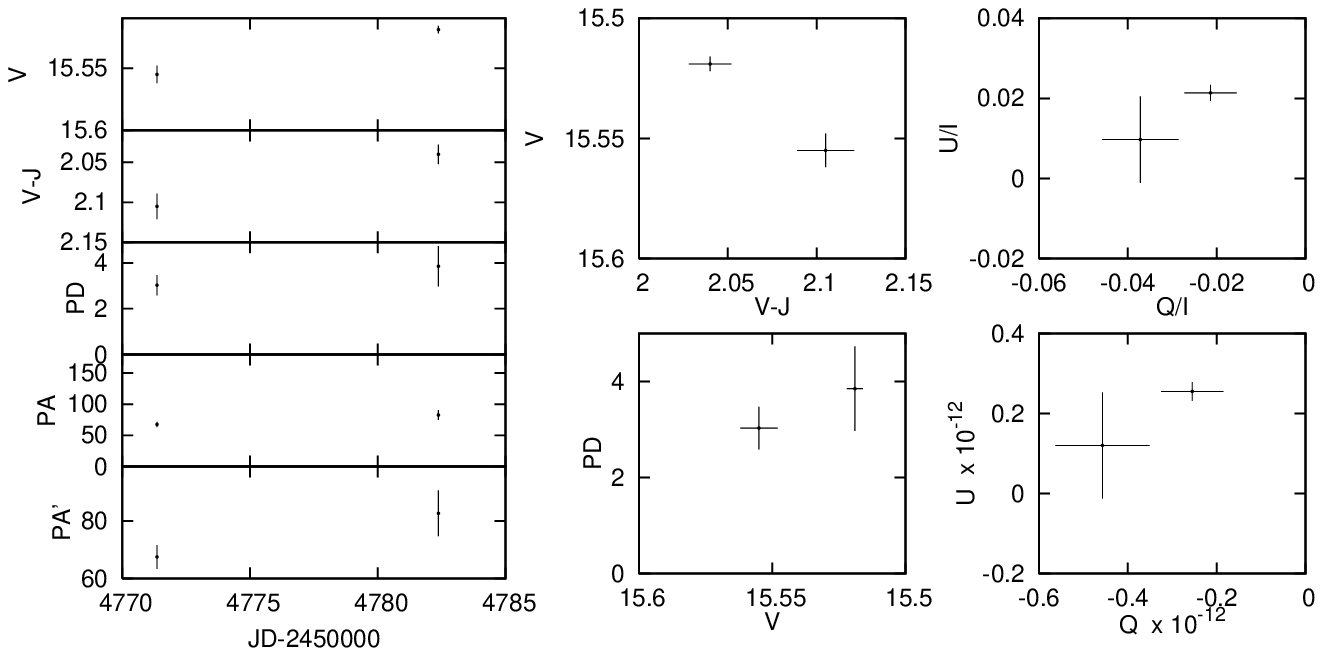}
 \FigureFile(160mm,170mm){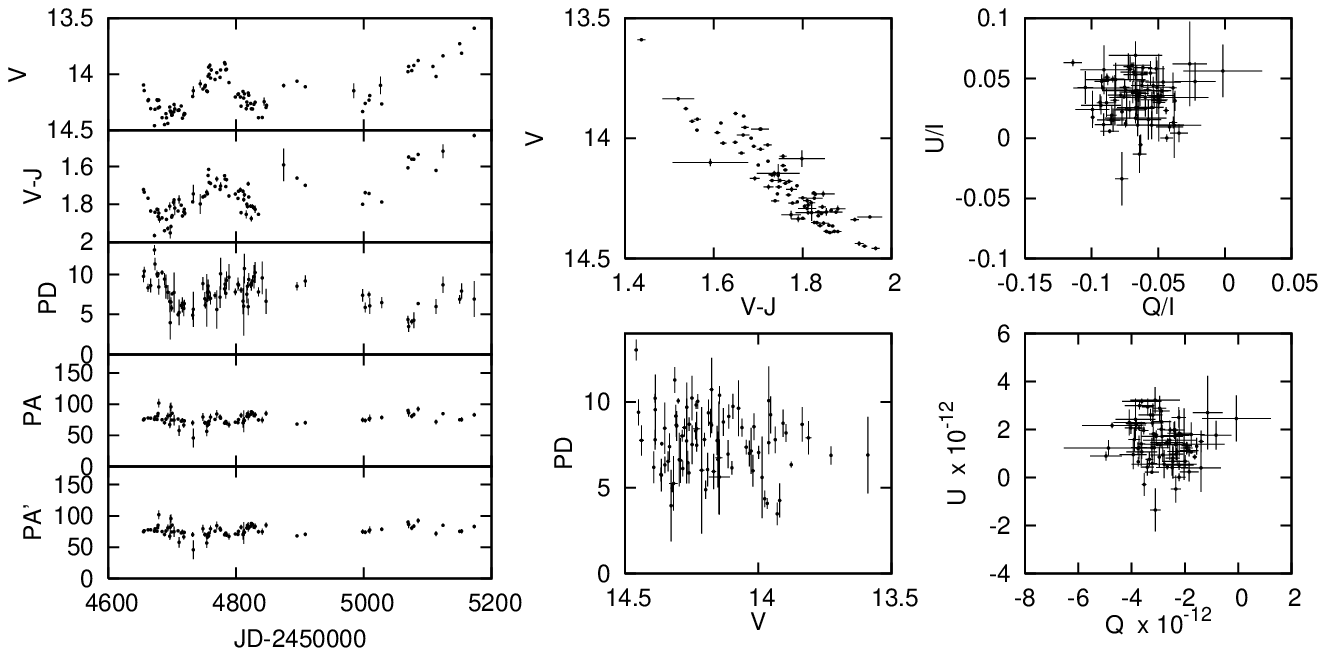}
 \caption{S5~0716$+$714 (top), 3EG~1052$+$571 (middle), and 3C~371
  (bottom).}
\end{figure*}
\begin{figure*}
 \FigureFile(160mm,170mm){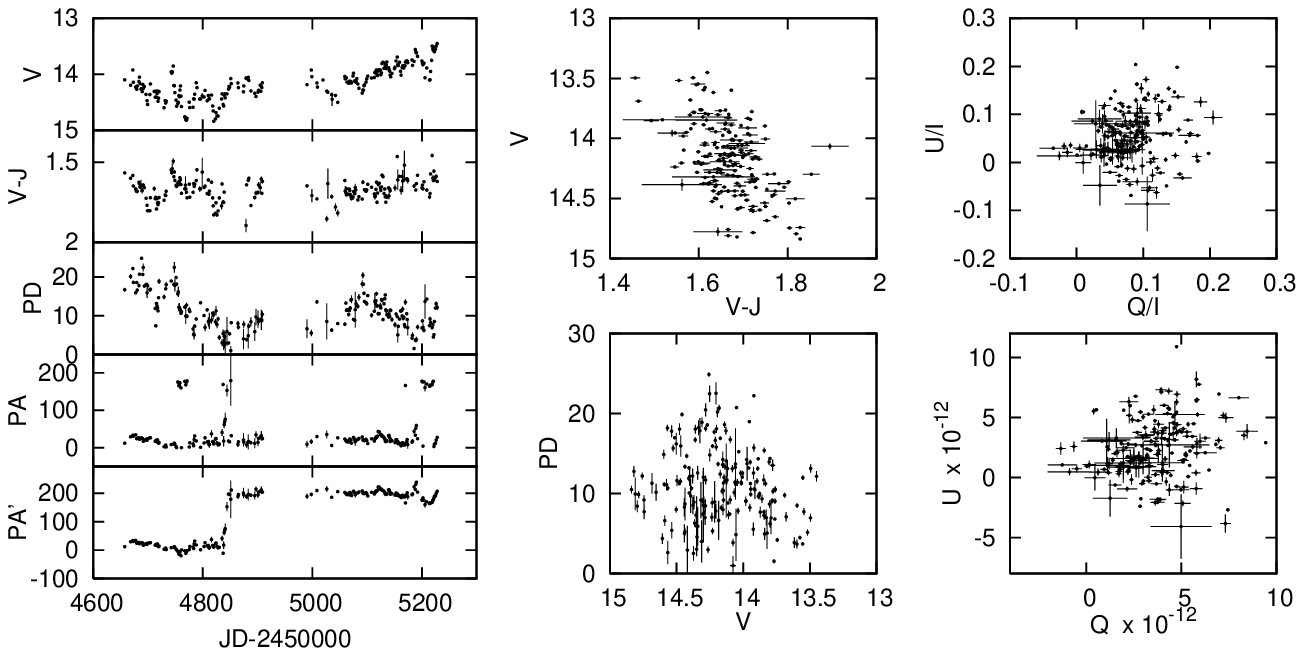}
 \FigureFile(160mm,170mm){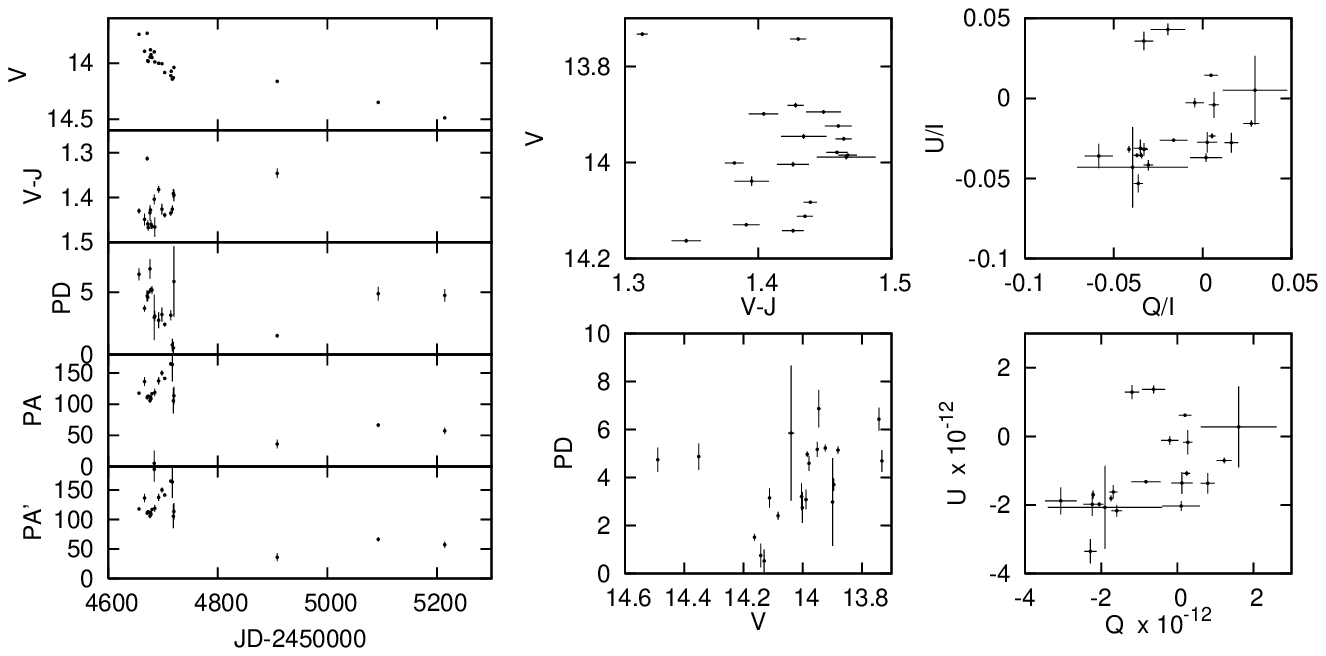}
 \FigureFile(160mm,170mm){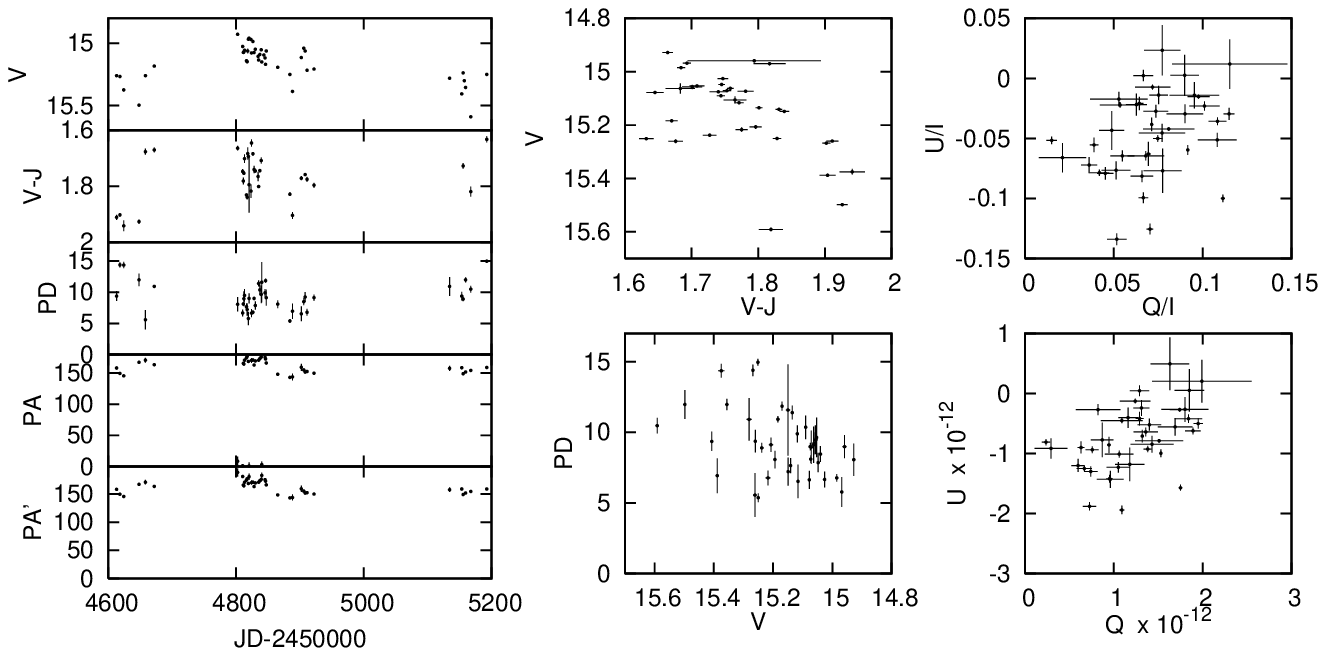}
 \caption{3C~66A (top), PG~1553$+$113 (middle), and ON~325
  (bottom).}
\end{figure*}
\begin{figure*}
 \FigureFile(160mm,170mm){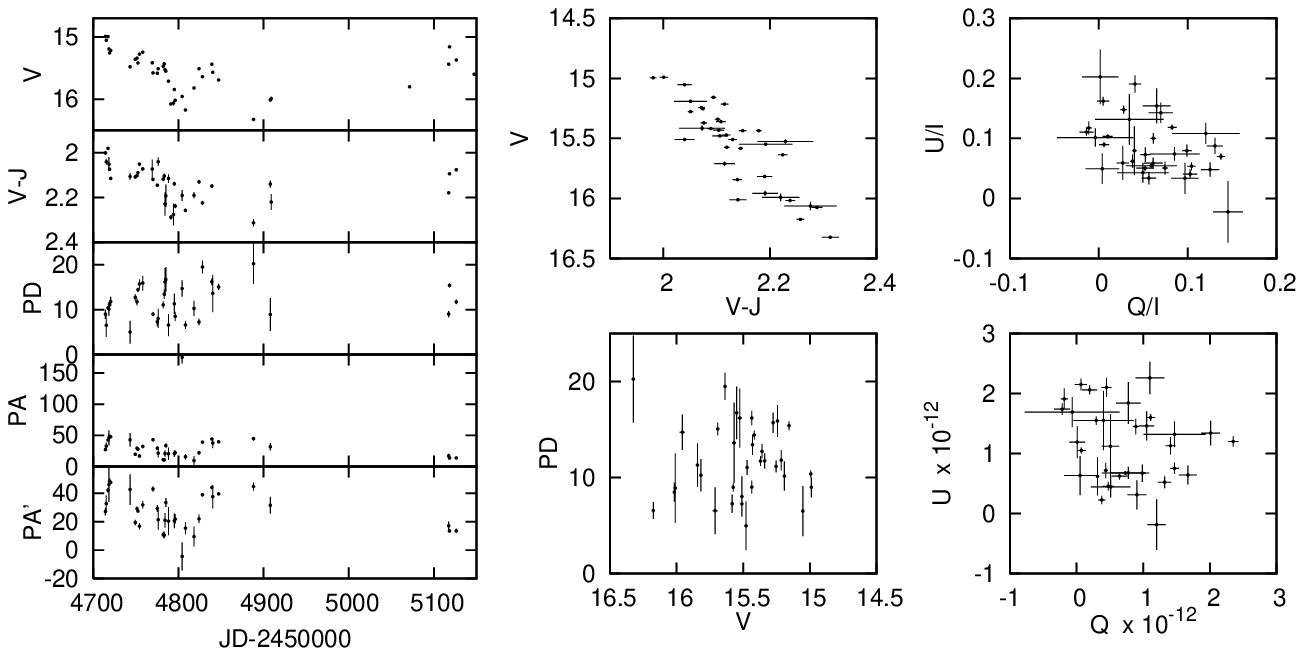}
 \FigureFile(160mm,170mm){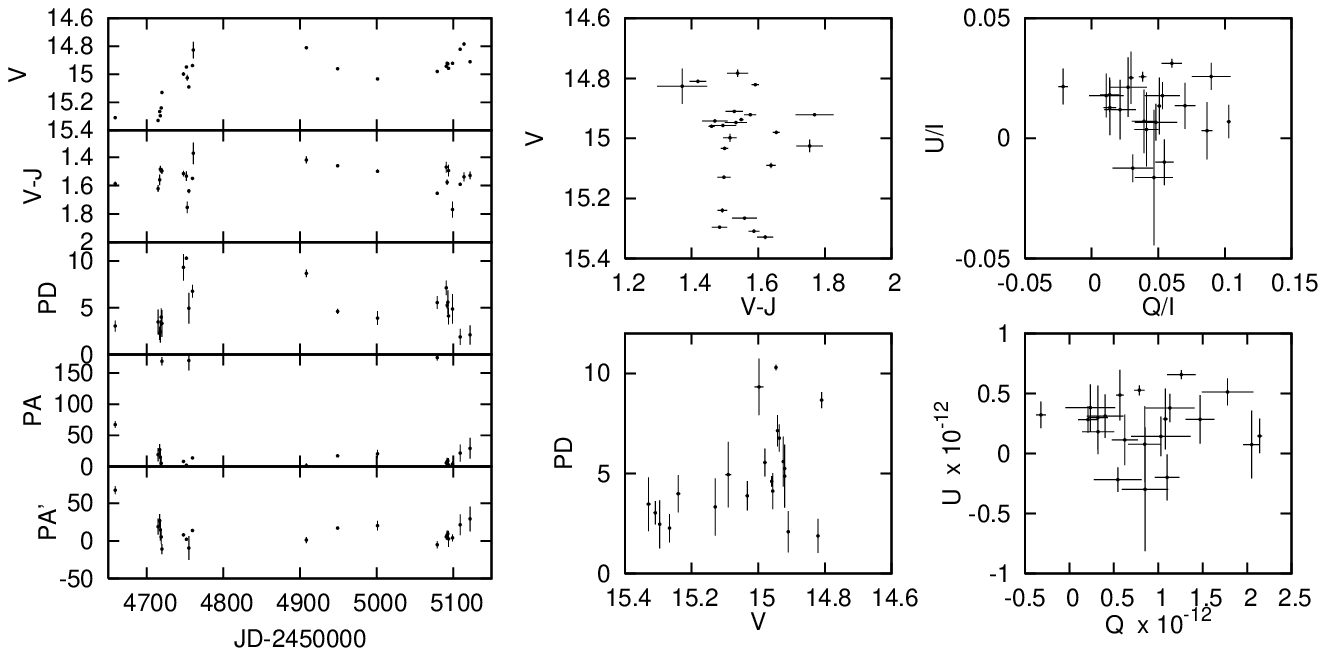}
 \FigureFile(160mm,170mm){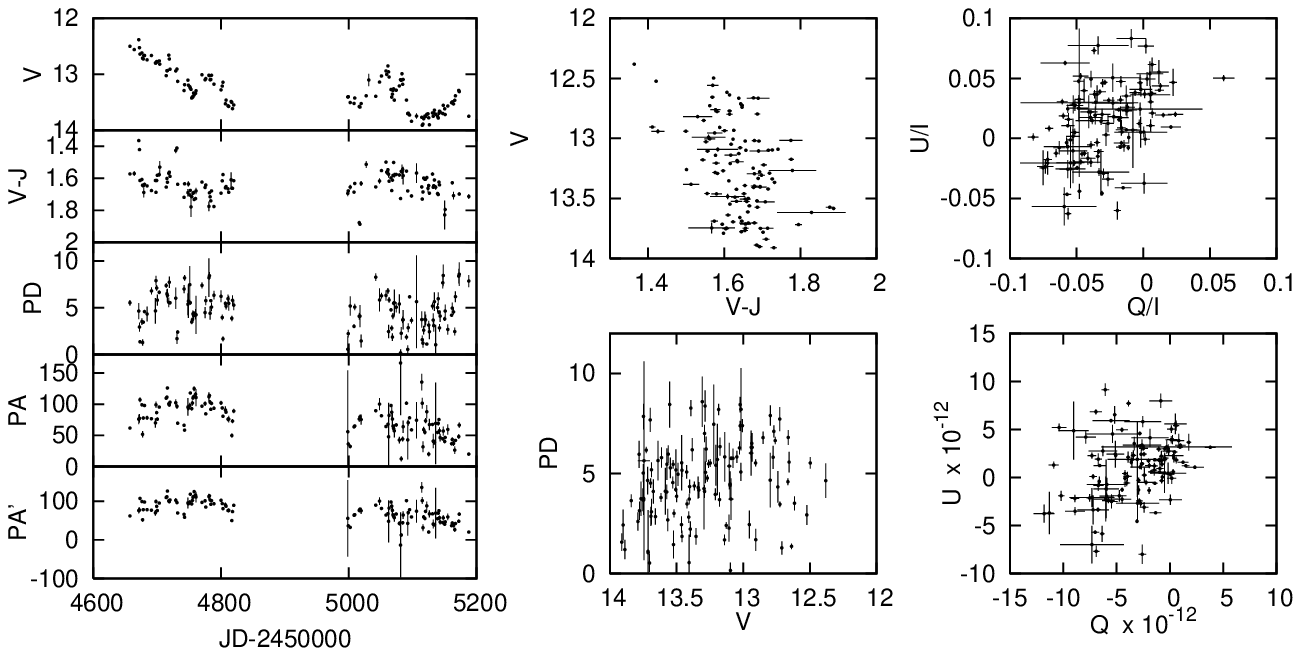}
 \caption{PKS~0422$+$004 (top), H~1722$+$119 (middle), and
  PKS~2155$-$304 (bottom).}
\end{figure*}
\begin{figure*}
 \FigureFile(160mm,170mm){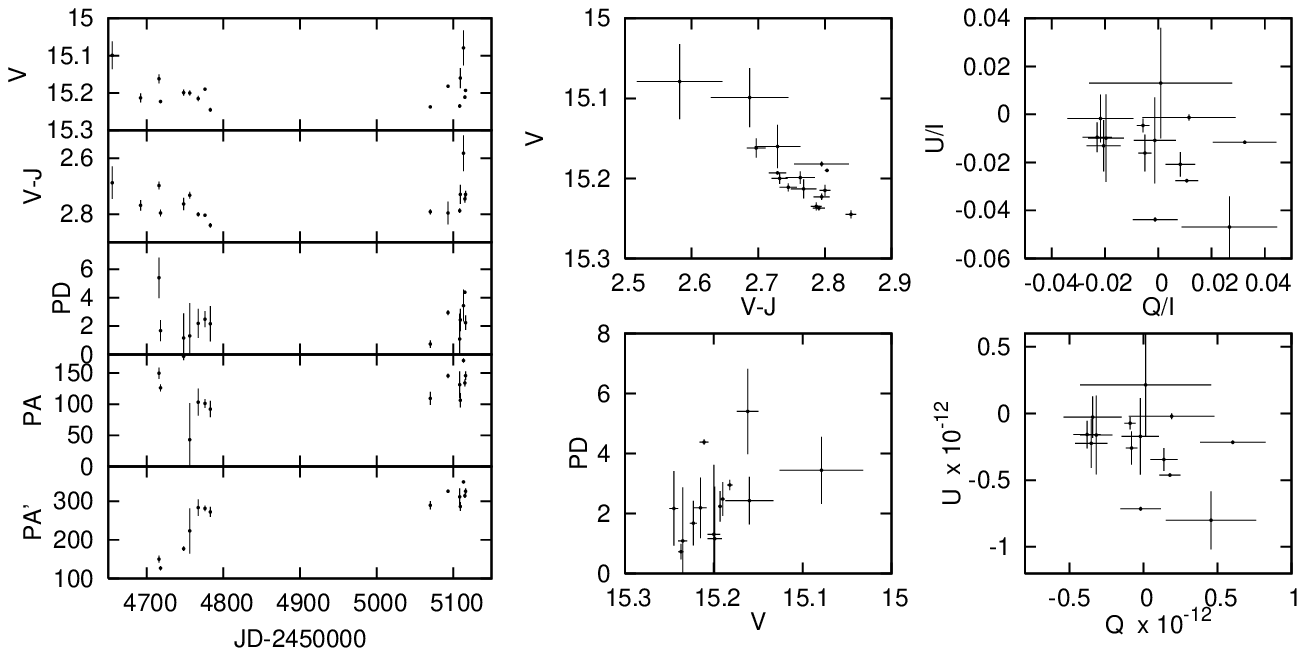}
 \FigureFile(160mm,170mm){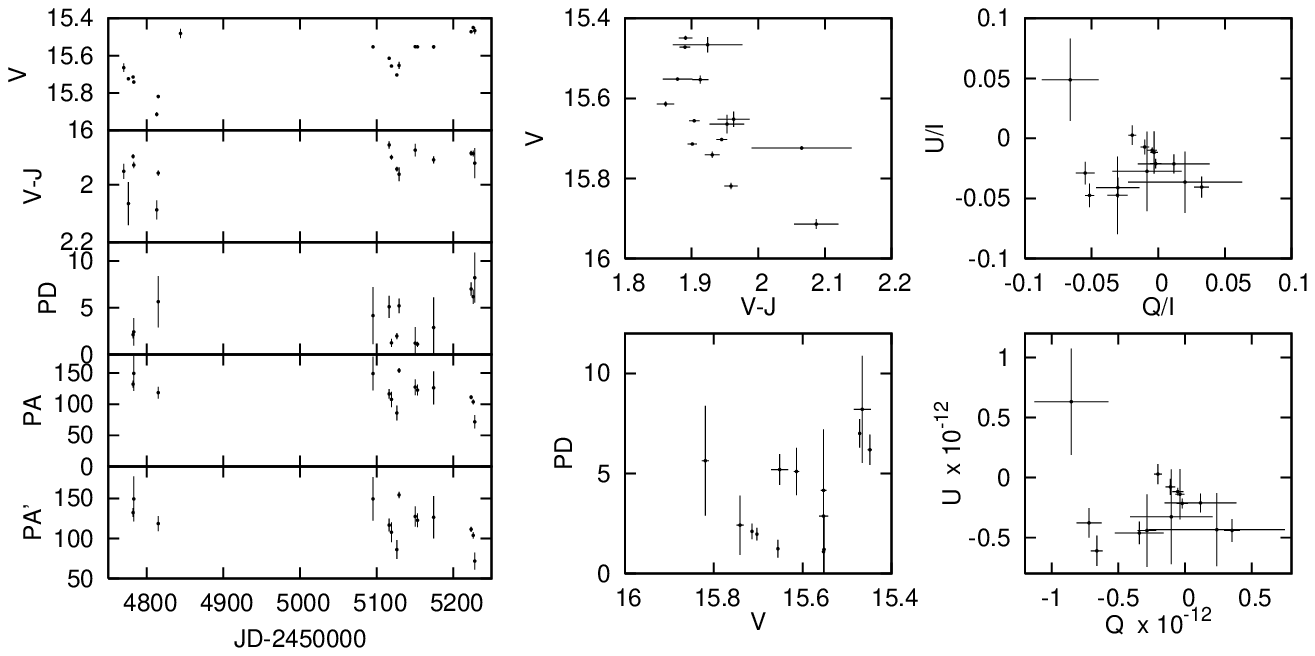}
 \FigureFile(160mm,170mm){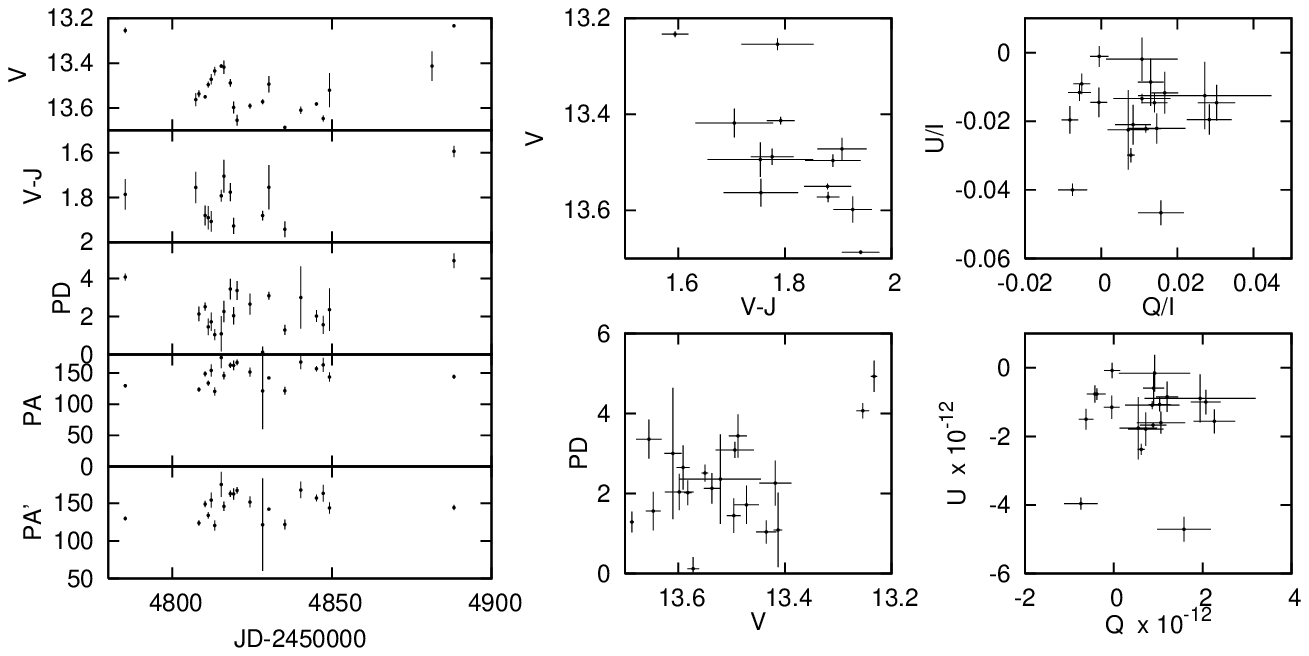}
 \caption{1ES~2344$+$514 (top), 1ES~0806$+$524 (middle), and Mrk~421 
  (bottom).}
\end{figure*}
\begin{figure*}
 \FigureFile(160mm,170mm){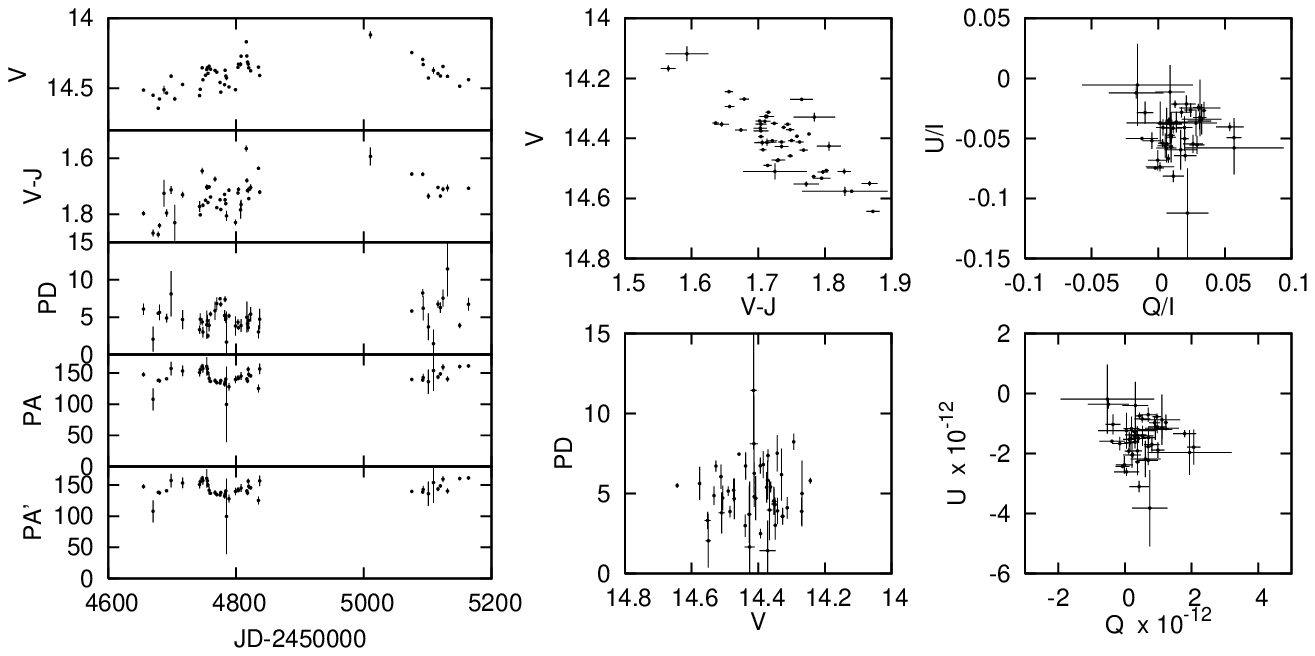}
 \FigureFile(160mm,170mm){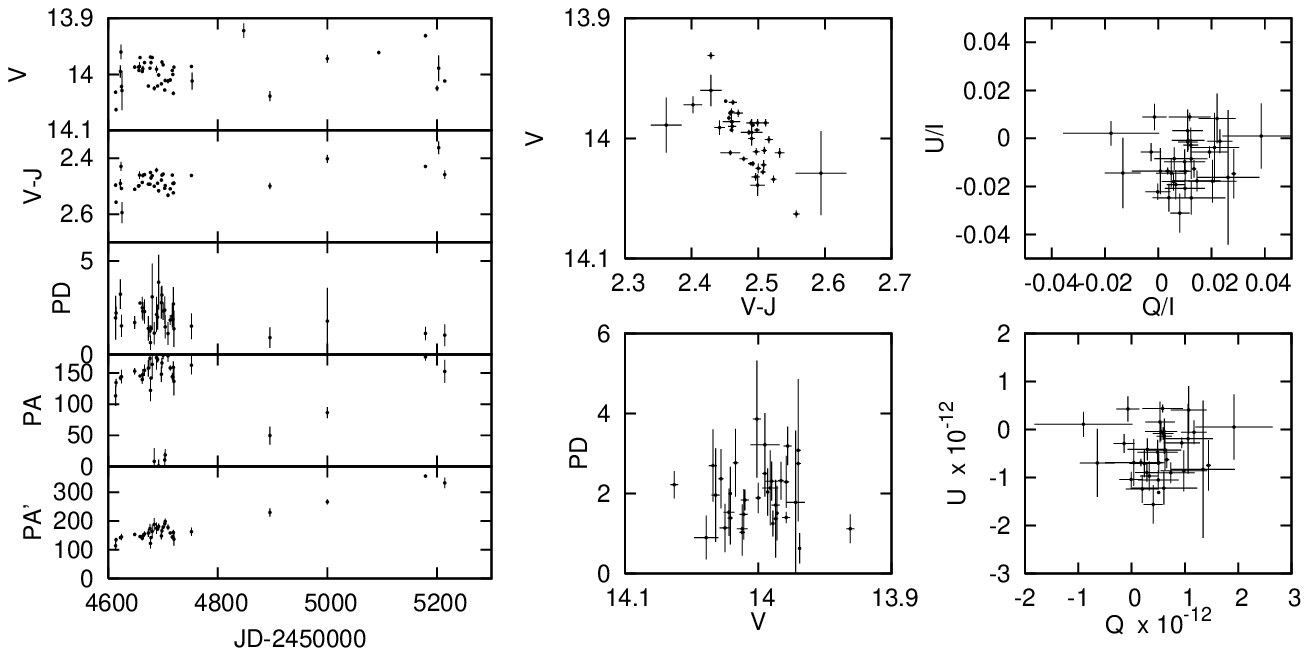}
 \FigureFile(160mm,170mm){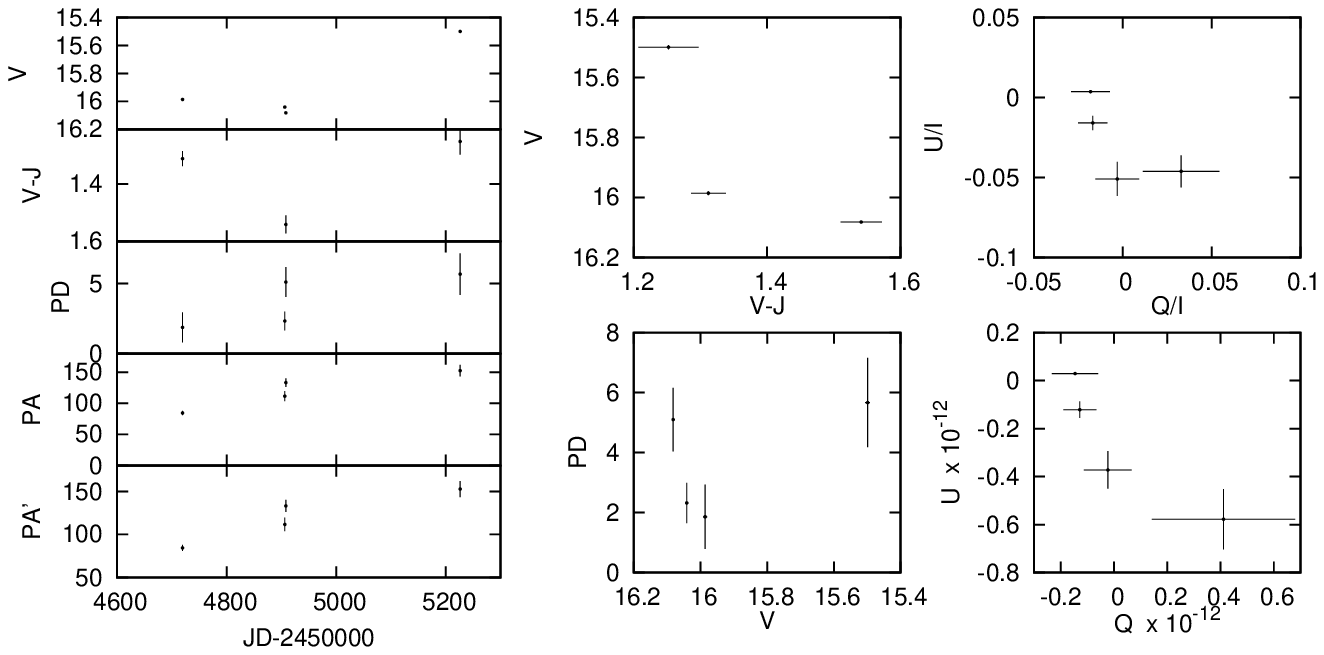}
 \caption{1ES~1959$+$650 (top), Mrk~501 (middle), and 1ES~0647$+$250 
  (bottom).}
\end{figure*}
\begin{figure*}
 \FigureFile(160mm,170mm){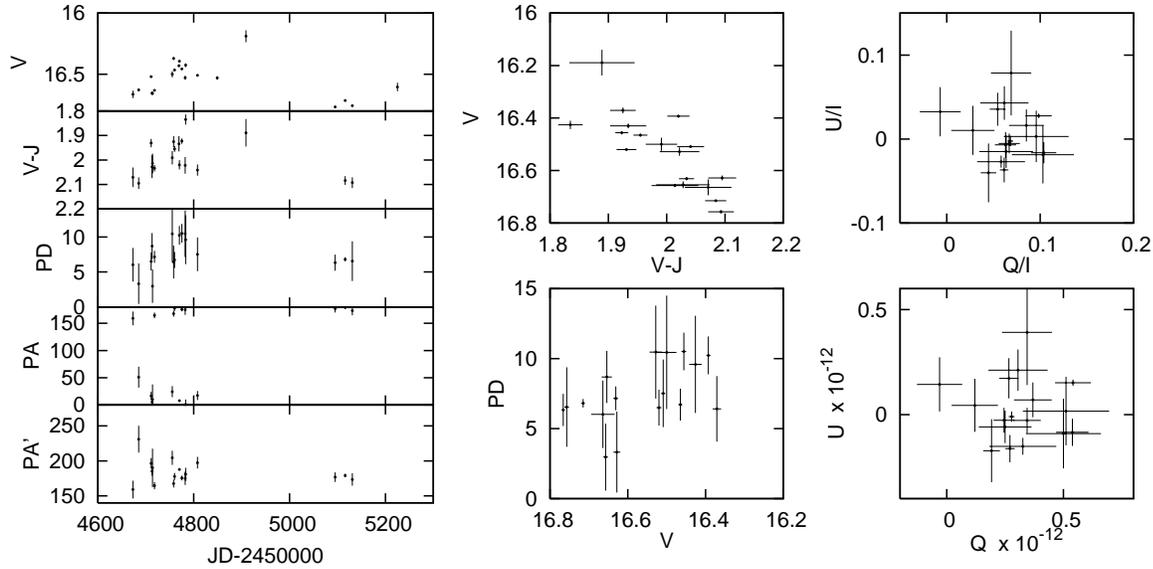}
 \caption{1ES~0323$+$022.}
\end{figure*}


\end{document}